\documentclass{aa}
\newcommand{\ergss}{\mathrm{erg~s^{-1}}}
\newcommand{\ergscms}{\mathrm{erg~cm^{-2}~s^{-1}}}

\newcommand{\xmmnewton}{\textit{XMM-Newton} }
\newcommand{\xmm}{\textit{XMM} }
\newcommand{\obs}[1]{\href{https://nxsa.esac.esa.int/nxsa-web/\#obsid=#1}{#1}}
\newcommand{\NRegTot}{60,127 }
\newcommand{\NRegUnique}{32,247 }
\newcommand{\NRegUniqueFsig}{4,083 }

\newcommand{\NBadObsIDs}{196}

\newcommand{\tSig}{($3 \sigma$) }
\newcommand{\fSig}{($5 \sigma$) }
\newcommand{\extras}{EXTraS }

\usepackage{float}
\usepackage{graphicx}
\usepackage{txfonts}
\usepackage[breaklinks, colorlinks, allcolors=blue]{hyperref}
\usepackage{longtable}
\usepackage{booktabs}
\usepackage{listings}
\usepackage{xcolor}
\usepackage{natbib}
\usepackage{epstopdf}
\usepackage{soul}
\usepackage{multirow,array}
\usepackage[normalem]{ulem}
\bibpunct{(}{)}{;}{a}{}{,} 

\begin{document} 
   \title{The EXOD search for faint transients in XMM-Newton observations}
   \subtitle{Part II. }
   \author{N. Khan\inst{1} \and 
   E. Quintin\inst{1,2} \and
   N. A. Webb\inst{1} \and
   R. Webbe\inst{1} \and
   M. Gupta\inst{3} \and
   I. Pastor-Marazuela\inst{4} \and
   F. Castellani\inst{5} \and
   A. D. Schwope\inst{6} \and
   I. Traulsen\inst{6} \and
   A. Nebot\inst{7}
   }
   \institute{Institute for Research in Astrophysics and Planetology (IRAP), CNRS, Toulouse, 31400, France \and 
   European Space Agency (ESA), European Space Astronomy Centre (ESAC), Camino Bajo del Castillo s/n, 28692 Villanueva de la Cañada, Madrid, Spain \and
   Astronomical Institute, Academy of Sciences, Boční II 1401, CZ-14131 Prague, Czech Republic \and
   Jodrell Bank Centre for Astrophysics, University of Manchester, Oxford Road, Manchester M13 9PL, UK \and
   Telespazio France, Space Systems and Operations (SSO), 31110, Toulouse \and
   Leibniz-Institut für Astrophysik Potsdam (AIP), An der Sternwarte 16, 14482 Potsdam, Germany \and
   Université de Strasbourg, CNRS, Observatoire astronomique de Strasbourg, UMR 7550, F-67000, Strasbourg, France
   }
   \date{}
 
  \abstract
   {The \xmmnewton observatory has accumulated a vast archive of over 17,000 X-ray observations over the last 25 years.
   However, the standard data processing pipelines may fail to detect certain types of
   transient X-ray sources due to their short-lived or dim nature. Identifying these transient sources 
   is important for understanding the full range of temporal X-ray behaviour, as well as 
   understanding the types of sources that could be routinely detected by future missions such as \textit{Athena}.}
   {This work aims to reprocess \xmmnewton archival observations using newly developed dedicated software in order to identify neglected and missed
   transient X-ray sources that were not detected by the existing pipeline.}
   {We use a new approach that builds upon previous methodologies, by transforming event lists
   into data cubes, which are then searched for transient variability in short time windows. Our method enhances the
   detection capabilities in the Poisson regime by accounting for the statistical properties
   of sparse count rates, and allowing for transient search in previously discarded periods of high background activity.}
   {Our reprocessing efforts identified 32,247 variable sources at the 3-sigma level and
   4,083 sources at the 5-sigma level in 12,926 XMM archival observations.
   We highlight four noteworthy sources: A candidate quasi-periodic eruption (QPE),
   a new magnetar candidate, a previously undetected Galactic hard X-ray burst and
   a possible X-ray counterpart to a Galactic radio pulsar.}
   {Our method demonstrates a new, fast, and effective way to process event list data from \textit{XMM-Newton},
   which is efficient in finding rapid outburst-like or eclipsing behaviour. This technique can be adapted for use with future telescopes,
   such as \textit{Athena}, and can be generalised to other photon counting instruments operating in the low-count Poisson regime.}
   \maketitle
\section{Introduction}\label{sec:intro}
Since its launch in December 1999, \xmmnewton \citep{2001_Jansen_A&A...365L...1J} has been detecting X-ray photons in the 0.2 - 12.0 keV energy range using the European Photon Imaging Camera (EPIC).
The most recent \xmm serendipitous source catalogue \citep{2020_Webb_A&A...641A.136W}, 4XMM-DR14, contains 13,864 archival observations,
1,035,832 detections of which 21,570 ($\sim$2\%) are flagged to be intra-observation variable.
Since many sources are detected more than once, these one million detections
correspond to 692,109 unique sources, of which 6,311 ($\sim$0.9 \%) are flagged as variable (this lower percentage is due to selection bias, as variable sources often have multiple observations). 

The \xmm pipeline detection routine first combines all the photons detected on 
each of the three instruments that compose the EPIC (pn, MOS1 and MOS2)
instrument to create an image, standardised detection algorithms are then applied to this image to detect sources \citep[see][for a review on source detection methods]{2012_Masias_MNRAS.422.1674M}. Sources with more than 100 EPIC counts are then further analysed
for transient behaviour using two methods: 1) a $\chi^2$ test comparing time series of the detection to a null hypothesis of a constant count rate, \citep{2009_Watson_A&A...493..339W}
and 2) a measure of the fractional variability amplitude $F_{\mathrm{var}}$ \citep{2016_Rosen_A&A...590A...1R}.

The $\chi^2$ statistic is suitable for binned data where the distribution of events in each 
time bin can be assumed to be Gaussian, it is not suitable for when the number of counts falls 
below $\sim$10~-~20 per bin (see \citealt{1979_Cash_ApJ...228..939C}). This is in part why the 
requirement of at least 100 total counts for the variability analysis
is specified. This prescription results in approximately two thirds of all detections (663,519 out of 1,035,832) not being analysed for transient behaviour. The standard pipeline variability
analysis is well suited for characterising the variability in detections with abundant counts and gradual variations, but is less suited to sources that are photon limited and show sharp
spikes or dips in their time series.

The awareness that short-term X-ray transients are lying undiscovered within the \xmm archive has motivated several studies to bring them to light.
To date, three major studies have systematically searched the entire \xmm catalogue, \extras \citep{2016_De_Luca_ASSP...42..291D, 2021_De_Luca_A&A...650A.167D}, a search for supernovae shock breakout \citep{2020_Alp_ApJ...896...39A}, and the first version of the algorithm presented in this paper \citep{2020_Pastor_Marazuela_A&A...640A.124P}. It is also worth mentioning
the STATiX pipeline \citep{2024_Ruiz_MNRAS.527.3674R} that is specifically designed for the detection of transients in \xmm data has not yet been applied to the data archive.

The method presented in this paper is a development of that presented in \cite{2020_Pastor_Marazuela_A&A...640A.124P}
but has been entirely re-written from the ground up, as part of the XMM2Athena project \citep{2023_Webb_AN....34420102W}.
We have made numerous changes to the way that transients are identified and importantly have
incorporated changes that allow us to probe the previously excluded time periods with background 
flaring caused by soft proton flares, which accounts for around $\sim 15\%$ of all observing exposure.

X-ray transients have a variety of origins, and when conducting broad searches, it is essential to consider all potential mechanisms.
In this section, we discuss some possible underlying causes of these transients,
along with their characteristic timescales and $0.2-12.0$ keV X-ray luminosities.

\noindent Stellar Flares: Magnetic reconnection in the atmospheres of stars can result in sudden
bursts of energy, which can be detected across the electromagnetic spectrum \citep[e.g.][]{2024_Kowalski_LRSP...21....1K}. 
Flares from cool stars (spectral type F-M) as well as Young Stellar Objects (YSO) are known to be particularly
abundant in the X-ray bands. Stellar flares have timescales of around $10^4 - 10^5$ seconds and have peak luminosities of
$\sim 10^{29} - 10^{32} \ \ergss$. \citep[e.g.][]{2003_Imanishi_PASJ...55..653I, 2015_Pye_A&A...581A..28P}

\noindent X-ray Binaries (XRBs): X-ray binaries, both high and low mass (HMXBs \& LMXBs) release X-rays
as material is accreted onto a compact object, either a neutron star (NS) or black hole (BH).
These sources are known to be variable on a wide range of timescales from seconds to months,
and have luminosities in the range
$\sim 10^{35} - 10^{39} \ \ergss$ \citep[e.g.][]{2004_van_der_klis_astro.ph.10551V, 2006_Fabbiano_ARA&A..44..323F}.

\noindent Cataclysmic Variables (CVs): Closely interacting binaries involving
accretion onto a white dwarf are powerful sources of X-rays \citep[see][]{1995_Warner_cvs..book.....W,2010_Kuulkers_csxs.book..421K}. 
CVs are often subclassified depending on the magnetic field strength of the white dwarf. Non-magnetic CVs
(sometimes called Dwarf Novae) have accretion discs extending down to the compact object and are known to
display large, often repeating outbursts over a few days before returning to quiescence. Magnetic CVs
are sometimes referred to as (intermediate) polars and may have truncated accretion discs where
X-rays can be emitted by a standing shock above the magnetic poles of the WD \citep[e.g.][]{1973_Aizu_PThPh..49.1184A, 2022_Page_hxga.book..107P}.
CVs span the luminosity range $\sim 10^{29} - 10^{35} \ \ergss$ \citep[see][]{2022_Suleimanov_MNRAS.511.4937S, 2024_Schwope_A&A...690A.243S}.
Additionally, these sources usually have orbital periods of order hundreds of minutes, which can be evident in 
the lightcurves of eclipsing sources \citep[e.g.][]{2001_Schwope_A&A...375..419S}.
A rare phenomena manifesting as a soft X-ray flash with luminosities of $\sim 10^{38} \ \ergss$ has
recently been observed in YZ Ret \citep{2022_Konig_Natur.605..248K}. This ``fireball'' phase is 
explained as the rapid expansion of the surrounding envelope of the WD, following the
runaway thermonuclear burning on its surface and is predicted to last several hours \citep[see][]{2015_Kato_ApJ...808...52K}.

\noindent Type I and II X-ray Bursts: Type I X-ray bursts are caused by runaway thermonuclear explosions on the surface of NSs. 
These bursts have been detected in over 121 sources \citep[e.g.][]{2020_Galloway_ApJS..249...32G, 2021_Galloway_ASSL..461..209G}.
The bursts have a characteristic profile of a fast rise ($\le 1 - 10 \ \mathrm{s}$) followed by an
exponential decay ($\sim 10 - 100 \ \mathrm{s}$).
Type I X-ray bursts peak luminosities cluster around $\sim 10^{38} \ \ergss$ which has motivated
their possible use as standard candles \citep[see][]{2003_Kuulkers_A&A...399..663K}.
Type II bursts are known to exist in only two sources and are
seen to repeat on the order of $\sim 10-20 \ \mathrm{s}$. Despite their existence being known for over 45 years, their nature still remains
unclear \citep[e.g.][]{2015_Bagnoli_MNRAS.449..268B}.

\noindent Magnetars (AXP, SGRs \& LGRs): Magnetars \citep{1992_Duncan_ApJ...392L...9D},
are young neutron stars with magnetic fields exceeding approximately $\gtrsim 10^{13} \ \mathrm{G}$,
they are known under a variety of different aliases such as anomalous X-ray pulsars (AXPs), Soft and Long Gamma Repeaters (SGRs \& LGRs) \citep[e.g.][]{2015_Turolla_RPPh...78k6901T}.
Magnetars exhibit a wide range of variability: short bursts (lasting $< 100 \ \mathrm{ms}$ with $L\sim10^{36} - 10^{43} \ \ergss$), outbursts (rise times of order $\sim 10- 100$s and reaching $10^{36} \ \ergss$)
and giant flares (seconds to minutes with $L\sim10^{44} - 10^{47} \ \ergss$) \citep[see][]{2017_Kaspi_ARA&A..55..261K}.
Pulse periods have been identified for 26 magnetars \citep[see][]{2014_Olausen_ApJS..212....6O}
and average around a few seconds, while their quiescent X-ray luminosities lie in the range of $\sim 10^{32}-10^{35} \ \ergss$.

\noindent Gamma Ray Bursts (GRBs): As their name suggests, the emission from GRBs primarily peak
at higher energies, however their afterglow emission can be detected in X-rays and lower energy bands.
Lasting from seconds to hours, GRBs are typically divided into short and long and are thought to respectively
arise from the mergers of neutron stars or the collapse of massive stars \citep[e.g.][]{2015_Kumar_PhR...561....1K}.
The observed X-ray luminosities for GRB afterglows decrease over time, but typically range from
$\sim 10^{43} - 10^{49} \ \ergss$ over a period of 5 minutes to 24 hours \citep[see][]{2012_DAvanzo_MNRAS.425..506D}.
GRBs can also be related to gravitational wave events.
For example, the detection of a transient X-ray counterpart to 
the NS-NS merger GW 170817, is suspected to be the
afterglow from an off-axis GRB \citep[see][]{2017_Troja_Natur.551...71T}.

\noindent Tidal Disruption Events (TDEs):
Stars in galactic nuclei can be captured or tidally disrupted by the supermassive black hole \citep[e.g.][]{1988_Rees_Natur.333..523R}. 
This process can result in luminous X-ray flares ($\sim10^{44} \ \ergss$) that may last months to years \citep[e.g.][]{2021_Saxton_SSRv..217...18S}.
A theoretical subclass of TDEs involves the tidal disruption of a white dwarf by an intermediate-mass black hole (IMBH) \citep[e.g.][]{2020_Maguire_SSRv..216...39M};
these would have considerably shorter evolution timescales than regular TDEs with the rise times possibly occurring on the timescales of hundred of seconds as opposed to days \citep[e.g][]{2013_Jonker_ApJ...779...14J}.

\noindent Quasi-Periodic Eruptions (QPEs):
Recently discovered large amplitude (peaking at $L \sim 10^{42} - 10^{43} \ \ergss$) soft QPEs from several galactic nuclei
(See: \citealt{2019_Miniutti_Natur.573..381M, 2020_Giustini_A&A...636L...2G, 2021_Arcodia_Natur.592..704A, 2021_Chakraborty_ApJ...921L..40C, 2022_Arcodia_A&A...662A..49A, 2023_Miniutti_A&A...674L...1M, 2023_Quintin_A&A...675A.152Q, 2023_Webbe_MNRAS.518.3428W, 2024_Arcodia_A&A...684A..64A}) has resulted in several 
suggestions for their origin, including repeated partial TDEs of stars by a SMBH, or interactions between stars and SMBH accretion discs \citep[e.g.][]{2024_Linial_MNRAS.527.4317L}. Recent discoveries in one source would appear to
strongly link QPEs and TDEs \citep[e.g.][]{2023_Quintin_A&A...675A.152Q,2024_Nicholl_Natur.634..804N}. The timescales of the outbursts and their
recurrence are around a few hours in the X-ray, however some galactic nuclei have been seen with longer recurrence times
(days-years) and are sometimes classified as repeating nuclear transients \citep[e.g.][]{2024_Guolo_NatAs...8..347G}.

\noindent Supernovae Shock Breakouts (SBOs): SBOs occur when the shock wave from
a supernova breaks through the outer layers of the star. This process can produce a bright X-ray/UV
flash in the time scales of seconds to fractions of an hour, they are among the earliest signals
that can be detected from supernovae and may provide key signatures on the structure of the progenitor star.
SBOs have observed peak luminosities in the range $\sim 10^{42} - 10^{46} \ \ergss$ and timescales of $30-3000$s.
\citep[e.g.][]{2017_Waxman_hsn..book..967W, 2020_Alp_ApJ...896...39A, 2022_Sun_ApJ...927..224S}

\noindent Fast X-ray Transients (FXTs): FXTs are extragalactic bursts of X-ray emission that last from hundreds of seconds to hours.
Possible origins include BH-NS mergers \citep[$L_{\mathrm{peak}} \sim 10^{44} - 10^{51} \ \ergss$; ][]{2014_Berger_ARA&A..52...43B}, tidal disruptions of white dwarfs by intermediate-mass black holes (WD-IMBH TDEs) \citep[$L_{\mathrm{peak}} \lesssim 10^{48} \ \ergss$; ][]{2020_Maguire_SSRv..216...39M} or supernovae SBOs (see above).
\citep[e.g.][]{2022_Quirola_A&A...663A.168Q, 2024_Eappachen_MNRAS.52711823E, 2025_Srivastav_ApJ...978L..21S}

\noindent Supergiant Fast X-ray Transients (SFXTs): Distinct from FXTs, SFXTs are Galactic HMXBs
with supergiant companions and have been observed to emit X-ray flares lasting $\sim 100 - 10,000 \ \mathrm{s}$
and have luminosities in the range of $\sim 10^{32} - 10^{38} \ \ergss$ \citep[see][]{2023_Romano_A&A...670A.127R}.
NS pulsations have been detected in around half of SFXTs, however the mechanism 
responsible for the transient X-ray emission is still of open debate.
\citep[e.g.][]{2013_Sidoli_arXiv1301.7574S, 2017_Sidoli_mbhe.confE..52S}

\noindent Blazar Flares: Blazars are AGN viewed directly down the
funnel of the relativistic jet, their X-ray luminosities lie in the range $\sim 10^{42.5} - 10^{47.5} \ \ergss$
\citep[see][]{2024_Zhang_MNRAS.529.3699Z}. Some blazars have shown fast variability in
X-ray bands such as the $\sim 2000$ second flare seen in NRAO 530 \citep{2006_Foschini_A&A...450...77F} 
or the dramatic flux changes occurring over a few hours seen in several other sources \citep[e.g.][]{2008_Lichti_A&A...486..721L, 2015_Hayashida_ApJ...807...79H,
2016_Kapanadze_ApJ...831..102K}.

\noindent Fast Radio Burst (FRBs): Though primarily observed in radio wavelengths, growing evidence links FRBs to magnetars and therefore X-ray activity. In particular, the detection of an FRB from the Galactic magnetar SGR 1935+2154 \citep[e.g.][]{2021_Ridnaia_NatAs...5..372R} was associated to an X-ray burst, suggesting that extragalactic FRBs may have transient X-ray counterparts. Other theories suggest that FRBs may originate from super-Eddington mass transfer binaries \citep[e.g.][]{2021_Sridhar_ApJ...917...13S},
commonly seen as ultraluminous X-ray sources (ULXs) \citep[e.g.][]{2023_King_NewAR..9601672K} which themselves are known to be highly 
variable in X-rays.

\noindent Gravitational Microlensing: Microlensing events occur when a massive object,
such as a star, black hole or even exoplanet, passes in front of a background source, causing
a characteristic temporary magnification of the distant source. Due to their rarity, searches for
microlensing events have primarily been focused on large scale optical surveys. Microlensing 
has been detected in X-rays in studies of strongly lens quasars such as RX J1131-1231
\citep[e.g.][]{2009_Chartas_ApJ...693..174C, 2012_Morgan_ApJ...756...52M} but to date, no serendipitous
X-ray micro-lensing events have ever been reported.

\noindent Unknown Processes: 
The aforementioned types of transients cover a wide range of objects, phenomena, and timescales. It is also possible that a wide archival search for faint and fast transients might uncover new types of astrophysical phenomena that may not currently have theoretical explanations.

\section{Methods}
\subsection{Overview}
The EPIC XMM Outburst Detector (EXOD) is designed to detect rapid, variable behaviour in photon counting data.
In order to do this, we transform a finite number of photon detections acquired over the course of an
observation into a meaningful metric for the quantification of variability.

Algorithms for variability detection in the X-rays
ultimately rely on a statistical comparison
between an expected signal ($\mu$) and the observed data ($N$).

Existing algorithms (see introduction) generally assume
the expected signal ($\mu$) to be constant over time.
In this paper, we implement a novel approach, 
in which the expected signal is modelled as a 
time-dependent spatial template for each time frame of the observation.
We then use a Bayesian framework to compare this signal to actual observed data.

Our approach allows for the quantification of the significance of a peak or eclipse in the low count (Poisson) regime,
probing the variability in both short time frames and previously discarded flaring bad time intervals (BTIs).

The full source code and documentation for EXOD may be found on
github\footnote{\href{https://github.com/nx1/EXOD2}{https://github.com/nx1/EXOD2}}. EXOD is implemented as a Python package and contains all the code required to detect transients from scratch.

EXOD is compatible with pn in \texttt{PrimeFullWindow}, \texttt{PrimeFullWindowExtended} and \texttt{PrimeLargeWindow} submodes, and all
MOS submodes excluding the two timing modes: \texttt{FastUncompressed} and \texttt{FastCompressed}.
The algorithm operates on the pipeline processed EPIC imaging mode event lists with the 
\texttt{`EVLI'} file ID. The only other required file for the algorithm is the EPIC summary
source list with the ID \texttt{`OBSMLI'}.

\subsection{Preprocessing} \label{sec:preprocessing}
Our first step is to apply standard filters\footnote{\href{https://heasarc.gsfc.nasa.gov/docs/xmm/hera_guide/node33.html}{https://heasarc.gsfc.nasa.gov/docs/xmm/hera\_guide/node33.html}} to clean the pipeline processed event files (\texttt{PATTERN<=4} for EPIC pn and \texttt{PATTERN<=12} for EPIC MOS, \texttt{200<PI<12000} for all EPIC instruments). This removes non-astrophysical photon events occurring over multiple detector pixels, and removes photons outside the 0.2 - 12.0 keV energy range.

We additionally remove all events associated with known warm-pixel events \citep{2020_Webb_A&A...641A.136W}, and bad rows \citep{2001_Struder_A&A...375L...5S}. In total, the number of unique warm pixels for each instrument are MOS1=10,078, MOS2=1,778 and pn=2,000.

Events associated with the central MOS CCD (1) when in partial modes are removed,
as is a 3 pixel margin from the edge of each pn CCD, this was found to reduce spurious 
signals.

\subsection{Event list binning}
Modern X-ray imaging spectrometers are capable of detecting individual X-ray photons in two spatial dimensions: $X$ and $Y$, a time dimension $t$ and an energy dimension $\nu$.
These quantities are measured at discrete intervals (bins), specified by the physical size of the individual detector elements on the CCD's recto-linear grid,
the readout speed of the instrument, and the method by which the integrated charge per pixel from an event is converted into an energy value.

\begin{figure}[t]
    \centering
    \includegraphics[scale=2.0]{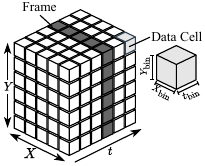}
    \caption{Schematic of an EXOD data cube with dimensions
    $X$, $Y$ and $t$.
    A single time frame is shaded in dark grey and a single
    data cell with dimensions $X_{\mathrm{bin}}$, $Y_{\mathrm{bin}}$, $t_{\mathrm{bin}}$ is shown.
    In practice, the data cube for a single event list can contain many millions of cells depending on the binning parameters.}
    \label{fig:DataCube}
\end{figure}

If we ignore the energy dimension, the photon detection parameter space may be visualised as a cube composed of discrete cells containing the number of photons detected in a given
$X,Y$ and $t$ interval (See Fig. \ref{fig:DataCube}). This data cube may be created from the photon event list by binning the data with a specified cell size. The size of each cell should
not be smaller than the specification of the detector, for EPIC-pn this would mean
($150~\mathrm{\mu m} \times 150\mathrm{\mu m} \times 73~\mathrm{ms}$)
\citep{2001_Struder_A&A...365L..18S},
and for MOS is
($40~\mathrm{\mu m} \times 40\mathrm{\mu m} \times 2.6~\mathrm{s}$)
\citep{2001_Turner_A&A...365L..27T}. These physical pixel sizes correspond to 4.1 and 1.1 arcsec$^2$ respectively.

Practically, creating a data cube with individual cells matching the detector resolution is computationally expensive and, when combining data 
from  multiple instruments (as we have done in this work) becomes impossible due to differences between the detectors.

Therefore, for the EXOD algorithm, we rebin the event list to have a spatial bin size in arcseconds ($X_{\mathrm{bin}}, \ Y_{\mathrm{bin}}$) and a temporal bin size ($t_{\mathrm{bin}}$) in seconds.
We acknowledge that there is a loss of spatial accuracy in the final detected positions using the re-binned photon events, however, this trade-off comes with the advantage of being able to reduce the temporal
bin size $t_{\mathrm{bin}}$ to probe shorter timescales. Additionally, by combining event lists from simultaneously observing instruments, we can obtain increased sensitivity in the low-count Poisson regime
for detecting transients. We find that a spatial binning of $X_{\mathrm{bin}} = Y_{\mathrm{bin}} = 20$ arcseconds serves as a reasonable compromise and corresponds to a fractional encircled energy of $\sim 0.8$ for the three instruments.

\subsection{Quantifying variability in the Poisson regime} \label{sec:Poission_var}
For a given cell of our data cube, we make the assumption that the photon events produced are the result of a Poisson process, and thus their probability function is given by:
\begin{equation}
    P(N, \mu) = \frac{e^{-\mu} \mu^{N}}{N!}
    \label{eq:Poisson_cdf}
\end{equation}
Where $N$ is the observed number of counts in a time and space interval,
and $\mu$ the expectation value of the Poisson distribution.
An important detail here is that $\mu$ is not the background expectation, but merely the expectation value of a given data cell, for example, a cell containing a non-variable source would have an expectation that is composed of the contribution from both the source and background ($\mu = \mu_{S} + \mu_{B}$).

At first glance this equation would suggest that if we know the expectation ($\mu$) of the Poisson distribution for a given data cell, one could simply calculate the probability, $P(N, \mu)$ of the $N$ observed events having been produced by a Poisson process, finally one could simply set a threshold
below which a cell would be flagged as significant.

However, as discussed in \cite{1991_Kraft_ApJ...374..344K}, classical attempts to derive confidence intervals
in the regime of non-zero background have several pitfalls, and instead the authors strongly recommend
using a Bayesian formalism to overcome these limitations.

%
%

\subsubsection{Bayesian formalism}
Adopting a Bayesian formalism, we define our null hypothesis model ($\theta_0$) as: ``there is no additional transient emission on top of the background and a possible constant source'',
such that the observed counts ($N$) are a Poisson realisation of the expectation ($\mu$).

Our alternative hypothesis ($\theta_1$) is that the observed counts
($N$) are a result of an additional transient emission on top of the expectation ($\mu$). This transient emission can also be negative (i.e. an eclipse instead of a peak), in which case the formalism is the same but with a negative amplitude, and thus some resulting formulas are slightly altered.

To compare two competing hypotheses, we define a Bayes factor $B$ (sometimes called the odds ratio)
as the ratio of the two marginal likelihoods:

\begin{equation}
    \begin{split}
    B = \underbrace{\frac{P(N~|~\theta_1)}{P(N~|~\theta_0)}}_{Likelihoods} = \underbrace{\frac{P(\theta_1~|~N)}{P(\theta_0~|~N)}}_{Posteriors}  \times \underbrace{\frac{P(\theta_0)}{ P(\theta_1)}}_{Priors \ = \ C} \times \underbrace{\frac{P(N)}{P(N)}}_{= \ 1}
    \end{split}
\end{equation}

\noindent where we have used Bayes' theorem 
to invert the conditional probabilities. The marginal distribution of observed counts, $P(N)$, (which would be given by $p(N) = \int P(s)p(N|s) ds$) end up cancelling out to 1.

We make the assumption of a flat prior for the ratio of $P(\theta_0) / P(\theta_1) = C$ and assume it is constant. It could be argued that this ratio is actually a function of meta-parameters (e.g. the timescale we are searching), but for the sake of simplicity and tractability of the computation, we keep it constant.

The likelihood $P(N~|~\theta_0)$, of having observed $N$ counts given the null hypothesis ($\theta_0$) is given simply by the Poisson probability function (eq. \ref{eq:Poisson_cdf}).

The posterior probability distribution, $P(N~|~\theta_1)$, of having observed $N$ counts given the presence of a real transient on top of the expectation $\mu$ is less straight-forward. Assuming the transient has an amplitude of $s$, the likelihood is given by the slightly modified expression (eq. \ref{eq:H_0_posterior2}):
\begin{equation}
    P(N~|~\mu + s) = \frac{e^{-(\mu+s)} (\mu+s)^{N}}{N!}
    \label{eq:H_0_posterior2}
\end{equation}

However, we do not know the value of $s$. The hypothesis $\theta_{1}$ takes into account all possible amplitudes. The likelihood of $\theta_{1}$ can thus be expressed by integrals over $s$. Assuming we are assessing the presence of a peak (and thus $s$>0), we can express it: 
\begin{equation}
    \begin{split}
        P(N~|~ \theta_1) &=  \int_{0}^{\infty} P(N~|~\mu + s) P(s)~ds \\
                         &=  \int_{0}^{\infty} e^{-(\mu+s)}\frac{(\mu+s)^N}{N!}P(s)~ds \\
                         &=  \int_{\mu}^{\infty} e^{-s'}\frac{s'^N}{N!}P(s'-\mu)~ds'
    \end{split}
\end{equation}
At this point, we have to assume a prior for the peak amplitude $P(s')$. The simplest is a flat prior, which can be moved out of the integral and will join the other prior of having a peak in the data. It is shown in \cite{1991_Kraft_ApJ...374..344K} that this assumption
results in negligible differences. In this case, we get
\begin{equation}
    \begin{split}
         P(N~|~\theta_1) &\propto \int_{\mu}^{\infty} e^{-s'}\frac{s'^N}{N!}~ds' \\
                         &\propto \frac{\Gamma(N+1,\mu)}{N!}\\
                         &\propto \frac{\Gamma(N+1,\mu)}{\Gamma(N+1)}\\
                         &\propto \mathcal{Q}(N+1, \mu)
    \end{split}
\end{equation}

\noindent Here, $\Gamma(N+1, \mu)$ is the incomplete upper gamma function, 
and $\mathcal{Q}(N+1, \mu)$ the regularized incomplete upper gamma function. 
We have used  $\Gamma(N+1) = N!$ for positive integers.

The Bayes factor for peaks can now be expressed:
\begin{equation}
    B_{Peak} = \frac{\mathcal{Q}(N+1, \mu)}{e^{-\mu}\mu^N / N!} \times K
\end{equation}
where $K$ is a combination of all flat priors.

Similarly, by integrating on negative values of $s$, we can obtain
the Bayes' factor for eclipses:

\begin{equation}
    B_{Eclipse} = \frac{\mathcal{P}(N+1, \mu)}{e^{-\mu}\mu^N / N!}  \times K'
\end{equation}

\noindent where $\mathcal{P}(N+1, \mu)$ is the regularized incomplete lower gamma function 
and $K'$ the combination of all flat priors (a priori different from $K$).

At this point, assuming we have an expected value $\mu$ for the counts in a given 3D pixel of our data cube, and the actual observed counts $N$, we have quantified the likelihood of having a transient in the form of $B_{Peak}$ and $B_{Eclipse}$, both still containing an unknown constant prior. From a practical standpoint, we now need to determine a threshold on these values above which a pixel presents a significant departure from $\mu$, i.e. can be  considered variable.

\subsubsection{Variability significance}\label{subsec:VarSignif}

\begin{figure}[t]
    \includegraphics[width=\columnwidth, trim={0 0.0cm 0 0.0cm}, clip]{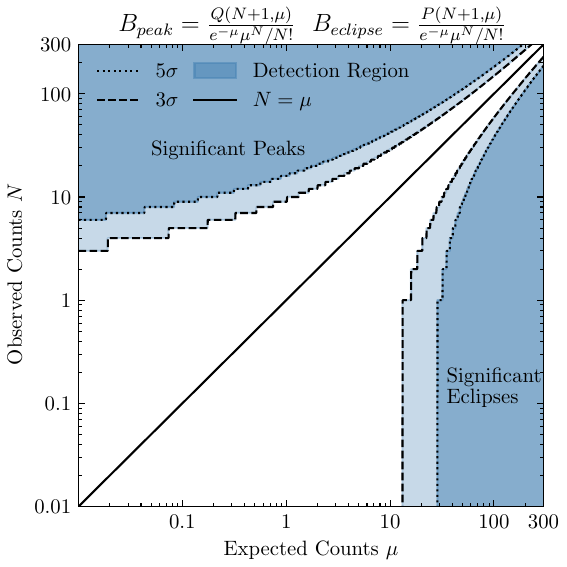}
    \caption{Observed counts ($N$) against expected counts ($\mu$) shown in the range 0.01-300 counts.
    The significant detection regions for peaks and eclipses are denoted
    by the dashed ($3\sigma$) and dotted ($5\sigma$) and shaded in blue.
    These thresholds correspond to the ``sigma equivalent'' thresholds (see text).}
    \label{fig:Bayes_Factor_threshold_3_sig}
\end{figure}


To determine the correct threshold, we can use the fact that at large values of $\mu$ (say $\mu > 1000$), 
a normal distribution with mean $\mu$ and standard deviation $\sigma = \sqrt{\mu}$ is an accurate 
approximation to the Poisson distribution.

In the large count regime, it is commonplace to calculate the significance of departure from the expectation through a sigma value ($\sigma$) given by:
\begin{equation}
\sigma = \frac{N-\mu}{\sqrt{N+\mu}}
\end{equation}

For example, if $\mu=1000$, to obtain a positive 3$\sigma$ departure from $\mu$, we would require an observed number of counts
of $N=1139$ or $N=870$, likewise for 5$\sigma$ this is $N=1237$ and $N=788$.
We can then substitute these values for $\mu$ and $N$ back into our formula for the Bayes' factor (fixing $K$ and $K'$ at a value of 1, since this normalization is arbitrary) to find values corresponding to a significant variability. For example, using the 3 sigma upper and lower values for $N$ we obtain:
\begin{equation}
    \begin{split}
    B_{Peak}(\mu=1000, N=1139)   &\approx 10^{5.94} \\
    B_{Eclipse}(\mu=1000, N=870) &\approx 10^{5.7}
    \end{split}
\end{equation}

\noindent And for 5 sigma, we obtain:
\begin{equation}
    \begin{split}
    B_{Peak}(\mu=1000, N=1237)   &\approx 10^{13.2} \\
    B_{Eclipse}(\mu=1000, N=788) &\approx 10^{12.38}
    \end{split}
\end{equation}

\noindent  We can then use these threshold values of $B$ globally, even in the small count regime, to obtain
a 3 or 5 sigma equivalent score (see Fig. \ref{fig:Bayes_Factor_threshold_3_sig}). We can verify visually that the regions of significant variability behave in a manner consistent with Poisson noise: for large counts, these regions converge with a symmetric normal departure from the diagonal $N=\mu$, while in low counts a relatively large number of counts is required to be considered a peak, and no eclipse can be detected.

This approach is designed to detect outbursts or eclipses occurring on similar timescales to the size of the time bins. This implies two key points: firstly, this method is not well-suited for identifying low-amplitude modulations within a light curve. Secondly, to detect faint bursts effectively, the bin size should closely match the timescale of the burst, making it essential to test across many time-binning parameters.

\subsubsection{Amplitude estimation}
The Bayesian formalism can also be used to estimate the parameter ($s$) of our model, i.e. the amplitude of the peak or eclipse. The precise computations can be found in Appendix \ref{app:ParamEstimate}. It is very similar to what is usually done to estimate the brightness or upper limit of X-ray sources \citep[e.g.][]{2022_Ruiz_MNRAS.511.4265R}.

The general idea is to use the intimate link between $\Gamma$ functions and Poisson statistics to provide a confidence interval for the amplitude of a transient (peak or eclipse), knowing only $N$ and $\mu$. After computation, we use

\begin{equation}
\begin{split}
    UL_{peak}(f)=\mathcal{P}^{-1}\Big(N+1,~ &\mathcal{P}(N+1,\mu) ~+~  \\ &{ f}\times\mathcal{Q}(N+1,\mu)\Big) - \mu
\end{split}
\end{equation}
and 
\begin{equation}
\begin{split}
    UL_{eclipse}(f) = \mu ~-~ \mathcal{P}^{-1}\Big(N+1,~ &\mathcal{P}(N+1,\mu) ~-~ \\ &{f}\times\mathcal{P}(N+1,\mu) \Big)
\end{split}
\end{equation}

\noindent where $\mathcal{Q}$ and $\mathcal{P}$ are the regularized incomplete upper and lower Gamma functions, and $ UL_{peak}(f)$ and $UL_{eclipse}(f)$ are the count upper limits with a probability of $f$ (i.e. the count values compared to which the actual amplitude has a probability $f$ to be smaller). Replacing $f$ with the fractions $\{0.16, 0.5, 0.84\}$, we can build the 1$\sigma$ confidence interval for the peak (or eclipse) amplitude.

From a practical standpoint, this computation is relatively expensive, so it is not possible to simply compute these values for the entire 3D grid and keep only the pixels for which the amplitude is non-zero at a given significance threshold. On the contrary, it is only used on the pixels that are good transient candidates, found by computing $B_{Peak}$ and $B_{Eclipse}$ and keeping values above the thresholds presented in subsection \ref{subsec:VarSignif}.

\subsection{Estimating the expected emission}\label{sec:Template_creation}

\subsubsection{General considerations}
In section \ref{sec:Poission_var}, we established a method for quantifying variability as
a function of the observed ($N$) and expected ($\mu$) counts.
We now require a method to obtain the expectation value ($\mu$) for each cell of our data cube.

We assume that two components are combined to create the expectation value ($\mu$).
The first is the background component which is ascribed to the galactic and
extra-galactic diffuse X-ray emission, as well as the time varying contribution from soft
proton emission originating from the sun. The second component is the contribution from the
sources in the field of view of the telescope, which are assumed to be constant over
the time of an observation. Any departure from the sum of these two components will be interpreted
as transient behaviour -- either a faint source appears from the background, or a known (brighter)
source departs from its expected-constant behaviour in a given time frame.

One of the novel features of our method is the ability to detect transients within 
flaring bad time intervals (BTIs), which are usually discarded.
These BTIs correspond to frames in which the background is much higher than for the rest of the frames.
This is typically defined using an arbitrary threshold on the high energy count rate over the field of view.
Other methods for fast transient detection have either neglected the variable background
\citep[e.g.][]{2020_Pastor_Marazuela_A&A...640A.124P,2010_Evans_ApJS..189...37E}, or used
background-estimating methods that accounted for them but required more data, and so forbid smaller time binning \citep{2024_Ruiz_MNRAS.527.3674R}.
We wanted a method that could both account for BTIs, and go down to instrumental time binning (i.e. about 5s).

During our tests, we observed that although the distribution of the background in a single \textit{XMM-Newton} observation has a spatial dependence due to effects such as
observing mode, gaps between the CCDs, dead pixels \& vignetting, this spatial dependence remains constant during the observation,
with only its amplitude changing over time. More precisely, this spatial dependence follows one template during Good Time Intervals (GTIs) and another, spatially flatter template during BTIs. This different spatial shape is due to the flatter vignetting profile of soft protons compared to X-ray photons \citep[e.g. Kuntz and Snowden 2023\footnote{XMM-CCF-REL-0399},][]{2024_fioretti} This means that a pair of 2D-templates describing the background in GTIs and BTIs can be simply scaled by a constant corresponding to the background amplitude in a given time frame in order to estimate the background for each data cell. This is the main observation that drove our background-computation method. The strength of this method is that, by using the 2D templates, we input prior spatial knowledge about the background, which allows us to go down to small time binning where the data is very sparse.

\subsubsection{Creating the expectation data cube} \label{sec:expectation_cube_creation}
The creation of the expectation ($\mu$) data cube is a multistep process and at a high level can be described as follows:
\begin{enumerate}
    \item Calculate the background template.
    \begin{enumerate}
        \item Mask the sources from the image.
        \item Calculate the number of counts in the background.
        \item Inpaint the sources (i.e. interpolate the background over the masked source areas).
        \item Divide the image by the total counts in the background.
    \end{enumerate}
    \item Calculate the constant-source template.
    \begin{enumerate}
        \item Subtract the background contribution from (1.c) from the image.
        \item Mask out the background.
        \item Divide the image by number of effective time frames in the cube.
    \end{enumerate}
    \item Combine the source and background template to create the estimation template.
    \item Create the expectation data cube by joining many estimation templates,
    each one scaled by the number of background counts in the specific frame.
\end{enumerate}

\noindent A diagram of this process can be found in Fig. \ref{fig:template_creation_diagram}, and further details are provided below.

First, since the vignetting is different between GTIs and BTIs, we split the observation in those two categories.
We split GTIs and BTIs based on thresholds applied to the total frame count rate in high energies (10–12 keV).
The threshold is set at 0.5 times the number of active instruments (e.g., 1.5 ct/s for pn, MOS1, and MOS2).
Unlike standard EPIC data processing, BTIs are retained and treated separately, so precise threshold computation
as is done in the standard pipeline is less critical.

Frames corresponding to BTIs and GTIs are independently summed to create separate images.
Source masks are generated for each image using the standard pipeline EPIC source list,
with extraction radii based on the 0.2 - 12.0 keV EPIC count rate
(20", 40", and 80" for rates <0.1, 0.1–1, and >1 counts~s$^{-1}$, respectively).
These masked regions are inpainted using \texttt{INPAINT\_NS} from \texttt{OpenCV} \citep{opencv_library}
to estimate the background's spatial structure. 
Images are normalized by dividing by their total counts, yielding the normalized background template.
Multiplying the normalized background template by the total photons contained in a single frame provides
the specific frame background template, this a very good approximation of background in the actual frame,
which is a Poisson realization of the frame background template.

We now seek to add the contribution from the sources in the field of view, which we model as being constant in time.
For this, we mask out the background from the GTI and BTI images, leaving behind only the sources.
We then calculate the net source emission by subtracting the background contribution (described in the previous paragraph)
from the observed source counts. Assuming the sources are constant, the average frame emission is calculated through dividing by the total number of frames, thus obtaining the frame source template
(which is identical for all frames in the data cube).

Summing the source and background templates in a given frame yields the frame expectation template.
Arranging all the frame templates in time then yields the expectation data cube, in which each
3D pixel (data cell) contains the expectation value ($\mu$) that accounts for both the background and source contribution.

Significant deviations between the observed ($N$) data cube and expectation ($\mu$) data cube are interpreted as transient behaviour: 
either a known point-like source is not constant, and thus it will deviate in data cells from its averaged emission,
or a faint transient event deviates from the background expectation in some data cells.

It is worth noting that we have only mentioned the constant emission from point-like sources.
Some significant issues appear in the presence of extended sources. They are too large to be correctly inpainted using the same method as for point sources, which results in a spurious over-estimation of the background. Additionally, since these sources are constant while the background is not, a large number of false positives can happen within their extent. 
We address these issues in post-processing (sec \ref{sec:post_processing}).

\subsection{Benchmarking the method}
Having created a method to quantify the statistical significance in any given cell in our data cube (sec \ref{subsec:VarSignif})
by means of comparing it to the expected emission (sec \ref{sec:Template_creation}),
we now need to assess the sensitivity of our method via simulations.

\begin{figure}[t]
    \centering
    \includegraphics[width=\columnwidth]{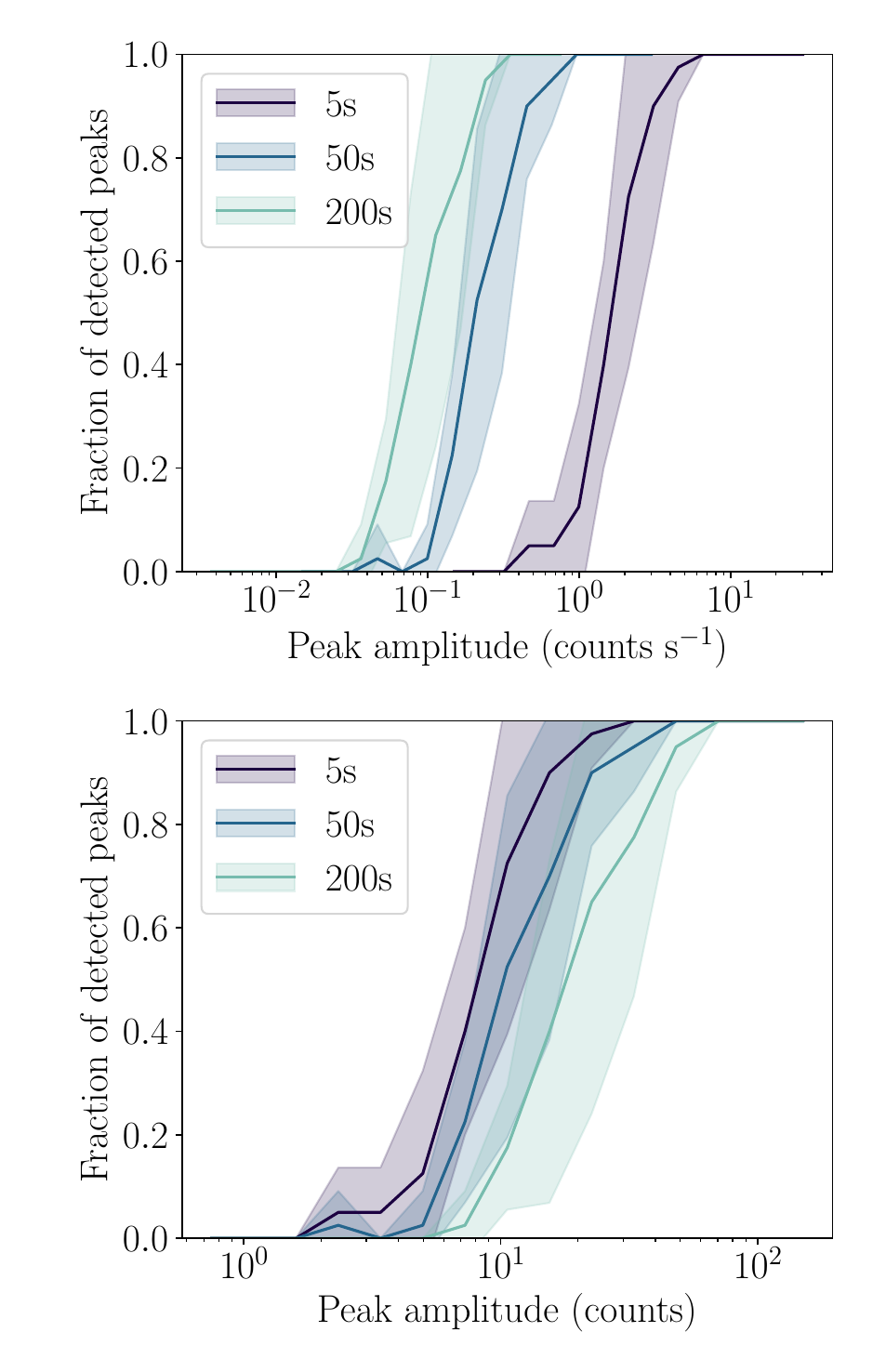}
    \caption{Fraction of synthetic peaks detected with $5\sigma$ confidence threshold, as a function of the amplitude of the peak and the time binning. The amplitude here is the theoretical amplitude, before the Poisson realisation. The shaded area represents the standard deviation across different observations. \textit{Top panel:} This completeness is depicted as a function of the input amplitude of the peak expressed in count rate. \textit{Bottom panel:} Same quantity but expressed as a function of the input counts.}
    \label{fig:benchmark_5sig}
\end{figure}

\subsubsection{Quantification of false negatives}
A commonly used approach for benchmarking detection tools is to use the end-to-end X-ray telescope simulation
package \texttt{SIXTE} \citep{2019_Dauser_A&A...630A..66D}. However, because the main contaminants of our method are 
elusive instrumental effects as well as the complex temporal and spatial evolution of the background, we have opted to use
real \textit{XMM-Newton} observations into which synthetic bursts are added.

Synthetic bursts are modelled by specifying an amplitude, 
spatio-temporal position and extent.
The spatial extent is modelled as a 2D normal distribution with a standard deviation
set to 6.6'' to match the FWHM PSF, and the temporal extent constrained to a single time bin.
The amplitude is then distributed across the data cells following these probability distributions.

We performed simulations by adding a single burst to a random position in an observation and then determining whether
the burst was detected at the 5 sigma level. We repeated the simulation over a grid of 10 observations, 3 time binnings (5s, 
50s and 200s), 15 burst amplitudes and 15 random burst positions. These values were then used to obtain constraints
on the detection fraction.

Fig. \ref{fig:benchmark_5sig} shows the mean and standard deviation of the detection fraction as a function of the 
burst amplitude in counts (bottom) and count rate (top), the different lines correspond to the three different time binnings.
Overall, our method is sensitive down to $\sim$10 photons, which is typically lower than the detection threshold for \textit{XMM-Newton} observations,
this is encouraging for the possibility of discovering new transients. With smaller time binning, fewer counts are necessary ($\sim$8 at 5s compared to 
$\sim$20 at 200s for a 50\% completeness), but these require much larger count rates to achieve such counts in much shorter time bins. A 90\% completeness at $5\sigma$ confidence level is reached at around 10$^{+7}_{-3}$, 20$\pm10$ and 30$\pm10$ counts for the 5s, 20s and 200s binning respectively (the error bars corresponding to the $1\sigma$ deviation of the detection rate over a sample of observations).

As expected, this method is more sensitive (by a factor of $\sim$3) for GTIs than for BTIs (due to the lower 
background levels), as can be seen in Fig. \ref{fig:benchmark_GTIvBTI}. However, the method is still sensitive 
enough to allow for the detection of new transients in BTIs which are usually ignored in automatic 
processing of the archives (which under our definition represents around $\sim$20\% of the archive).

An interesting side result of the benchmark presented in Fig. \ref{fig:benchmark_5sig} is that it provides us with an estimate of
the maximum time binning at which our method is relevant. Indeed, looking at the completeness of the 200s-timebin peaks as a function
of the amplitude in terms of counts, we see that even around 10 photons, a relatively low fraction of the peaks are detected.
For an amplitude of 20 photons, the 1$\sigma$ lower limit is at less than 10\% completeness -- but this is also enough photons for
the traditional source detection pipeline to pick up this source (although 100 counts are required to perform a variability assessment).

It is important to note that this sensitivity was measured at a 5$\sigma$ level, which is expected to provide us with few false positives.
The 3$\sigma$ completeness benchmark is shown in Fig. \ref{fig:benchmark_3sig}, with as expected a higher completeness (or equivalent completeness at lower peak amplitudes). For instance, the 90\% completeness counts drop to 8$\pm2$, 12$\pm5$ and 20$\pm10$ for the 5s, 20s and 200s binning respectively. We however expect higher rates of false positives at a 3$\sigma$ level compared to 5$\sigma$ once applied to real data.

\subsubsection{Quantification of false positives}\label{sec:false_positives}
Having quantified the detection fraction (rate of false negatives), it is important to try to quantify the rate of spurious signals (false positives).
There are several possible sources of false positives: the first is due to statistical variations, due to the large number of samples drawn from the underlying Poisson distribution, random fluctuations in the data may mimic true signals. The second comes from instrumental effects that are not fully accounted for in our method.
The third source of spurious detections can arise from an inaccurate estimate of a data cell's background rate.

While assessing the rate of false positives due to instrumental effects and inaccurate background estimates remains challenging, the rate of false positives arising from statistical fluctuations can be quantified through simulations.

To estimate the number of false positives caused by statistical fluctuations, we performed the following steps.
First, through the process described in sec. \ref{sec:expectation_cube_creation} we created data cubes of the
expectation value for 20 random \textit{XMM-Newton} observations across three time bin durations: (5s, 50s, 200s).
For each of these data cubes, we generated 120 Poisson realizations, thereby simulating a sample of 2400 observations
at each time bin. These simulated observations were then processed through the EXOD detection pipeline, and the number
of spurious detections arising from samples drawn in the tail of the Poisson distribution were counted.

At the $3\sigma$ detection threshold, we found that across all time binnings around
$\sim$ 200 / 2,400 observations ($\sim 8.3\%$) led to spurious transient events,
while at the $5 \sigma$ threshold we did not detect any spurious detections in the simulated observations.
When extrapolated to the full archive of 15,000 observations we predict a contamination of 1,250 false detections
at the $3 \sigma$ threshold and only a handful of false positives at the $5\sigma$ level.

Do note that this only corresponds to false positives due to Poisson fluctuations,
and other types of contamination will be assessed in section \ref{sec:hot_reg}.

\subsection{Post-processing \& filtering} \label{sec:post_processing}
As is the nature of conducting a search for signals in such a large dataset, one is naturally confronted with many
false positives which may arise from a variety of sources.

One source is simply from the extremely large number of free trials over which the search was conducted (see sec \ref{sec:false_positives}).
Over all our runs we considered approximately $\sim 2.3$ trillion data cells, this means that encountering events with extremely low probabilities such as to $P \sim 10^{-10}$ are not inconceivable, this would be equivalent to observing a count of $N=10$ given an expectation of $\mu = 1$ from a Poisson distribution.

Other sources of spurious signals can be caused by detector effects such as hot pixels
(sec. \ref{sec:hot_reg}) or uneven exposure across the
CCDs (See appendix A in \citealt{2021_De_Luca_A&A...650A.167D}) or optical overloading due to
bright stars. Bright extended sources such as supernova remnants or galaxy clusters may also cause
false positive transient alerts, as well as the brightest point-sources, for which the PSF is so extended that our inpainting algorithm fails. We additionally notice that spurious signals are often found near the start and end of
observations due to instrumental effects.
Optical loading is present in a few detections such as 0823432701, while moving target observations such as those made of
Jupiter or comet 21P/Giacobini-ZinneR 0864840301 can cause spurious signals.
\subsubsection{Detection Quality Flags}
Table \ref{tab:flags} shows the breakdown of the number of detections by various post-processing flags and their percentages, based on 3$\sigma$ and 5$\sigma$ significance levels. The table lists isolated flares (defined as alerts
that were preceded and followed by 0 count bins), last-bin alerts (where a significant alert occurred in the last bin of the observation), and other critical conditions, including problematic observations, observations with more than 20 detections, and regions occurring in
hot areas (sec. \ref{sec:hot_reg}).
\begin{table}[h!]
\caption{Number of post-processing of detection flags and percentages as a total of the 3 and 5 sigma detections.}
\centering
\begin{tabular}{lrr}
\hline
\textbf{Quality Flag}           & \textbf{3$\sigma$ (60,127)} & \textbf{5$\sigma$ (11,273)} \\
\hline
\# Isolated Flares        & 27,916 (46.4\%)  & 1,746 (15.5\%)  \\
\# Last Bin Alerts        & 1,103 (1.8\%)    & 176 (1.6\%)     \\
\# < 5 Counts Max in LC   & 17,785 (29.6\%)  & 26 (0.2\%)      \\
\# Obs /w >20 Detections  & 3,990 (6.6\%)    & 759 (6.7\%)     \\
\# in Hot Areas           & 4,890 (8.1\%)    & 908 (8.1\%)     \\ 
\# in~\NBadObsIDs~Problematic Obs     & 7,658 (12.7\%)   & 1,515 (13.4\%)  \\
\hline
\end{tabular}
\label{tab:flags}
\end{table}

\subsubsection{Identification of instrumental hot areas} \label{sec:hot_reg}
As previously mentioned in section \ref{sec:preprocessing}, previously known hot pixels on
the three detectors had been already removed during the pre-processing steps
when loading the event lists. However, because we are operating in previously
unexplored temporal regimes, it is possible that a number of previously unknown
detector quirks can contribute to false positive alerts.

To explore this effect, we took all regions detected by EXOD and transformed the
coordinates into detector coordinates. Fig. \ref{fig:spatial_dist} shows the spatial 
distribution of detections in the 0.2-12.0 keV band split by the three different time 
binnings. We observe that at the 5s binning, there are a number of areas that contain a much higher density
of regions which are likely due to intrinsic detector effects, this excludes the
region that is near the centre of the detector which is where the target of
the observation is usually positioned. At the 50 and 200s binning there are
fewer of these over-dense regions observed, however we observe an over-abundance 
of detections on the semicircular segment on the MOS chip that does not overlap with the pn.
A line of detected regions extending over one of the pn CCDs can be seen and is due to
an instrumental read-out streak caused by bright targets.

We identified 35 over-dense areas at the 5s binning, 5 areas at 50s and 3 areas at 200s.
In total 4,890 / \NRegTot (8.13\%) \tSig of the detected regions were flagged in the final
catalogue for being potentially instrumental false-positives.

\section{Results}
\begin{figure}[h]
    \centering
    \includegraphics[scale=0.71]{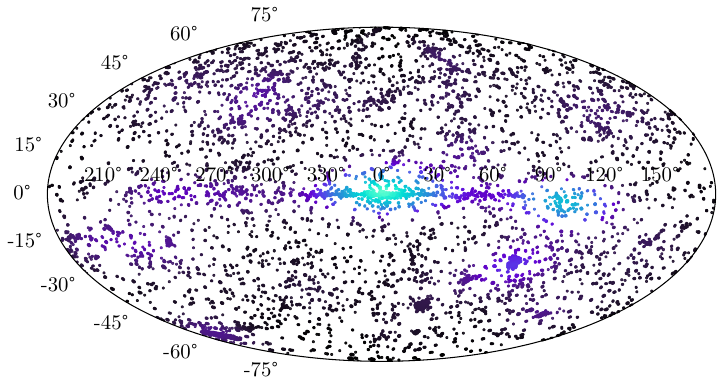}
    \caption{Position of the \NRegUnique \tSig unique variable regions in galactic coordinates.
    The colour map represents number density of transient regions.}
    \label{fig:unique_sources}
\end{figure}

\subsection{Overview}
EXOD was run over 15,105 observations in the XMM archive.
Runs were carried out for 3 different timescales: $t_{\mathrm{bin}} = 5, 50$ and $200$ seconds
and three different energy bands: $0.2-2.0, 2.0-12.0$ and $0.2-12.0$ for a total of 9 different
run subsets. This resulted in 135,945 ($15,105 \times 9$) total runs for a combined total of
$\sim87$ billion processed photon events in $\sim 2.3$ trillion data cells.

116,310/135,945 ($\sim85\%$) runs ran successfully, accounting for approximately, 12,923/15,105 observations.
The remaining unsuccessful runs can be broken down into: 1,228 observations did not have source lists likely due to
very short exposures, 905 observations were in timing mode, 22 observations were in unsupported submodes, and finally,
27 observations failed due to edge-cases.

The spatial binning was fixed at $XY_{\mathrm{bin}} = 20''$, while the global count
rate for identifying bad time intervals was set to either $0.5, 1.0$ or $1.5$ ct/s depending on if there were 1, 2 or 3 simultaneously observing instruments (pn, MOS1, MOS2).
A detection threshold ``sigma equivalent'' of 3 was set for the runs, however we saved the Bayes factors so that this threshold could be later adjusted. In this section, we will denote values arising from the 3 and 5 sigma populations by \tSig and \fSig.

\subsection{Detection statistics}

\begin{table}[h]
\caption{Number of detected regions for different run parameters at the 3 sigma (top) and 5 sigma (bottom)
level.}
\setlength{\extrarowheight}{2pt}
\centering
Total number of detections for 3$\sigma$ population
\begin{tabular}{rrrrr}
\hline
\textbf{Energy Band}& $t_{\mathrm{bin}} = 5$ & $t_{\mathrm{bin}} = 50$ & $t_{\mathrm{bin}} = 200$ & \textbf{Total} \\
\hline
$0.2 - 12.0$ keV    & 12,159  & 7,142   & 12,311  & \textbf{31,612} \\
$0.2 - 2.0$ keV     & 10,861  & 5,843   & 6,147   & \textbf{22,851} \\
$2.0 - 12.0$ keV    & 959    & 1,274   & 3,431   & \textbf{5,664}  \\
\hline
\textbf{Total}    & \textbf{23,979} & \textbf{14,259} & \textbf{21,889} & \textbf{60,127} \\
\end{tabular}
Total number of detections for 5$\sigma$ population
\begin{tabular}{rrrrr}
\hline
\textbf{Energy Band}& $t_{\mathrm{bin}} = 5$ & $t_{\mathrm{bin}} = 50$ & $t_{\mathrm{bin}} = 200$ & \textbf{Total} \\
\hline
$0.2 - 12.0$ keV    & 974   & 1,589  & 2,967  & \textbf{5,530} \\
$0.2 - 2.0$ keV     & 947   & 1,348  & 2,089  & \textbf{4,384} \\
$2.0 - 12.0$ keV    & 126   & 399    & 834    & \textbf{1,359} \\
\hline
\textbf{Total}    & \textbf{2,047} & \textbf{3,336} & \textbf{5,890} & \textbf{11,273} \\
\end{tabular}

\label{tab:run_counts}
\end{table}

Across all 135,954 runs, we detected a total of \NRegTot \tSig or 11,273 \fSig variable regions.
Table \ref{tab:run_counts} shows the number of detected regions separated into
the three energy bands and three time binnings investigated, as well as the number of detected regions for 3 sigma (top)
and 5 sigma (bottom).

To identify unique regions across our runs,
we converted the RA ($\alpha$) and DEC ($\delta$) to Cartesian coordinates on a unit sphere using
$x=\cos{\delta} \times \cos{\alpha}$, $y=\cos{\delta} \times \sin{\alpha}$ and $z=\sin{\delta}$. We then
used a k-d tree, a commonly used spatial clustering algorithm \citep{1975_Bentley}, to cluster together
all detected regions within 20''. This clustering resulted in \NRegUnique \tSig or \NRegUniqueFsig \fSig unique regions.

\begin{table}[h]
\caption{Number of unique (clustered within 20'') detected regions for different run parameters at the 3 sigma (top) and 5 sigma (bottom) level.\\
}
\centering
\setlength{\extrarowheight}{2pt}
Total number of unique detections for 3$\sigma$ population
\begin{tabular}{rrrrr}
\hline
\textbf{Energy Band} & $t_{\mathrm{bin}} = 5$ & $t_{\mathrm{bin}} = 50$ & $t_{\mathrm{bin}} = 200$ & \textbf{Total} \\
\hline
$0.2 - 12.0$ keV & 11,950 & 6,572  & 11,185 & \textbf{25,578} \\
$0.2 - 2.0$ keV  & 10,706 & 5,401  & 5,399  & \textbf{17,322} \\
$2.0 - 12.0$ keV & 897    & 1,087  & 3,057  & \textbf{4,313}  \\
\hline
\textbf{Total}     & \textbf{15,104} & \textbf{9,070}  & \textbf{13,884} & \textbf{32,247} \\
\end{tabular}
Total number of \textbf{unique} detections for 5$\sigma$ population
\begin{tabular}{rrrrr}
\hline
\textbf{Energy Band} & $t_{\mathrm{bin}} = 5$ & $t_{\mathrm{bin}} = 50$ & $t_{\mathrm{bin}} = 200$ & \textbf{Total} \\
\hline
$0.2 - 12.0$ keV   & 898 & 1,294 & 2,496 & \textbf{3,448} \\
$0.2 - 2.0$ keV    & 898 & 1,142 & 1,737 & \textbf{2,667} \\
$2.0 - 12.0$ keV   & 91  & 294   & 651   & \textbf{755} \\
\hline
\textbf{Total}     & \textbf{1,162} & \textbf{1,556} & \textbf{2,853} & \textbf{4,083} \\
\end{tabular}
\tablefoot{Because the clustering occurs within each subpopulation separately, the row totals no longer add up across the different time bins or energy bands,
as unique values are counted within each subpopulation independently.}

\label{tab:run_counts_unique}
\end{table}

Table \ref{tab:run_counts_unique} shows the number of unique regions across our grid of run parameters, while Fig. \ref{fig:unique_sources} shows the sky distribution of these unique regions in galactic coordinates, where the colormap denotes the spatial density.

A clustering radius of 20'' was chosen to match the spatial binning used when creating the data cubes in EXOD,
and was further motivated by the observation of a kink at this value seen when plotting the
number of clusters against the clustering radius.


The percentage of peak detections found in BTIs is 22\% at $3 \sigma$ (26\% at $5 \sigma$), which is slightly higher
than the overall fraction of BTI exposure that is around $\sim 20\%$. This means that there is a slight bias for peak detections in BTIs, which could be due to failures in accounting for the increased background -- however, the fact that there is only a few percentage difference in the fractions also shows that most of the peaks picked up by EXOD during BTIs are genuine.

EXOD found 4,115 unique sources containing at least one
eclipsing (dipping) bin at the 3$\sigma$ threshold and 1,710 at 5$\sigma$ threshold. While many of these eclipses are genuine, they tend to suffer more greatly from instrumental effects than peaks. In particular, for eclipsing (dipping) detections there is a much stronger bias towards detections in BTIs with around $\sim 60\%$ of eclipsing detections being caught during BTIs for both the 3 and 5 sigma thresholds. This over-representation confirms that a part of them are due to limitations of our algorithm. Manual inspection of the eclipses revealed two main contamination scenarios. In the first one, if any of the CCDs turns off during a period of flaring background, this leads to a reduction of the observed counts and potentially a spurious eclipse. In the second scenario, some extended or particularly bright sources are not adequately dealt with using the filtering procedure described in section \ref{sec:expectation_cube_creation} which leads
to spurious eclipses during BTIs.



\subsection{Crossmatching with external catalogues}


To enhance the characterization of EXOD sources, we crossmatched against two existing \textit{XMM-Newton} catalogues 4XMM-DR14 \citep{2020_Webb_A&A...641A.136W} ($\sim650$k sources) and the XMM-OM Serendipitous Source Survey Catalogue (XMM-OM-SUSS6.0) ($\sim6.6$m sources) \citep{2012_Page_MNRAS.426..903P, 2023_Page_yCat.2378....0P}.
Additionally, we crossmatched sources from EXOD with the SIMBAD astronomical database ($\sim 18.8$m sources),
\citep{2000_Wenger_A&AS..143....9W}, GAIA DR3 \citep{2023_Gaia_A&A...674A...1G} ($\sim 1.8$b sources) and the GLADE+ galaxy catalogue
\citep{2022_Dalya_MNRAS.514.1403D} ($\sim 23.2$m sources), we also conduct comparisons to \extras (sec. \ref{ref:extras_comparison}) and 
a catalogue of fast radio bursts (sec. \ref{sec:frb}).

We selected the nearest match within 20'' which are added as columns in the final EXOD catalogue,
cross-matches further than this radius are not included in the catalogue.

\begin{figure}[H]
    \centering
    \includegraphics{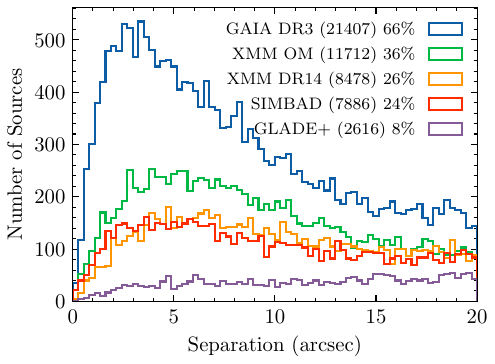}
    \caption{Distribution of separations obtained from cross-matching the \NRegUnique EXOD unique regions
    against several catalogues. The number in brackets shows the number of cross-matches that are found within 20 arcseconds
    for each catalogue respectively, the percentage of the total unique regions \tSig is also shown.}
    \label{fig:cmatch_seperations}
\end{figure}

The distribution of separations is shown in Fig. \ref{fig:cmatch_seperations}, the number in brackets
denoting the number of sources found within 20'' and the associated percentage of the \NRegUnique unique
regions, we find the majority of matches peaking at an absolute separation of 3.33''. By considering only the successful
matches against the 4XMM-DR14 catalogue, we estimate our positional error
to be of $9.75\pm5.22$ arcseconds (1 sigma).

This error is consistent with a simple prediction that can be made from the spatial binning of the data
cube, assuming that the true position of a source can be uniformly distributed anywhere within
a pixel of 20x20 arcseconds, the $1\sigma$ error for both RA and Dec would be given by
$20 / \sqrt{12} \approx 5.8''$ and thus the absolute separation error can be estimated by
$\sqrt{2 \times 5.8^2} \approx 8.16''$.

\begin{figure}
    \includegraphics{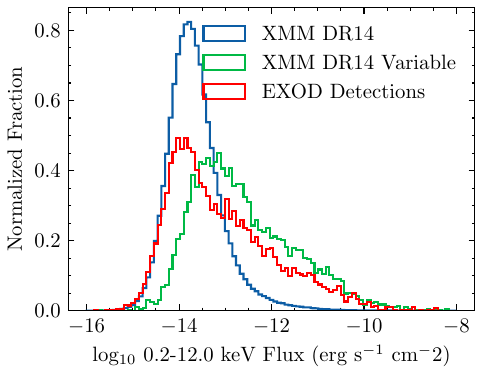}
    \caption{Flux comparison between the whole XMM DR14 Catalogue (blue),
    Variable sources in the DR14 catalogue (green),
    and EXOD sources with a successful 4XMM-DR14 crossmatch (red).}
    \label{fig:Flux_comparison}
\end{figure}

By taking the successfully crossmatched sources with 4XMM-DR14, we can compare the fluxes found
between the transients detected by EXOD and the standard \textit{XMM-Newton} pipeline. From Fig. \ref{fig:Flux_comparison},
we see that EXOD is capable of finding transient behaviour at lower flux levels than the standard pipeline.
We note that this comparison only contains fluxes for previously detected pipeline sources,
meaning that \textbf{in reality the red histogram extends to lower fluxes}.
EXOD does not currently calculate the energy flux for each detection, it only provides the user with a raw photon count. This is not corrected for instance for response change across the off-axis angle, neither does it make any spectral assumption required to convert photon flux into catalog-comparable energy flux. This is an avenue of improvement for the post-processing of EXOD alerts.

\begin{figure}
    \includegraphics{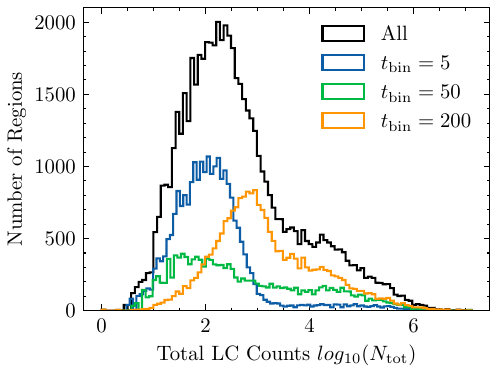}
    \caption{Histogram of the log total number of counts in the lightcurve for all EXOD 
    detections \tSig.}
    \label{fig:total_lc_count_hist}
\end{figure}

Fig. \ref{fig:total_lc_count_hist} shows the $\mathrm{log}_{10}$ of total number of counts in each detected EXOD
lightcurve, shown for the full population and also split by the three different time binnings investigated.
The threshold on the number of counts for variability analysis for the standard XMM-Newton pipeline is set at 100
counts, we can see that we have a considerable number of lightcurves with fewer counts than this in our sample,
further justifying the ability of EXOD to search down to lower detection limits.

\subsubsection{Crossmatch with CHIME FRB catalogue}\label{sec:frb}
We also crossmatched the EXOD sources with the CHIME FRB catalogue \citep{2021_chime_ApJS..257...59C},
which includes 536 fast radio bursts (FRBs) detected by the Canadian Hydrogen Intensity Mapping Experiment (CHIME).
FRBs are extremely bright and rapid radio bursts lasting only a few milliseconds, with their origins still under debate.
Notably, \citet{2020_chime_Natur.587...54C} reported an FRB from the galactic magnetar SGR 1935+2154,
which coincided with a burst of X-ray activity \citep{2020_Palmer_ATel13675....1P}. If this FRB had occurred in an external galaxy,
it would share similar properties to known extragalactic FRBs, suggesting that some FRBs may be associated with X-ray transients from
highly magnetized neutron stars. Our crossmatch identified 67 EXOD sources with positions consistent with 23 CHIME/FRB sources.

We find many of the bursts could be conceivably explained by other astrophysical phenomena based on their SIMBAD crossmatches, or result from
spurious detections or associations, the latter which are more common due to be the large localisation uncertainties for sources in the CHIME
catalogue which are around $0.26 \pm 0.1$ degrees ($925 \pm 367$ arcseconds).
Manual inspection of the EXOD matches, did not reveal any fast flaring behaviour similar to those
seen in SGR 1935+2154, nor did we find any transients with hard spectra similar to those seen in other transient magnetars.

The weakest detections in EXOD contain 3 counts in a 5-second bin, in other words, they reach a peak count rate of $\sim 0.6$ when averaged
over a single 5-second bin. Using this value, we can obtain a crude estimate as to what depth was searched by EXOD for
FRB counterparts, using the \texttt{WebPIMMS} v4.14, we converted a count rate of 0.6 in XMM/pn Thin 5' region to 0.2-12.0 keV flux.
We used a power-law model of $\sim 2.6$, this number obtained from the average value of 16 magnetars in the McGill online
magnetar catalogue \citep{2014_Olausen_ApJS..212....6O}. We used a hydrogen column density of $N_{H} = 5 \times 10^{20} \ \mathrm{cm^{-2}}$
as an approximate value for the sky outside the galactic plane. We obtain an absorbed flux of $\sim 9\times 10^{-13} \ \ergss$ or $\sim 1.4\times 10^{-12} \ \ergss$ if unabsorbed.

\begin{table*}[h]
\caption{Comparison of the number of unique sources in EXOD and \extras for different data subsets in EXOD and EXTraS (see sec \ref{ref:extras_comparison}).}
\centering
\begin{tabular}{l|rrrr}
\hline
                          & EXOD \tSig     & EXOD \tSig     & EXOD \fSig      & EXOD \fSig \\
Metric                    & $\cap$ \extras (full) & $\cap$ \extras (var) & $\cap$ \extras (full) & $\cap$ \extras  (var)     \\
\hline
\textbf{Overlapping Observations} & 4,189                   & 2,466                  & 1,427                   & 1,114         \\
\textbf{EXOD Sources}             & 14,594                  & 10,489                 & 2,101                   & 1,672         \\
\textbf{\extras Sources}          & 225,019                 & 6,442                  & 92,861                  & 3,104         \\
\textbf{EXOD w/ Crossmatch}       & 3,585 (24.6\%)         & 1,506 (14.4\%)        & 1,207 (57.5\%)         & 780 (46.7\%) \\
\textbf{\extras w/ Crossmatch}    & 3,231 (1.4\%)          & 1,319 (20.5\%)        & 1,061 (1.1\%)          & 695 (22.4\%) \\
\hline
\end{tabular}
\tablefoot{\textbf{Overlapping Obs}: The number of unique observations containing detections in both EXOD and \extras data subsets.\\
\textbf{EXOD / \extras Sources}: The total number of unique sources detected in the overlapping observations.\\
\textbf{* Crossmatch} : Number of sources and percentage of total with (w/) intersecting catalogue.}
\label{tab:extras_comparison}
\end{table*}

\subsubsection{Comparison to \extras} \label{ref:extras_comparison}
To date, one of the most exhaustive studies for variability in the \textit{XMM-Newton} archive has been
the \extras project: (Exploring the X-ray transient and variable sky) \cite{2021_De_Luca_A&A...650A.167D}.
The analysis undertaken in this paper contains partial overlap with \extras but also is distinct in a number
of ways. One difference is the data set, in \extras only data up to 3XMM-DR4 is used,
meaning only 7,437 observations up to December 2012 were included. By using 4XMM-DR14 our analysis includes
an additional 10 years of data up to the end of December 2023.
Another key difference is the temporal domain covered, as the shortest new transient discovered by \extras had
a bin width of only 315s, which is greater than the largest binning (200s) we consider in this study.
Despite some of these differences, we can still obtain a preliminary comparison between the transients
detected with EXOD and those detected by EXTraS.

Table \ref{tab:extras_comparison} contains the results from the crossmatch performed between the sources in EXOD 
\tSig \& \fSig and the sources in \extras (full population and variable only\footnote{Variable sources in \extras
were defined as those with \texttt{UB\_LC500\_CO\_PVAL} < $1\times10^{-5}$, see 
\citep{2021_De_Luca_A&A...650A.167D}} population).

The number of overlapping observations is the intersection between 
the 8,670 observations \tSig (2,787 at $5 \sigma$) that contained EXOD detections and the 7,007 (3,312 variable only)
observations that contained \extras detections. The number of EXOD and \extras sources that are 
contained in these overlapping observations is shown in rows two and three of Table \ref{tab:extras_comparison}.
The final two rows of the table display the number of sources and percentage, with (w/)
crossmatches within 20'' for both EXOD and EXTraS.

We see that for the EXOD \tSig population $\sim 25\%$ of sources have a crossmatch in the full \extras 
catalogue. As expected, this value is similar to the percentage obtained between 4XMM-DR14 and EXOD \tSig
(see legend in Fig. \ref{fig:cmatch_seperations}) as the sources in EXTraS closely mirror the sources in the standard
pipeline catalogue. This number increases to $\sim 57 \%$ if we only include the $5 \sigma$ EXOD detections.

Looking at the intersection between EXOD and the EXTraS variable sources (columns 2 and 4), we see that only 
around $\sim 15\%$ of EXOD sources \tSig are associated with a variable source in EXTraS, and conversely
only $\sim 20\%$ of variable EXTraS sources are associated with an EXOD source. This value does increase considerably
for the \fSig EXOD population to around $\sim 45\%$. Despite the overlap between the catalogues, we can still safely say that the variable 
sources detected by EXOD and EXTraS are substantially different from one another to warrant the use of our method
alongside existing methodologies in the literature.

These observations suggest that the EXOD and \extras variable sources differ significantly.
This divergence likely arises from the differing definitions of variability between EXOD and \extras and
also from our analysis targetting shorter time intervals. These results validate our methods' ability
to uncover new transients, distinct from those identified by \extras and the standard \textit{XMM-Newton} pipeline,
justifying its use as an additional tool in X-ray variability studies.

\subsubsection{Estimating the fraction of spurious crossmatches}

When crossmatching catalogues from different wavelengths, a significant fraction of matches can be spurious.
Indeed, taking the extreme case of crossmatching two completely independent sets of positions and uncertainties,
a certain number (given roughly by the product of the sky densities of both catalogues) will overlap and thus be wrongfully associated.
Taking the obtained crossmatched fraction at face value would thus lead to the wrong physical conclusion,
in this example, that the two sets of coordinates are not independent.

Given the high source densities of the external catalogues we
are crossmatching against, combined with the relatively large positional uncertainty of EXOD sources, we expect a certain fraction of our crossmatches to be spurious.
While it may be possible to distinguish the real and spurious associations on a case-by-case basis,
this is a relatively intensive process, and so we quantified these values
through the use of Monte Carlo simulations.

To estimate the effect of spurious matches, we first take the catalogue we are crossmatching against and shift
all positions by a specific amount in RA and Dec. This shifting serves to preserve the statistical properties
(such as the sky density) of the secondary catalogue while essentially creating a new catalogue that should be completely
independent of the sources in EXOD. Crossmatching with this shifted catalogue should yield fewer associations than the unshifted crossmatch,
the difference (excess) in the number of associations thus indicates the number of true, non-spurious matches.
Repeating this process many times allows us to quantify the statistical significance of the number of true matches.

Because the sky distribution of sources is not isotropic, (i.e. there is a higher density in the Galactic plane, LMC etc...),
shifting the secondary catalogue too far would result in very dramatic changes in the number of associations which
would be not representative of the underlying spurious crossmatch values.
Therefore, we chose to shift the coordinates of the secondary catalogue by up to $\pm 15'$ which is approximately
half the EPIC field of view, this value is small enough that local statistical properties such as the sky density are maintained but large enough to suppress true associations.

We focus on the 23,769 / 32,247 sources \tSig in EXOD that did not have counterparts
in 4XMM-DR14 -- these are the sources with the greatest uncertainty in their authenticity.
We perform the aforementioned test with the SIMBAD, Gaia, and OM catalogues.
For each catalogue, we shift their positions by 10 values in both RA and Dec ($\pm 15'$),
each time measuring the number of associations.
We then compare the distribution of these 100 samples against the number of 
matches obtained from the unshifted catalogue.
The results of this experiment are shown in Fig. \ref{fig:Crossmatch_Shift}),
and summarized in Table \ref{tab:cmatch_false_pos}.

\begin{table}[H]
\caption{
Estimation of real and spurious crossmatch associations.
}
\centering
\begin{tabular}{lrrr}
\hline
\textbf{Catalogue} & \textbf{Unshifted} & \textbf{Real} & \textbf{Spurious}\\
\hline
SIMBAD  & 3,489   & $\sim$1,560 (45\%) & $\sim$1,928  (55\%)\\
Gaia    & 14,307  & $\sim$2,433 (17\%) & $\sim$11,873 (83\%)\\
OM      & 6,810   & $\sim$3,503 (51\%) & $\sim$3,306  (49\%)\\
\hline
\end{tabular}
\tablefoot{
The estimation of real and spurious crossmatch associations is performed exclusively on the subset of sources that do not have a 4XMM-DR14 crossmatch (i.e., new sources). It is important to note that a high proportion of spurious crossmatch associations does not indicate that the detected sources themselves are spurious.}

\label{tab:cmatch_false_pos}
\end{table}

As can be seen in Table \ref{tab:cmatch_false_pos} and Fig \ref{fig:Crossmatch_Shift}, the unshifted crossmatches
significantly exceed the shifted crossmatches for all catalogues. Our simulation suggests that for the subset of EXOD sources 
with no counterparts in 4XMM-DR14, approximately half of the SIMBAD and OM associations are real and around $\sim 20\%$ of
the Gaia associations are real, this is expected as Gaia contains almost 1000 times the number of sources as SIMBAD and
the OM catalogues.

There are two main takeaways: first, the fraction of spurious matches is large, so performing this test was essential.
Secondly, even for sources not present in the DR14 catalogue, the excess is very significant. This means that these
sources are most likely actual astrophysical sources. This is an essential result: it confirms that EXOD is able to
detect new faint astrophysical sources, below the standard detection levels of the catalogues. Finally, the large fraction
of spurious matches does not necessarily mean that these EXOD sources are spurious transients and instrumental effects,
but rather that they do not have a multi-wavelength counterpart in the catalogues we have used.


\subsection{Breakdown of SIMBAD crossmatch results}
Table \ref{tab:simbad_cmatch} contains the results of the crossmatch of 
various data subsets with SIMBAD.
The table is sorted by the number of each SIMBAD object classification
for the full EXOD \tSig dataset and object types that had at least 15 sources.

The other columns in the table show the results from the same crossmatch
but performed with different starting datasets, the number below the name
of the dataset denotes the total number of sources in the population.
The \textbf{DR14} and \textbf{DR14 (var)} columns denote the results from
crossmatching the whole 4XMM-D14 catalogue and only the variable sources
(defined as those with \texttt{SC\_VAR\_FLAG=True}) respectively.
The \textbf{EXOD~$\cap$~DR14~($\neg$~var)} and \textbf{EXOD~$\cap$~DR14~(var)} correspond
to the intersection of sources from EXOD with the non-variable and variable sources in
4XMM-DR14 respectively.
Finally, the \textbf{SIMBAD} column shows the total number of each object classification
in the SIMBAD catalogue, this number is helpful in evaluating any possible bias in the crossmatching
process.

The first observation is that only a minority of sources ($\sim25\%$) have possible SIMBAD counterparts,
this is expected as the SIMBAD database only contains $\sim 18.8$m sources across many wavelengths
and is skewed towards bright sources that have been studied in detail, while many of the
sources we have in EXOD are right at the limit of the X-ray detection limit, are new, and will not
have multi-wavelength counterparts.

Of the sources that do have counterparts, we see the largest proportion ($\sim 4\%$) arising from stars.
This makes sense as most stellar flares detected by EXOD will arise from nearby stars which are likely to have a SIMBAD entry.
These stellar flares are also responsible for many other of the stellar classifications,
for example T Tauri stars are a class of young stars and are known to be 
strongly variable in the X-ray and optical bands \citep{1999_Feigelson_ARA&A..37..363F}.

The second most common object type in the EXOD sample would appear to be galaxies
which have a similar abundance to stars with around 1200, but are noticeably lacking in sources identified
as variable in the standard pipeline with only 54. Manual inspection of the lightcurves for these sources
suggests that some of these classifications are correct, such as the known 
QPEs in 2MASX J02344872-441 (eRO-QPE1) \citep{2021_Arcodia_Natur.592..704A} and GSN 069 
\citep{2019_Miniutti_Natur.573..381M}. We also expect a number of these objects to be incorrect associations, owing to the relatively dense distribution of galaxies in SIMBAD. In this case, a faint source in our sample can be associated with a galaxy if there is one nearby even if it is physically independent. It is also possible that some of these associations are true, in which case further studies of short term bursts in galaxies will need to be done to identify the nature of these objects. This is beyond the scope of this current work.

Eclipsing binaries and Long Period Variable stars are also over-represented in EXOD due to large numbers of these object types existing in SIMBAD, which are a result of large catalogues of these systems being created by surveys such as Gaia \citep{2023_Mowlavi_A&A...674A..16M, 2023_Lebzelter_A&A...674A..15L} and others.
However, the contamination rate for these classifications is lower compared to galaxies,
as some of these sources do display genuine transient X-ray behaviour.

Continuing down the list, we find sub-types of galaxies and stars.
We also find extended objects that are not expected to produce X-ray bursts, e.g. ``Part of Cloud'' or ''HII region''.
These are likely due either to spurious associations, or maybe extended X-ray emission that leads to issues in the
EXOD pipeline as previously mentioned.

Known X-ray emitting categories in SIMBAD such as cataclysmic variables (CVs), HMXBs and the general
``X-ray Source'' category contain fewer spurious associations as expected. This is because it is
unlikely (though not impossible) that a spurious EXOD transient would be located at the position of a known X-ray
source.

By looking at the intersection between EXOD sources and variable DR14 sources [EXOD $\cap$ DR14 (var)],
we observe 2,568 source. The highest fraction of common detections between the
two are stars, AGNs, and CVs.

The intersection between the sources detected with EXOD and the sources
in 4XMM-DR14 that are not variable [EXOD $\cap$ DR14 ($\neg$ var)] amounts to 5,910 sources. These sources indicate that EXOD may be more sensitive to certain types of variability than the standard XMM 
pipeline.
This column shows that a number of AGN-related types such as Seyferts and Quasars are identified as being
variable in EXOD but not detected as such in 4XMM-DR14. The reason for this is due to the stochastic, aperiodic
variability commonly seen in these sources \citep{1987_McHardy_Natur.325..696M} leading to significant
departures from the expected emission in a given time bin, while the standard $\chi^2$ variability test
in the XMM pipeline is less sensitive to this type of flickering variability.

\begin{table*}
\caption{A selection of sources detected by EXOD showing novel or previously unreported variability.}
\centering
\begin{tabular}{lrrccl}
\hline
Source Name        & RA ($\alpha$) & DEC ($\delta$)& Mean Flux    & Obs ID      & Classification    \\
                   & "h:m:s"       & "d:m:s"       & $\ergscms$ &            &                   \\
\hline

4XMM J235440.7-373019  & 23:54:40.76   & -37:30:19.4  & $1.01 \times 10^{-14}$     & \obs{0884250101} & Candidate QPE             \\ 
4XMM J175136.9-275858  & 17:51:36.91   & -27:58:58.98   & $5.59 \times 10^{-10}$   & \obs{0886121001} & Candidate Magnetar        \\ 
EXOD J174037.7-304544  & 17:40:37.71   & -30:45:44.0    & $\sim 10^{-9}$           & \obs{0301730101} & Candidate Type I burst    \\ 
EXOD J174603.8-284931  & 17:46:03.84   & -28:49:30.6    & $\sim10^{-10}$           & \obs{0724210501} & PSR J1746-2849            \\ 
\hline
\end{tabular}
\tablefoot{Sources previously detected in 4XMM-DR14 are given by their 4XMM names,
while new sources are given by their EXOD identifiers and their detection coordinates.}
\label{tab:source_list}
\end{table*}

\section{Examples of variable sources detected with EXOD}
In this section, we will focus on a small subset of the sources that were detected by EXOD
that show intriguing behaviour. Table \ref{tab:source_list} lists the sources discussed in
this section and gives the 4XMM identifier if they are present in the 4XMM-DR14 catalogue,
while if they are new sources detected by EXOD
we provide their EXOD identifier as per IAU naming conventions.
For pipeline detected sources, the mean flux from the 4XMM-DR14 catalogue is shown. Note that due to the out bursting nature
of these sources, the mean flux calculated over the whole observation will not well describe
the peak flux or flux change during the burst. We also provide a tentative identification for the source.

\subsection{A candidate QPE in 4XMM J235440.7-373019?}
EXOD detected a transient at RA=23h54m40.8s DEC=-37d30m21s in obsid \obs{0884250101}
taken on 27th May 2021 during a 53ks observation of the exoplanet LTT 9779b.

\begin{figure}[H]
    \includegraphics{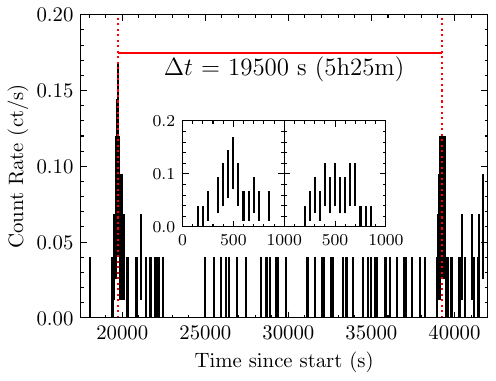}
    \caption{0.2-12 keV lightcurve of the candidate QPE 4XMM J235440.7-373019 binned at 50s originating from the galaxy WISEA J235440.81-373020.6.
    The inset figures show a zoom of $\pm 500$s on the first (left) and second burst (right).}
    \label{fig:exod_qpe_lc}
\end{figure}

The source shows two distinct soft bursts lasting $\sim 1000$s ($\sim 15$m)
each separated by $\sim 19500$s ($\sim$ 5h25m)
(see Fig. \ref{fig:exod_qpe_lc} and is associated with the galaxy WISEA J235440.81-373020.6.
When binned at 200s the mean count rate during the $\sim 20$ks preceding the first
burst was found to be $\sim0.002$ ct/s while the first burst reached a maximum count rate of $\sim 0.28$ ct/s
This represents a factor $\sim 100$ increase in the source count rate.

Visual inspection of images taken by the VLT Survey Telescope 
\citep{2015_Shanks_MNRAS.451.4238S} suggest that this galaxy could potentially be two
overlapping galaxies, however higher resolution imaging and or spectroscopy is required to 
confirm this.

The source is present in the 4XMM-DR14 catalogue with a mean flux of
$1.01 \times 10^{-14} \pm 3.57 \times 10^{-15} \ \mathrm{erg \ s^{-1}}$
and was also previously identified as variable.

\begin{figure}[H]
    \includegraphics{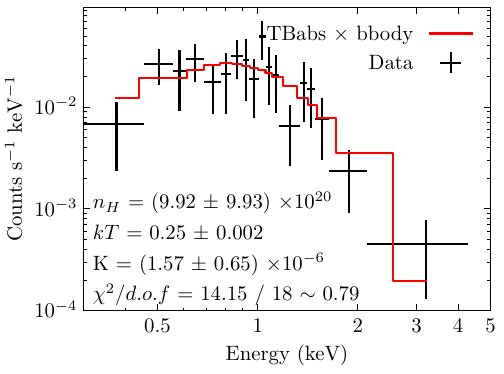}
    \caption{Spectrum of candidate QPE, 4XMM J235440.7-373019 extracted during the two outbursts. Values of $n_{H}$ are given in cm$^{-2}$, and $kT$ in keV.}
    \label{fig:exod_qpe_spec}
\end{figure}

The spectra of the bursts (Fig. \ref{fig:exod_qpe_spec}) appear to be consistent with a black body of temperature $kT$=0.25$\pm 0.002$ keV with a scaling constant $K=(1.57 \pm 0.65) \times 10^{-6}$, and column density of $n_{H} = (9.92 \pm 9.93) \times 10^{20} \ \mathrm{cm^{-2}}$ and a
fit statistic of $\chi^{2}_{d.o.f.} = 14.15 / 18 \sim 0.79$. 
In the absence of any accurate distance measurements, a crude luminosity estimate of $10^{41} - 10^{42} \ \mathrm{erg \ s^{-1}}$ is obtained assuming distances in the range 100-300 Mpc.
The hydrogen column density at the position of the source obtained via the HEASARC $n_{H}$ 
tool\footnote{\href{https://heasarc.gsfc.nasa.gov/cgi-bin/Tools/w3nh/w3nh.pl}{https://heasarc.gsfc.nasa.gov/cgi-bin/Tools/w3nh/w3nh.pl}}
is $\sim 9.7 \times 10^{19} \ \mathrm{cm^{-2}}$.

These values are broadly similar to known QPEs, however with only two observed bursts and a scarcity of the available
data, further investigations are needed to confirm the nature of the source.

\subsection{A magnetar candidate 4XMM J175136.9-275858?}
EXOD detected a hard transient outburst towards the end of observation \obs{0886121001},
at the coordinates 17h51m36.91s -27:58:58.98. The source had previously been detected in the XMM 
catalogue as 4XMM J175136.9-275858 and was also flagged as variable. The source is located towards the Galactic
centre, and no known optical counterparts are found within the X-ray position's error circle.

The source would appear to share many similar spectral and timing properties to known
magnetars. Specifically, the hard spectrum in both quiescence and during the $\sim 1$ks outburst,
combined with the estimated luminosity of $L \sim 10^{33} \ \ergss$ in quiescence and $L \sim 10^{35} \ \ergss$
during the outburst (assuming a distance to the Galactic centre of $\sim 7$ kpc) is a factor 100 above the brightest stellar flares.
SXFTs can also be ruled out as any supergiant companion at these distances should be clearly visible in
other wavelengths. A search for pulsations was conducted but proved inconclusive,
and a full investigation into this source can be found in \cite{2025_Webbe_MNRAS}.

\subsection{A previously unknown hard X-ray burst near the tornado}

On the 5th March 2006, \xmmnewton made a 61ks observation (obsid: \obs{0301730101}) of G357.7-0.1,
a dual-lobed radio and X-ray structure referred to as the ``Tornado'' \citep{1985_Helfand_Natur.313..118H, 2003_Gaensler_ApJ...594L..35G}.
Approximately $\sim 37$ ks into the observation, EXOD detected a large X-ray burst in all three time
binnings (5s, 50s \& 200s) and in both the full (0.2 - 12.0 keV) and hard (2.0 - 12.0 keV) energy bands.
The source was extremely off-axis and thus only in the field of view of MOS2 (thick filter) during the outburst.
The burst is faintly visible in the pipeline processed image, however was not identified as a source by the standardized pipeline.
The observation was reprocessed from the raw CCF files using a circular extraction region for both the source and background, with approximately half of the region being inside the CCD in both cases. The RMF and ARF were created from the spectra extracted in these regions using the \texttt{rmfgen} and \texttt{arfgen} respectively.

\begin{figure}[H]
    \centering
    \includegraphics{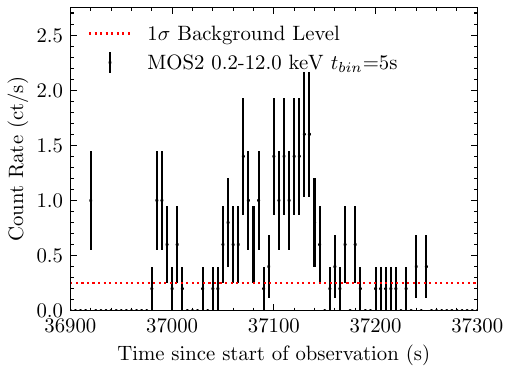}
    \caption{0.2-12.0 keV MOS2 lightcurve of EXOD J174037.7-304544 binned at 5 seconds.}
    \label{fig:tornado_lc}
\end{figure}

\begin{figure}[H]
    \centering
    \includegraphics[width=\columnwidth]{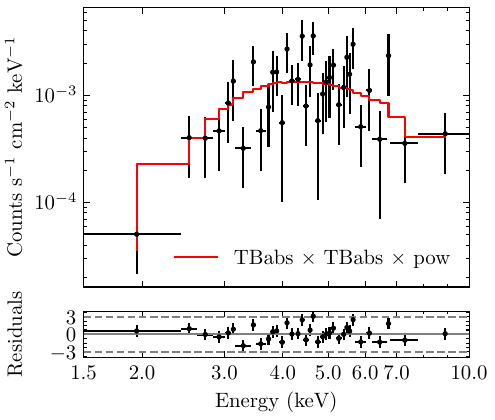}
    \caption{Spectrum of EXOD J174037.7-304544, extracted during the outburst,
    binned at minimum of 3 counts per bin and fit using a doubly absorbed power law.}
    \label{fig:tornado_spec}
\end{figure}

By manually estimating the centroid of the source, we obtain coordinates of 17:40:38.9590 -30:45:54.724 and estimate
approximately 100 counts present within the burst.

Visual inspection of the field in the optical wavelengths shows no plausible counterparts, due to the high
extinction present in the Galactic plane at these wavelengths. In the Near-Infrared (2MASS colour 10806 - 23552 \r{A})
\citep{2006_Skrutskie_AJ....131.1163S} several plausible counterparts are observed, the nearest of these is
denoted 17403893-3045540 at a separation of 0.79'' and has magnitudes of $J = 18.3$, $H = 14.8$, $K_s = 13.2$,
the next closest is 17403844-3045584 at a separation of 7.64'' with $J = 17.183$, $H = 12.806$, $K_s = 10.533$.

The post-processed lightcurve is shown in Fig. \ref{fig:tornado_lc}, which is binned at 5 second intervals
with the $1\sigma$ background assuming a Poisson process at 0.25 ct/s shown as a red dotted line.
The burst is very bright, $\sim 1.8$ raw counts per second at its peak, this value
is estimated to be closer to $\sim 8$ counts per second once absolute corrections are included
(vignetting and PSF). We note that the shape of the burst appears to not
have the typical fast rise / exponential decay that is observed in Type-I bursts, but somewhat
more of a symmetrical profile, although with only $\sim 100$ total counts it is difficult
to assess the true shape of the burst.

We fit the spectrum extracted during the burst (see Fig. \ref{fig:tornado_spec}) with several models in
\texttt{xspec} (see Table \ref{tab:tornado_fits}) using the \texttt{cstat} statistic to account for the low number of counts.

We used both a single and double absorbed power-law model, commonly applied to XRBs, neutron stars or Type-I X-ray bursts,
as well as two models that model the emission from hot, diffuse gas that may be more commonly associated with flaring
young stellar objects. Based on the fits in Table \ref{tab:tornado_fits} it was not possible to determine which
model was preferred. The 0.2 - 12.0 keV absorbed flux derived using the doubly absorbed power law
model gave $1 \times 10^{-9} \ \ergscms$, with a 1-sigma uncertainty range of
$2 \times 10^{-10}$ to $1 \times 10^{-8} \ \ergscms$. The hydrogen column density at the position of the source
via the HEASARC $n_{H}$ tool is $\sim 1.3 \times 10^{22} \ \mathrm{cm^{-2}}$.

Assuming a distance of 8 kpc (approximately the distance to the Galactic centre) we obtain a luminosity of
$1.9 \pm 0.4 \times 10^{36} \ \ergss$.

This luminosity equates to approximately $\sim 0.01 \ L_{\mathrm{Edd}}$ (Eddington) for a 1.4 $M_{\odot}$ NS.
However, the observed spectrum is significantly harder than is usually seen
in type-I X-ray bursts which usually have temperatures of around 0.2 keV \citep{2020_Galloway_ApJS..249...32G}.

Assuming a closer distance of 100 pc, the luminosity is around $\sim 1.2 \times 10^{33} \ \ergss$,
with the 1 sigma range being between $\sim 2.4 \times 10^{32} - 1.2 \times 10^{34} \ \ergss$.
This would place it above the luminosity range for stellar flares, but within the luminosity
range of XRBs, CVs and SFXTs (see sec. \ref{sec:intro}).


\subsection{A possible X-ray counterpart to a galactic radio pulsar}
EXOD J174603.8-284931 is an example of one of the fainter sources that EXOD can detect that are not detected by the standard pipeline. Its lightcurve is consistent with a steady background (at a rate of about $2\times10^{-3}$ photons per second in EXOD pixels), apart from one 5-second-long time bin containing 4 hard photons. This corresponds to a 3.8$\sigma$ detection in the EXOD framework. Out of these photons, three were detected by EPIC~pn in adjacent pixels and time bins (less than one second apart), and the fourth photon was detected by EPIC~MOS2, all with energies between 4 and 7 keV. This indicates that it is unlikely to be an instrumental effect (which tend to happen in a single instrument and in a single pixel). While not extremely significant in itself, this detection is interesting because of its multi-wavelength context.

Indeed, this detection is consistent with the position of PSR~J1746-2849 (at 11"), a known radio pulsar in the Galactic centre. The statistics of
the pulsars detected by EXOD are shown in Table \ref{tab:simbad_cmatch}: out of 3739 SIMBAD pulsars, EXOD detected 23. From those, 22 are in DR14, and 7 are flagged as variable by DR14. EXOD J174603.8-284931 is the only EXOD pulsar that is not in DR14 -- which is understandable given its faintness. The very low counts prevent us from any deep spectral study, however we can roughly estimate the luminosity of the burst (using the WebPIMMS tool\footnote{\href{https://heasarc.gsfc.nasa.gov/cgi-bin/Tools/w3pimms/w3pimms.pl}{https://heasarc.gsfc.nasa.gov/cgi-bin/Tools/w3pimms/w3pimms.pl}}). The radio-measured dispersion measure of PSR~J1746-2849 is significant, at $DM\approx1330$~pc~cm$^{-3}$ \citep{2023_abbate}, corresponding to $N_{H}\approx4\times10^{22}$~cm$^{-2}$ \citep[using the conversion formula from][]{2013He}. Assuming a rate of about 4 counts/s in the 4-7 keV band (which is what was observed), and an absorbed power-law spectrum with $\Gamma=1.7$ and $N_{H}=4\times10^{22}$~cm$^{-2}$, this leads to a 0.2-12~keV flux of around $10^{-10}$ erg~s$^{-1}$~cm$^{-2}$, and thus a luminosity of $\sim10^{35}$~erg~s$^{-1}$. This is consistent with reconnection in the magnetosphere of the pulsar, with a surface magnetic field of $B\approx4\times10^{12}$~G from the ATNF catalog \citep{ATNF2005}.

Although tentative, this detection of a possible X-ray counterpart to a radio pulsar proves the potential of EXOD to detect faint sources, beyond the capabilities of the usual \textit{XMM-Newton} pipeline, and thus provides us with potentially new scientific discoveries from existing archives.

\subsection{Eclipse detections}\label{sec:eclipsing_detections}
Manual inspection of the eclipsing / dipping detections from EXOD
yielded many previously known real astrophysical phenomena that were previously identified as variable by the standard XMM-Newton pipeline. Examples include the modulations observed in the magnetic CV RX J0838.7-2827 \citep[see][]{2017_Rea_MNRAS.471.2902R} and the eclipsing LMXB AX J1745.6-2901 \citep{1996_Maeda_PASJ...48..417M} detected in 26 different observations.

One example of an eclipsing detection
that was not previously identified as variable by the standard pipeline is XMMSL1 J063045.9-603110.
This source was initially detected in a slew observation \citep{2011_Read_ATel.3811....1R}, and displayed a 5$\sigma$ threshold eclipse
during the last $\sim 4000$ seconds of observation 0679381201. 
Initially theorized to be a TDE \citep{2016_Mainetti_A&A...592A..41M} the source was 
reclassified as a Nova following optical spectroscopy by \cite{2017_Oliveira_AJ....153..144O}. This specific eclipse was also picked up by the EXTraS project
\citep{2016_De_Luca_ASSP...42..291D, 2022_Kovacevic_A&A...659A..66K}.

\subsection{BTI detections and a candidate FXT}
Since the standard pipeline removes Bad Time Intervals from the data products, any transients that happen during those will be missed, even if they are bright. We show in  Table \ref{tab:BTI_detections} some examples of transients detected by EXOD during periods of high background that were not identified as variable by the standard pipeline -- most of these are stellar flares in already X-ray emitting stars, but the last source is completely new.

\begin{table}[h]
    \caption{Examples of transients detected by EXOD during periods of high background activity (BTIs) not identified
    as variable by the standard pipeline.}
    \centering
    \begin{tabular}{lll} 
        \toprule
        \textbf{Source Name} & \textbf{Observation} & \textbf{Classification} \\
        \midrule
        4XMM J175134.7-294554 & 0307110101 & Stellar Flare\\
        4XMM J084729.0-525304 & 0201910101 & Stellar Flare\\
        4XMM J054159.2-705711 & 0840820101 & Stellar Flare\\
        4XMM J054239.3-082628 & 0503560301 & Stellar Flare\\
        4XMM J162308.9-225744 & 0801830101 & Stellar Flare\\
        4XMM J205757.5+003458 & 0693221501 & Stellar Flare\\
        4XMM J183237.7-092308 & 0654480101 & Stellar Flare\\
        EXOD J034111.1-284816 & 0653770101 & Candidate FXT \\
        \bottomrule
    \end{tabular}
    \label{tab:BTI_detections}
\end{table}

\begin{figure}[H]
    \centering
    \includegraphics[]{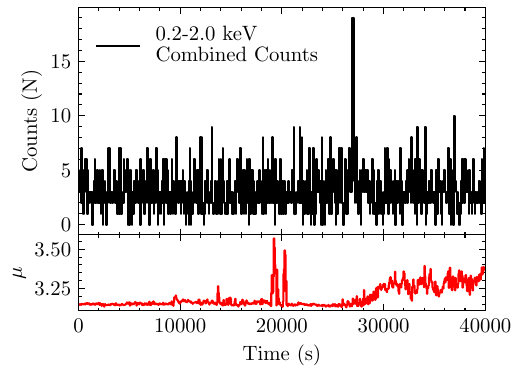}
    \caption{Top: 0.2 - 2.0 keV lightcurve for EXOD J034111.1-284816 binned at 50 seconds, the significant
    burst is around the $\sim 26$ks mark and lasts around $\sim 75$ seconds.
    The counts are combined from all three instruments MOS1, MOS2 \& pn. The bottom panel shows the background
    expectation counts as over the course of observation 0653770101.}
    \label{fig:J034111_lc}
\end{figure}

EXOD J034111.1-284816 displayed a soft (0.2-2.0 keV) burst lasting around
$\sim 75$s, approximately 26ks into observation 0653770101 right during a Bad Time Interval, peaking at around 0.5 counts per second (Fig. \ref{fig:J034111_lc}). Assuming a soft thermal emission (with for instance a black-body temperature of 100 eV), this leads to a peak flux of the order of magnitude of $10^{-13}$ erg s$^{-1}$.
Our crossmatching routine revealed no nearby sources in 4XMM-DR14, SIMBAD or GAIA, although
it may be associated with the AllWise source J034111.03-284815.8 and two faint PanSTARRS sources at a r-band magnitude of about 21 -- all of them falling in the error region.
It is difficult to confirm the nature of this object without more data, but the timing and spectral properties as well as the large X-rays--to--optical ratio makes it a credible FXT candidate.

\section{Discussion}\label{sec:discussion}

\subsection{Strengths and weaknesses}
EXOD demonstrates several key strengths that make it a powerful tool for the search
of rapid X-ray transients. The Bayesian framework, which relies on comparisons with an informed
template, proves to be robust even in the presence of a both temporally and spatially varying 
background. The statistical specification for the quantification of sensitivity in
the Poisson regime enables the detection of transient events in very short time windows
that are dominated by shot noise, and permits searching down to deeper limits than other methodologies
that assume a Gaussian distribution of counts \citep{2012_Masias_MNRAS.422.1674M}.
EXOD also allows the detection of both positive and negative departures from the expected emission,
that is to say it is simultaneously sensitive to both outbursts and eclipses, the latter of which are
often overlooked in classical source detection algorithms.
Finally, EXOD is computationally efficient with average processing time for a single observation in this work taking $\sim 3.3$ seconds 
over 12 cores with our largest bottleneck being memory allocation of the data cubes. The rapid computational time allows for the exploration of many parameter combinations
such as binning and timescales, which make it well suited for large scale searches for transient events.

EXOD does have certain limitations, as previously mentioned. EXOD may find spurious detections
in the presence of extended sources with diffuse emission.
As was mentioned at the end of Section \ref{sec:expectation_cube_creation}, extended sources lead to spurious transients detection within their extent. Despite this
, it is still possible
for EXOD to detect true signals within these diffuse regions, hence we only flag these
detections and do not throw them out. EXOD is also not particularly effective in
identifying long-term low-amplitude variability trends, as it relies on a local comparison in a specific
time bin. However, this type of variability is identified in the \textit{XMM-Newton} pipeline analysis using the chi-squared tests.
Similarly, EXOD is not directly sensitive to structured variability such as repeating signals,
however this could be implemented as a post-processing step. EXOD remains vulnerable
to unpredictable forms of instrumental noise such as hot pixel events or optical loading,
but we have flagged all currently known issues in post-processing.
Finally, since many of our detections are fast and faint, containing only a few photons in a single bin,
standard X-ray analysis methods such as the production of spectra or analysis of light curves is often
unfeasible. This means that we must turn to analysis of the transient population as a whole and rely
on external information such as those obtained via crossmatching in order to overcome these limitations.

Despite these weaknesses, EXOD can and has been successful in identifying reliable variability in sources that were previously
considered non-variable, as well as finding new sources in previously discarded times of high
background.

\subsection{Planned future developments for EXOD}
\subsubsection{Searching at the position of known objects.}
Instead of attempting to pick out all significant signals in the field of view,
a possible alternative approach to finding new transients would be to simply extract 
photon events at the position of known sources. For example, if one was interested in 
finding weak TDE or QPEs, then simply extracting photon events at the position of known
galaxies could prove fruitful.

\subsubsection{Investigation of other energy bands}
This study has only investigated the variability in three broad XMM energy ranges 0.2-12.0, 0.2-2.0 \& 2.0-12.0 keV.
Due to the computational efficiency of EXOD it would not be unreasonable to perform searches over smaller, targeted bands
to search for instances of variability arising from specific spectral emission or absorption lines.
This may be especially useful when extending our method to future instruments with very high X-ray spectral resolutions
such as \textit{Athena}.

\subsubsection{Adaptive time binning}
EXOD currently only tests for significance in individual time bins set by the $t_{\mathrm{bin}}$ parameter, which for this work has been set to 5, 50 and 200s.
It is possible however to miss transients that are not at timescales that are explicitly searched for.
For example, a transient with a duration of 10 seconds may be too faint to be evaluated as significant at 50 seconds binning,
but also may have its photons split over two bins at 5 second binning and therefore not reach the significance threshold in either bin.

One possible solution would be to have an initially lower detection threshold (say $2 \sigma$),
which would first identify these low significance events, then extract the photon events in a small spacial and temporal region surrounding the
candidate, finally we can then rapidly iterate to find out which time binning (and possibly other parameters) would provide the largest significance for the event.
This method would allow for greater flexibility in the types of detected transients and would avoid having to re-run the whole pipeline on the entire archive each
time a new set of parameters was investigated.

\subsubsection{Improved position estimation}
In a similar fashion to the temporal dimension, it is possible that some post-processing steps might improve the precision of the detections in the spatial dimension. For now, the position of a source is taken as the center of the corresponding 20'$\times$20" cell. We could for instance compute the centroid of the photons in the flaring frame in order to get a better estimate of the position of the source.



\subsubsection{Advanced light curve post-processing}
Recent advances in machine learning methods such as neural networks and their variants, have proven
to be effective methods in the modelling, classification and even prediction of future variability trends. Many of these supervised machine learning methods however require well labelled training sets in
order to work effectively. Unsupervised methods such as clustering or anomaly detection may
also prove to be effective in flagging novel variability patterns. To date we have performed some
preliminary experiments using these methods, and we have reason to believe that the dataset 
created through EXOD has strong potential for exploitation via these methods.

\section{Conclusions}
In this paper, we have presented the development of the EPIC XMM Outburst Detector (EXOD).
Through the creation of a novel detection method, we have improved the algorithm's sensitivity
in the low-count (Poisson) regime and are able to probe previously discarded bad time intervals (BTIs).
We highlighted four noteworthy sources, including a possible candidate quasi-periodic eruption (QPE)
found with EXOD that may be of particular interest to the scientific community.

EXOD was run successfully on 12,926 observations in the \xmmnewton archive, which includes data up to December 2023.
We investigated three different energy bands: a full (0.2-12.0 keV), a soft (0.2-2.0 keV) and a hard band (2.0-12.0 keV)
as well as three different time bins (5s, 50s \& 200s), with the aim of capturing a wide range of transient phenomena.

EXOD initially yielded \NRegUnique unique variable regions in 12,926 observations
at the $3 \sigma$ detection level and \NRegUniqueFsig unique regions at the $5 \sigma$ level.

At the $3 \sigma$ detection level, 8,478, or around one quarter 
of the sources detected by EXOD are associated with a previously known source in
the latest \xmmnewton catalogue (4XMM-DR14). At the
$5 \sigma$ detection level, there are 1,766 / 4,083 ($\sim 43\%$).
Conversely, the number of new sources (without a 4XMM-DR14 counterpart) are 
23,769 ($\sim 74\%$) at the $3 \sigma$ level and 2,317 ($\sim 57\%$) at the $5 \sigma$ level.

4XMM-DR14 contains 6,311 variable sources, while the total number of sources detected by EXOD at
the three sigma level is approximately five times this number. Of the 8,478 EXOD sources
with an associated XMM pipeline detection, 2,568 ($\sim 30\%$) had been previously 
flagged as variable in the standard pipeline, while 5,910 ($\sim 70\%$) had not.

Because EXOD is specifically adapted for the search of rapid transient behaviour and
also sensitive to transient events at the limit of the detector sensitivity,
it is expected that the population detected by EXOD only contains partial overlap with the
standard catalogue as both of these regions of parameter space were known to be overlooked
during the creation of the standardized pipeline.

Future papers will focus on the detailed characterisation of newly identified transients,
as well as further refinement of the detection algorithm in order to enhance its detection
sensitivity and reduce the number of false-positive alerts.

\section*{Data access}
We provide two catalogues that were created using EXOD on the \textit{XMM-Newton} archive,
one for detections and one for unique (spatially clustered) sources.
They are provided as Flexible Image Transport System (FITS) files. The EXOD detections catalogue
contains 60,127 rows and 27 columns, while the unique source catalogue contains 32,247 rows and 40 columns.
These Tables are available in electronic form at the CDS via anonymous ftp to cdsarc.u-strasbg.fr (130.79.128.5) or via \href{http://cdsweb.u-strasbg.fr/cgi-bin/qcat?J/A+A/}{http://cdsweb.u-strasbg.fr/cgi-bin/qcat?J/A+A/} \citep{2000_Ochsenbein_A&AS..143...23O}, as well as on the EXOD site: \href{https://xmm-ssc.irap.omp.eu/exod/}{https://xmm-ssc.irap.omp.eu/exod/} which contains further documentation on the tables.

\begin{acknowledgements}
This project has received funding from the European
Union’s Horizon 2020 research and innovation programme under grant agreement
number 101004168, the XMM2ATHENA project \citep{2023_Webb_AN....34420102W}.
NAW and RW also acknowledge support from the CNES.

IT acknowledges support by Deutsches Zentrum f\"ur Luft- und Raumfahrt (DLR) through grant 50\,OX\,2301.

This research has made use of data obtained from the 4XMM
\xmmnewton serendipitous source catalogue compiled by the 10 institutes of
the \xmmnewton Survey Science Centre selected by ESA \citep{2020_Webb_A&A...641A.136W}. This research has
also made use of the SIMBAD database, operated at CDS, Strasbourg, France \citep{2000_Wenger_A&AS..143....9W}.

This research has made use of the NASA/IPAC Extragalactic Database (NED), which is funded by the National Aeronautics and Space Administration and operated by the California Institute of Technology.  \citep{1991_Helou_ASSL..171...89H}
\end{acknowledgements}

\bibliographystyle{aa}
\bibliography{references}

\begin{thebibliography}{110}
\expandafter\ifx\csname natexlab\endcsname\relax\def\natexlab#1{#1}\fi

\bibitem[{Abbate {et~al.}(2023)Abbate, Noutsos, Desvignes, Wharton, Torne,
  Kramer, Eatough, Karuppusamy, Liu, Shao, \& Wongphechauxsorn}]{2023_abbate}
Abbate, F., Noutsos, A., Desvignes, G., {et~al.} 2023, Monthly Notices of the
  Royal Astronomical Society, 524, 2966

\bibitem[{{Aizu}(1973)}]{1973_Aizu_PThPh..49.1184A}
{Aizu}, K. 1973, Progress of Theoretical Physics, 49, 1184

\bibitem[{{Alp} \& {Larsson}(2020)}]{2020_Alp_ApJ...896...39A}
{Alp}, D. \& {Larsson}, J. 2020, \apj, 896, 39

\bibitem[{{Arcodia} {et~al.}(2024){Arcodia}, {Liu}, {Merloni}, {Malyali},
  {Rau}, {Chakraborty}, {Goodwin}, {Buckley}, {Brink}, {Gromadzki},
  {Arzoumanian}, {Buchner}, {Kara}, {Nandra}, {Ponti}, {Salvato}, {Anderson},
  {Baldini}, {Grotova}, {Krumpe}, {Maitra}, {Miller-Jones}, \&
  {Ramos-Ceja}}]{2024_Arcodia_A&A...684A..64A}
{Arcodia}, R., {Liu}, Z., {Merloni}, A., {et~al.} 2024, \aap, 684, A64

\bibitem[{{Arcodia} {et~al.}(2021){Arcodia}, {Merloni}, {Nandra}, {Buchner},
  {Salvato}, {Pasham}, {Remillard}, {Comparat}, {Lamer}, {Ponti}, {Malyali},
  {Wolf}, {Arzoumanian}, {Bogensberger}, {Buckley}, {Gendreau}, {Gromadzki},
  {Kara}, {Krumpe}, {Markwardt}, {Ramos-Ceja}, {Rau}, {Schramm}, \&
  {Schwope}}]{2021_Arcodia_Natur.592..704A}
{Arcodia}, R., {Merloni}, A., {Nandra}, K., {et~al.} 2021, \nat, 592, 704

\bibitem[{{Arcodia} {et~al.}(2022){Arcodia}, {Miniutti}, {Ponti}, {Buchner},
  {Giustini}, {Merloni}, {Nandra}, {Vincentelli}, {Kara}, {Salvato}, \&
  {Pasham}}]{2022_Arcodia_A&A...662A..49A}
{Arcodia}, R., {Miniutti}, G., {Ponti}, G., {et~al.} 2022, \aap, 662, A49

\bibitem[{{Bagnoli} {et~al.}(2015){Bagnoli}, {in't Zand}, {D'Angelo}, \&
  {Galloway}}]{2015_Bagnoli_MNRAS.449..268B}
{Bagnoli}, T., {in't Zand}, J.~J.~M., {D'Angelo}, C.~R., \& {Galloway}, D.~K.
  2015, \mnras, 449, 268

\bibitem[{Bentley(1975)}]{1975_Bentley}
Bentley, J.~L. 1975, Commun. ACM, 18, 509–517

\bibitem[{{Berger}(2014)}]{2014_Berger_ARA&A..52...43B}
{Berger}, E. 2014, \araa, 52, 43

\bibitem[{Bradski(2000)}]{opencv_library}
Bradski, G. 2000, Dr. Dobb's Journal of Software Tools

\bibitem[{{Cash}(1979)}]{1979_Cash_ApJ...228..939C}
{Cash}, W. 1979, \apj, 228, 939

\bibitem[{{Chakraborty} {et~al.}(2021){Chakraborty}, {Kara}, {Masterson},
  {Giustini}, {Miniutti}, \& {Saxton}}]{2021_Chakraborty_ApJ...921L..40C}
{Chakraborty}, J., {Kara}, E., {Masterson}, M., {et~al.} 2021, \apjl, 921, L40

\bibitem[{{Chartas} {et~al.}(2009){Chartas}, {Kochanek}, {Dai}, {Poindexter},
  \& {Garmire}}]{2009_Chartas_ApJ...693..174C}
{Chartas}, G., {Kochanek}, C.~S., {Dai}, X., {Poindexter}, S., \& {Garmire}, G.
  2009, \apj, 693, 174

\bibitem[{{CHIME/FRB Collaboration} {et~al.}(2021){CHIME/FRB Collaboration},
  {Amiri}, {Andersen}, {Bandura}, {Berger}, {Bhardwaj}, {Boyce}, {Boyle},
  {Brar}, {Breitman}, {Cassanelli}, {Chawla}, {Chen}, {Cliche}, {Cook},
  {Cubranic}, {Curtin}, {Deng}, {Dobbs}, {Dong}, {Eadie}, {Fandino}, {Fonseca},
  {Gaensler}, {Giri}, {Good}, {Halpern}, {Hill}, {Hinshaw}, {Josephy},
  {Kaczmarek}, {Kader}, {Kania}, {Kaspi}, {Landecker}, {Lang}, {Leung}, {Li},
  {Lin}, {Masui}, {McKinven}, {Mena-Parra}, {Merryfield}, {Meyers}, {Michilli},
  {Milutinovic}, {Mirhosseini}, {M{\"u}nchmeyer}, {Naidu}, {Newburgh}, {Ng},
  {Patel}, {Pen}, {Petroff}, {Pinsonneault-Marotte}, {Pleunis},
  {Rafiei-Ravandi}, {Rahman}, {Ransom}, {Renard}, {Sanghavi}, {Scholz}, {Shaw},
  {Shin}, {Siegel}, {Sikora}, {Singh}, {Smith}, {Stairs}, {Tan}, {Tendulkar},
  {Vanderlinde}, {Wang}, {Wulf}, \& {Zwaniga}}]{2021_chime_ApJS..257...59C}
{CHIME/FRB Collaboration}, {Amiri}, M., {Andersen}, B.~C., {et~al.} 2021,
  \apjs, 257, 59

\bibitem[{{CHIME/FRB Collaboration} {et~al.}(2020){CHIME/FRB Collaboration},
  {Andersen}, {Bandura}, {Bhardwaj}, {Bij}, {Boyce}, {Boyle}, {Brar},
  {Cassanelli}, {Chawla}, {Chen}, {Cliche}, {Cook}, {Cubranic}, {Curtin},
  {Denman}, {Dobbs}, {Dong}, {Fandino}, {Fonseca}, {Gaensler}, {Giri}, {Good},
  {Halpern}, {Hill}, {Hinshaw}, {H{\"o}fer}, {Josephy}, {Kania}, {Kaspi},
  {Landecker}, {Leung}, {Li}, {Lin}, {Masui}, {McKinven}, {Mena-Parra},
  {Merryfield}, {Meyers}, {Michilli}, {Milutinovic}, {Mirhosseini},
  {M{\"u}nchmeyer}, {Naidu}, {Newburgh}, {Ng}, {Patel}, {Pen},
  {Pinsonneault-Marotte}, {Pleunis}, {Quine}, {Rafiei-Ravandi}, {Rahman},
  {Ransom}, {Renard}, {Sanghavi}, {Scholz}, {Shaw}, {Shin}, {Siegel}, {Singh},
  {Smegal}, {Smith}, {Stairs}, {Tan}, {Tendulkar}, {Tretyakov}, {Vanderlinde},
  {Wang}, {Wulf}, \& {Zwaniga}}]{2020_chime_Natur.587...54C}
{CHIME/FRB Collaboration}, {Andersen}, B.~C., {Bandura}, K.~M., {et~al.} 2020,
  \nat, 587, 54

\bibitem[{{D{\'a}lya} {et~al.}(2022){D{\'a}lya}, {D{\'\i}az}, {Bouchet},
  {Frei}, {Jasche}, {Lavaux}, {Macas}, {Mukherjee}, {P{\'a}lfi}, {de Souza},
  {Wandelt}, {Bilicki}, \& {Raffai}}]{2022_Dalya_MNRAS.514.1403D}
{D{\'a}lya}, G., {D{\'\i}az}, R., {Bouchet}, F.~R., {et~al.} 2022, \mnras, 514,
  1403

\bibitem[{{Dauser} {et~al.}(2019){Dauser}, {Falkner}, {Lorenz}, {Kirsch},
  {Peille}, {Cucchetti}, {Schmid}, {Brand}, {Oertel}, {Smith}, \&
  {Wilms}}]{2019_Dauser_A&A...630A..66D}
{Dauser}, T., {Falkner}, S., {Lorenz}, M., {et~al.} 2019, \aap, 630, A66

\bibitem[{{D'Avanzo} {et~al.}(2012){D'Avanzo}, {Salvaterra}, {Sbarufatti},
  {Nava}, {Melandri}, {Bernardini}, {Campana}, {Covino}, {Fugazza},
  {Ghirlanda}, {Ghisellini}, {La Parola}, {Perri}, {Vergani}, \&
  {Tagliaferri}}]{2012_DAvanzo_MNRAS.425..506D}
{D'Avanzo}, P., {Salvaterra}, R., {Sbarufatti}, B., {et~al.} 2012, \mnras, 425,
  506

\bibitem[{{De Luca} {et~al.}(2021){De Luca}, {Salvaterra}, {Belfiore},
  {Carpano}, {D'Agostino}, {Haberl}, {Israel}, {Law-Green}, {Lisini},
  {Marelli}, {Novara}, {Read}, {Rodriguez-Castillo}, {Rosen}, {Salvetti},
  {Tiengo}, {Vianello}, {Watson}, {Delvaux}, {Dickens}, {Esposito}, {Greiner},
  {H{\"a}mmerle}, {Kreikenbohm}, {Kreykenbohm}, {Oertel}, {Pizzocaro}, {Pye},
  {Sandrelli}, {Stelzer}, {Wilms}, \& {Zagaria}}]{2021_De_Luca_A&A...650A.167D}
{De Luca}, A., {Salvaterra}, R., {Belfiore}, A., {et~al.} 2021, \aap, 650, A167

\bibitem[{{De Luca} {et~al.}(2016){De Luca}, {Salvaterra}, {Tiengo},
  {D'Agostino}, {Watson}, {Haberl}, \& {Wilms}}]{2016_De_Luca_ASSP...42..291D}
{De Luca}, A., {Salvaterra}, R., {Tiengo}, A., {et~al.} 2016, in Astrophysics
  and Space Science Proceedings, Vol.~42, The Universe of Digital Sky Surveys,
  ed. N.~R. {Napolitano}, G.~{Longo}, M.~{Marconi}, M.~{Paolillo}, \&
  E.~{Iodice}, 291

\bibitem[{{Duncan} \& {Thompson}(1992)}]{1992_Duncan_ApJ...392L...9D}
{Duncan}, R.~C. \& {Thompson}, C. 1992, \apjl, 392, L9

\bibitem[{{Eappachen} {et~al.}(2024){Eappachen}, {Jonker},
  {Quirola-V{\'a}squez}, {Mata S{\'a}nchez}, {Inkenhaag}, {Levan}, {Fraser},
  {Torres}, {Bauer}, {Chrimes}, {Stern}, {Graham}, {Smartt}, {Smith},
  {Ravasio}, {Zabludoff}, {Yue}, {Stoppa}, {Malesani}, {Stone}, \&
  {Wen}}]{2024_Eappachen_MNRAS.52711823E}
{Eappachen}, D., {Jonker}, P.~G., {Quirola-V{\'a}squez}, J., {et~al.} 2024,
  \mnras, 527, 11823

\bibitem[{{Evans} {et~al.}(2010){Evans}, {Primini}, {Glotfelty}, {Anderson},
  {Bonaventura}, {Chen}, {Davis}, {Doe}, {Evans}, {Fabbiano}, {Galle}, {Gibbs},
  {Grier}, {Hain}, {Hall}, {Harbo}, {He}, {Houck}, {Karovska}, {Kashyap},
  {Lauer}, {McCollough}, {McDowell}, {Miller}, {Mitschang}, {Morgan},
  {Mossman}, {Nichols}, {Nowak}, {Plummer}, {Refsdal}, {Rots}, {Siemiginowska},
  {Sundheim}, {Tibbetts}, {Van Stone}, {Winkelman}, \&
  {Zografou}}]{2010_Evans_ApJS..189...37E}
{Evans}, I.~N., {Primini}, F.~A., {Glotfelty}, K.~J., {et~al.} 2010, \apjs,
  189, 37

\bibitem[{{Fabbiano}(2006)}]{2006_Fabbiano_ARA&A..44..323F}
{Fabbiano}, G. 2006, \araa, 44, 323

\bibitem[{{Feigelson} \& {Montmerle}(1999)}]{1999_Feigelson_ARA&A..37..363F}
{Feigelson}, E.~D. \& {Montmerle}, T. 1999, \araa, 37, 363

\bibitem[{{Fioretti} {et~al.}(2024){Fioretti}, {Mineo}, {Lotti}, {Molendi},
  {Lanzuisi}, {Amato}, {Macculi}, {Cappi}, {Dadina}, {Ettori}, \&
  {Gastaldello}}]{2024_fioretti}
{Fioretti}, V., {Mineo}, T., {Lotti}, S., {et~al.} 2024, \aap, 691, A229

\bibitem[{{Foschini} {et~al.}(2006){Foschini}, {Pian}, {Maraschi}, {Raiteri},
  {Tavecchio}, {Ghisellini}, {Tosti}, {Malaguti}, \& {Di
  Cocco}}]{2006_Foschini_A&A...450...77F}
{Foschini}, L., {Pian}, E., {Maraschi}, L., {et~al.} 2006, \aap, 450, 77

\bibitem[{{Gaensler} {et~al.}(2003){Gaensler}, {Fogel}, {Slane}, {Miller},
  {Wijnands}, {Eikenberry}, \& {Lewin}}]{2003_Gaensler_ApJ...594L..35G}
{Gaensler}, B.~M., {Fogel}, J.~K.~J., {Slane}, P.~O., {et~al.} 2003, \apjl,
  594, L35

\bibitem[{{Gaia Collaboration} {et~al.}(2023){Gaia Collaboration}, {Vallenari},
  {Brown}, {Prusti}, {de Bruijne}, {Arenou}, {Babusiaux}, {Biermann},
  {Creevey}, {Ducourant}, {Evans}, {Eyer}, {Guerra}, {Hutton}, {Jordi},
  {Klioner}, {Lammers}, {Lindegren}, {Luri}, {Mignard}, {Panem}, {Pourbaix},
  {Randich}, {Sartoretti}, {Soubiran}, {Tanga}, {Walton}, {Bailer-Jones},
  {Bastian}, {Drimmel}, {Jansen}, {Katz}, {Lattanzi}, {van Leeuwen}, {Bakker},
  {Cacciari}, {Casta{\~n}eda}, {De Angeli}, {Fabricius}, {Fouesneau},
  {Fr{\'e}mat}, {Galluccio}, {Guerrier}, {Heiter}, {Masana}, {Messineo},
  {Mowlavi}, {Nicolas}, {Nienartowicz}, {Pailler}, {Panuzzo}, {Riclet}, {Roux},
  {Seabroke}, {Sordo}, {Th{\'e}venin}, {Gracia-Abril}, {Portell}, {Teyssier},
  {Altmann}, {Andrae}, {Audard}, {Bellas-Velidis}, {Benson}, {Berthier},
  {Blomme}, {Burgess}, {Busonero}, {Busso}, {C{\'a}novas}, {Carry}, {Cellino},
  {Cheek}, {Clementini}, {Damerdji}, {Davidson}, {de Teodoro}, {Nu{\~n}ez
  Campos}, {Delchambre}, {Dell'Oro}, {Esquej}, {Fern{\'a}ndez-Hern{\'a}ndez},
  {Fraile}, {Garabato}, {Garc{\'\i}a-Lario}, {Gosset}, {Haigron}, {Halbwachs},
  {Hambly}, {Harrison}, {Hern{\'a}ndez}, {Hestroffer}, {Hodgkin}, {Holl},
  {Jan{\ss}en}, {Jevardat de Fombelle}, {Jordan}, {Krone-Martins}, {Lanzafame},
  {L{\"o}ffler}, {Marchal}, {Marrese}, {Moitinho}, {Muinonen}, {Osborne},
  {Pancino}, {Pauwels}, {Recio-Blanco}, {Reyl{\'e}}, {Riello}, {Rimoldini},
  {Roegiers}, {Rybizki}, {Sarro}, {Siopis}, {Smith}, {Sozzetti}, {Utrilla},
  {van Leeuwen}, {Abbas}, {{\'A}brah{\'a}m}, {Abreu Aramburu}, {Aerts},
  {Aguado}, {Ajaj}, {Aldea-Montero}, {Altavilla}, {{\'A}lvarez}, {Alves},
  {Anders}, {Anderson}, {Anglada Varela}, {Antoja}, {Baines}, {Baker},
  {Balaguer-N{\'u}{\~n}ez}, {Balbinot}, {Balog}, {Barache}, {Barbato},
  {Barros}, {Barstow}, {Bartolom{\'e}}, {Bassilana}, {Bauchet}, {Becciani},
  {Bellazzini}, {Berihuete}, {Bernet}, {Bertone}, {Bianchi}, {Binnenfeld},
  {Blanco-Cuaresma}, {Blazere}, {Boch}, {Bombrun}, {Bossini}, {Bouquillon},
  {Bragaglia}, {Bramante}, {Breedt}, {Bressan}, {Brouillet}, {Brugaletta},
  {Bucciarelli}, {Burlacu}, {Butkevich}, {Buzzi}, {Caffau}, {Cancelliere},
  {Cantat-Gaudin}, {Carballo}, {Carlucci}, {Carnerero}, {Carrasco},
  {Casamiquela}, {Castellani}, {Castro-Ginard}, {Chaoul}, {Charlot}, {Chemin},
  {Chiaramida}, {Chiavassa}, {Chornay}, {Comoretto}, {Contursi}, {Cooper},
  {Cornez}, {Cowell}, {Crifo}, {Cropper}, {Crosta}, {Crowley}, {Dafonte},
  {Dapergolas}, {David}, {David}, {de Laverny}, {De Luise}, {De March}, {De
  Ridder}, {de Souza}, {de Torres}, {del Peloso}, {del Pozo}, {Delbo},
  {Delgado}, {Delisle}, {Demouchy}, {Dharmawardena}, {Di Matteo}, {Diakite},
  {Diener}, {Distefano}, {Dolding}, {Edvardsson}, {Enke}, {Fabre}, {Fabrizio},
  {Faigler}, {Fedorets}, {Fernique}, {Fienga}, {Figueras}, {Fournier},
  {Fouron}, {Fragkoudi}, {Gai}, {Garcia-Gutierrez}, {Garcia-Reinaldos},
  {Garc{\'\i}a-Torres}, {Garofalo}, {Gavel}, {Gavras}, {Gerlach}, {Geyer},
  {Giacobbe}, {Gilmore}, {Girona}, {Giuffrida}, {Gomel}, {Gomez},
  {Gonz{\'a}lez-N{\'u}{\~n}ez}, {Gonz{\'a}lez-Santamar{\'\i}a},
  {Gonz{\'a}lez-Vidal}, {Granvik}, {Guillout}, {Guiraud},
  {Guti{\'e}rrez-S{\'a}nchez}, {Guy}, {Hatzidimitriou}, {Hauser}, {Haywood},
  {Helmer}, {Helmi}, {Sarmiento}, {Hidalgo}, {Hilger}, {H{\l}adczuk}, {Hobbs},
  {Holland}, {Huckle}, {Jardine}, {Jasniewicz}, {Jean-Antoine Piccolo},
  {Jim{\'e}nez-Arranz}, {Jorissen}, {Juaristi Campillo}, {Julbe}, {Karbevska},
  {Kervella}, {Khanna}, {Kontizas}, {Kordopatis}, {Korn}, {K{\'o}sp{\'a}l},
  {Kostrzewa-Rutkowska}, {Kruszy{\'n}ska}, {Kun}, {Laizeau}, {Lambert},
  {Lanza}, {Lasne}, {Le Campion}, {Lebreton}, {Lebzelter}, {Leccia}, {Leclerc},
  {Lecoeur-Taibi}, {Liao}, {Licata}, {Lindstr{\o}m}, {Lister}, {Livanou},
  {Lobel}, {Lorca}, {Loup}, {Madrero Pardo}, {Magdaleno Romeo}, {Managau},
  {Mann}, {Manteiga}, {Marchant}, {Marconi}, {Marcos}, {Marcos Santos},
  {Mar{\'\i}n Pina}, {Marinoni}, {Marocco}, {Marshall}, {Martin Polo},
  {Mart{\'\i}n-Fleitas}, {Marton}, {Mary}, {Masip}, {Massari},
  {Mastrobuono-Battisti}, {Mazeh}, {McMillan}, {Messina}, {Michalik}, {Millar},
  {Mints}, {Molina}, {Molinaro}, {Moln{\'a}r}, {Monari}, {Mongui{\'o}},
  {Montegriffo}, {Montero}, {Mor}, {Mora}, {Morbidelli}, {Morel}, {Morris},
  {Muraveva}, {Murphy}, {Musella}, {Nagy}, {Noval}, {Oca{\~n}a}, {Ogden},
  {Ordenovic}, {Osinde}, {Pagani}, {Pagano}, {Palaversa}, {Palicio},
  {Pallas-Quintela}, {Panahi}, {Payne-Wardenaar}, {Pe{\~n}alosa Esteller},
  {Penttil{\"a}}, {Pichon}, {Piersimoni}, {Pineau}, {Plachy}, {Plum}, {Poggio},
  {Pr{\v{s}}a}, {Pulone}, {Racero}, {Ragaini}, {Rainer}, {Raiteri}, {Rambaux},
  {Ramos}, {Ramos-Lerate}, {Re Fiorentin}, {Regibo}, {Richards}, {Rios Diaz},
  {Ripepi}, {Riva}, {Rix}, {Rixon}, {Robichon}, {Robin}, {Robin}, {Roelens},
  {Rogues}, {Rohrbasser}, {Romero-G{\'o}mez}, {Rowell}, {Royer}, {Ruz Mieres},
  {Rybicki}, {Sadowski}, {S{\'a}ez N{\'u}{\~n}ez}, {Sagrist{\`a} Sell{\'e}s},
  {Sahlmann}, {Salguero}, {Samaras}, {Sanchez Gimenez}, {Sanna},
  {Santove{\~n}a}, {Sarasso}, {Schultheis}, {Sciacca}, {Segol}, {Segovia},
  {S{\'e}gransan}, {Semeux}, {Shahaf}, {Siddiqui}, {Siebert}, {Siltala},
  {Silvelo}, {Slezak}, {Slezak}, {Smart}, {Snaith}, {Solano}, {Solitro},
  {Souami}, {Souchay}, {Spagna}, {Spina}, {Spoto}, {Steele},
  {Steidelm{\"u}ller}, {Stephenson}, {S{\"u}veges}, {Surdej}, {Szabados},
  {Szegedi-Elek}, {Taris}, {Taylor}, {Teixeira}, {Tolomei}, {Tonello}, {Torra},
  {Torra}, {Torralba Elipe}, {Trabucchi}, {Tsounis}, {Turon}, {Ulla}, {Unger},
  {Vaillant}, {van Dillen}, {van Reeven}, {Vanel}, {Vecchiato}, {Viala},
  {Vicente}, {Voutsinas}, {Weiler}, {Wevers}, {Wyrzykowski}, {Yoldas}, {Yvard},
  {Zhao}, {Zorec}, {Zucker}, \& {Zwitter}}]{2023_Gaia_A&A...674A...1G}
{Gaia Collaboration}, {Vallenari}, A., {Brown}, A.~G.~A., {et~al.} 2023, \aap,
  674, A1

\bibitem[{{Galloway} {et~al.}(2020){Galloway}, {in't Zand}, {Chenevez},
  {W{\"o}rpel}, {Keek}, {Ootes}, {Watts}, {Gisler}, {Sanchez-Fernandez}, \&
  {Kuulkers}}]{2020_Galloway_ApJS..249...32G}
{Galloway}, D.~K., {in't Zand}, J., {Chenevez}, J., {et~al.} 2020, \apjs, 249,
  32

\bibitem[{{Galloway} \& {Keek}(2021)}]{2021_Galloway_ASSL..461..209G}
{Galloway}, D.~K. \& {Keek}, L. 2021, in Astrophysics and Space Science
  Library, Vol. 461, Timing Neutron Stars: Pulsations, Oscillations and
  Explosions, ed. T.~M. {Belloni}, M.~{M{\'e}ndez}, \& C.~{Zhang}, 209--262

\bibitem[{{Giustini} {et~al.}(2020){Giustini}, {Miniutti}, \&
  {Saxton}}]{2020_Giustini_A&A...636L...2G}
{Giustini}, M., {Miniutti}, G., \& {Saxton}, R.~D. 2020, \aap, 636, L2

\bibitem[{{Guolo} {et~al.}(2024){Guolo}, {Pasham}, {Zaja{\v{c}}ek}, {Coughlin},
  {Gezari}, {Sukov{\'a}}, {Wevers}, {Witzany}, {Tombesi}, {van Velzen},
  {Alexander}, {Yao}, {Arcodia}, {Karas}, {Miller-Jones}, {Remillard},
  {Gendreau}, \& {Ferrara}}]{2024_Guolo_NatAs...8..347G}
{Guolo}, M., {Pasham}, D.~R., {Zaja{\v{c}}ek}, M., {et~al.} 2024, Nature
  Astronomy, 8, 347

\bibitem[{{Hayashida} {et~al.}(2015){Hayashida}, {Nalewajko}, {Madejski},
  {Sikora}, {Itoh}, {Ajello}, {Blandford}, {Buson}, {Chiang}, {Fukazawa},
  {Furniss}, {Urry}, {Hasan}, {Harrison}, {Alexander}, {Balokovi{\'c}},
  {Barret}, {Boggs}, {Christensen}, {Craig}, {Forster}, {Giommi},
  {Grefenstette}, {Hailey}, {Hornstrup}, {Kitaguchi}, {Koglin}, {Madsen},
  {Mao}, {Miyasaka}, {Mori}, {Perri}, {Pivovaroff}, {Puccetti}, {Rana},
  {Stern}, {Tagliaferri}, {Westergaard}, {Zhang}, {Zoglauer}, {Gurwell},
  {Uemura}, {Akitaya}, {Kawabata}, {Kawaguchi}, {Kanda}, {Moritani}, {Takaki},
  {Ui}, {Yoshida}, {Agarwal}, \& {Gupta}}]{2015_Hayashida_ApJ...807...79H}
{Hayashida}, M., {Nalewajko}, K., {Madejski}, G.~M., {et~al.} 2015, \apj, 807,
  79

\bibitem[{{He} {et~al.}(2013){He}, {Ng}, \& {Kaspi}}]{2013He}
{He}, C., {Ng}, C.~Y., \& {Kaspi}, V.~M. 2013, \apj, 768, 64

\bibitem[{{Helfand} \& {Becker}(1985)}]{1985_Helfand_Natur.313..118H}
{Helfand}, D.~J. \& {Becker}, R.~H. 1985, \nat, 313, 118

\bibitem[{{Helou} {et~al.}(1991){Helou}, {Madore}, {Schmitz}, {Bicay}, {Wu}, \&
  {Bennett}}]{1991_Helou_ASSL..171...89H}
{Helou}, G., {Madore}, B.~F., {Schmitz}, M., {et~al.} 1991, in Astrophysics and
  Space Science Library, Vol. 171, Databases and On-line Data in Astronomy, ed.
  M.~A. {Albrecht} \& D.~{Egret}, 89--106

\bibitem[{{Imanishi} {et~al.}(2003){Imanishi}, {Nakajima}, {Tsujimoto},
  {Koyama}, \& {Tsuboi}}]{2003_Imanishi_PASJ...55..653I}
{Imanishi}, K., {Nakajima}, H., {Tsujimoto}, M., {Koyama}, K., \& {Tsuboi}, Y.
  2003, \pasj, 55, 653

\bibitem[{{Jansen} {et~al.}(2001){Jansen}, {Lumb}, {Altieri}, {Clavel}, {Ehle},
  {Erd}, {Gabriel}, {Guainazzi}, {Gondoin}, {Much}, {Munoz}, {Santos},
  {Schartel}, {Texier}, \& {Vacanti}}]{2001_Jansen_A&A...365L...1J}
{Jansen}, F., {Lumb}, D., {Altieri}, B., {et~al.} 2001, \aap, 365, L1

\bibitem[{{Jonker} {et~al.}(2013){Jonker}, {Glennie}, {Heida}, {Maccarone},
  {Hodgkin}, {Nelemans}, {Miller-Jones}, {Torres}, \&
  {Fender}}]{2013_Jonker_ApJ...779...14J}
{Jonker}, P.~G., {Glennie}, A., {Heida}, M., {et~al.} 2013, \apj, 779, 14

\bibitem[{{Kapanadze} {et~al.}(2016){Kapanadze}, {Dorner}, {Vercellone},
  {Romano}, {Aller}, {Aller}, {Hughes}, {Reynolds}, {Kapanadze}, \&
  {Tabagari}}]{2016_Kapanadze_ApJ...831..102K}
{Kapanadze}, B., {Dorner}, D., {Vercellone}, S., {et~al.} 2016, \apj, 831, 102

\bibitem[{{Kaspi} \& {Beloborodov}(2017)}]{2017_Kaspi_ARA&A..55..261K}
{Kaspi}, V.~M. \& {Beloborodov}, A.~M. 2017, \araa, 55, 261

\bibitem[{{Kato} {et~al.}(2015){Kato}, {Saio}, \&
  {Hachisu}}]{2015_Kato_ApJ...808...52K}
{Kato}, M., {Saio}, H., \& {Hachisu}, I. 2015, \apj, 808, 52

\bibitem[{{King} {et~al.}(2023){King}, {Lasota}, \&
  {Middleton}}]{2023_King_NewAR..9601672K}
{King}, A., {Lasota}, J.-P., \& {Middleton}, M. 2023, \nar, 96, 101672

\bibitem[{{K{\"o}nig} {et~al.}(2022){K{\"o}nig}, {Wilms}, {Arcodia}, {Dauser},
  {Dennerl}, {Doroshenko}, {Haberl}, {H{\"a}mmerich}, {Kirsch}, {Kreykenbohm},
  {Lorenz}, {Malyali}, {Merloni}, {Rau}, {Rauch}, {Sala}, {Schwope},
  {Suleimanov}, {Weber}, \& {Werner}}]{2022_Konig_Natur.605..248K}
{K{\"o}nig}, O., {Wilms}, J., {Arcodia}, R., {et~al.} 2022, \nat, 605, 248

\bibitem[{{Kova{\v{c}}evi{\'c}} {et~al.}(2022){Kova{\v{c}}evi{\'c}},
  {Pasquato}, {Marelli}, {De Luca}, {Salvaterra}, \&
  {Belfiore}}]{2022_Kovacevic_A&A...659A..66K}
{Kova{\v{c}}evi{\'c}}, M., {Pasquato}, M., {Marelli}, M., {et~al.} 2022, \aap,
  659, A66

\bibitem[{{Kowalski}(2024)}]{2024_Kowalski_LRSP...21....1K}
{Kowalski}, A.~F. 2024, Living Reviews in Solar Physics, 21, 1

\bibitem[{{Kraft} {et~al.}(1991){Kraft}, {Burrows}, \&
  {Nousek}}]{1991_Kraft_ApJ...374..344K}
{Kraft}, R.~P., {Burrows}, D.~N., \& {Nousek}, J.~A. 1991, \apj, 374, 344

\bibitem[{{Kumar} \& {Zhang}(2015)}]{2015_Kumar_PhR...561....1K}
{Kumar}, P. \& {Zhang}, B. 2015, \physrep, 561, 1

\bibitem[{{Kuulkers} {et~al.}(2003){Kuulkers}, {den Hartog}, {in't Zand},
  {Verbunt}, {Harris}, \& {Cocchi}}]{2003_Kuulkers_A&A...399..663K}
{Kuulkers}, E., {den Hartog}, P.~R., {in't Zand}, J.~J.~M., {et~al.} 2003,
  \aap, 399, 663

\bibitem[{{Kuulkers} {et~al.}(2010){Kuulkers}, {Norton}, {Schwope}, \&
  {Warner}}]{2010_Kuulkers_csxs.book..421K}
{Kuulkers}, E., {Norton}, A., {Schwope}, A., \& {Warner}, B. 2010, in Compact
  Stellar X-ray Sources, ed. W.~{Lewin} \& M.~{van der Klis}, 421

\bibitem[{{Lebzelter} {et~al.}(2023){Lebzelter}, {Mowlavi}, {Lecoeur-Taibi},
  {Trabucchi}, {Audard}, {Garc{\'\i}a-Lario}, {Gavras}, {Holl}, {Jevardat de
  Fombelle}, {Nienartowicz}, {Rimoldini}, \&
  {Eyer}}]{2023_Lebzelter_A&A...674A..15L}
{Lebzelter}, T., {Mowlavi}, N., {Lecoeur-Taibi}, I., {et~al.} 2023, \aap, 674,
  A15

\bibitem[{{Lichti} {et~al.}(2008){Lichti}, {Bottacini}, {Ajello}, {Charlot},
  {Collmar}, {Falcone}, {Horan}, {Huber}, {von Kienlin}, {L{\"a}hteenm{\"a}ki},
  {Lindfors}, {Morris}, {Nilsson}, {Petry}, {R{\"u}ger}, {Sillanp{\"a}{\"a}},
  {Spanier}, \& {Tornikoski}}]{2008_Lichti_A&A...486..721L}
{Lichti}, G.~G., {Bottacini}, E., {Ajello}, M., {et~al.} 2008, \aap, 486, 721

\bibitem[{{Linial} \& {Quataert}(2024)}]{2024_Linial_MNRAS.527.4317L}
{Linial}, I. \& {Quataert}, E. 2024, \mnras, 527, 4317

\bibitem[{{Maeda} {et~al.}(1996){Maeda}, {Koyama}, {Sakano}, {Takeshima}, \&
  {Yamauchi}}]{1996_Maeda_PASJ...48..417M}
{Maeda}, Y., {Koyama}, K., {Sakano}, M., {Takeshima}, T., \& {Yamauchi}, S.
  1996, \pasj, 48, 417

\bibitem[{{Maguire} {et~al.}(2020){Maguire}, {Eracleous}, {Jonker}, {MacLeod},
  \& {Rosswog}}]{2020_Maguire_SSRv..216...39M}
{Maguire}, K., {Eracleous}, M., {Jonker}, P.~G., {MacLeod}, M., \& {Rosswog},
  S. 2020, \ssr, 216, 39

\bibitem[{{Mainetti} {et~al.}(2016){Mainetti}, {Campana}, \&
  {Colpi}}]{2016_Mainetti_A&A...592A..41M}
{Mainetti}, D., {Campana}, S., \& {Colpi}, M. 2016, \aap, 592, A41

\bibitem[{{Manchester} {et~al.}(2005){Manchester}, {Hobbs}, {Teoh}, \&
  {Hobbs}}]{ATNF2005}
{Manchester}, R.~N., {Hobbs}, G.~B., {Teoh}, A., \& {Hobbs}, M. 2005, \aj, 129,
  1993

\bibitem[{{Masias} {et~al.}(2012){Masias}, {Freixenet}, {Llad{\'o}}, \&
  {Peracaula}}]{2012_Masias_MNRAS.422.1674M}
{Masias}, M., {Freixenet}, J., {Llad{\'o}}, X., \& {Peracaula}, M. 2012,
  \mnras, 422, 1674

\bibitem[{{McHardy} \& {Czerny}(1987)}]{1987_McHardy_Natur.325..696M}
{McHardy}, I. \& {Czerny}, B. 1987, \nat, 325, 696

\bibitem[{{Miniutti} {et~al.}(2023){Miniutti}, {Giustini}, {Arcodia}, {Saxton},
  {Chakraborty}, {Read}, \& {Kara}}]{2023_Miniutti_A&A...674L...1M}
{Miniutti}, G., {Giustini}, M., {Arcodia}, R., {et~al.} 2023, \aap, 674, L1

\bibitem[{{Miniutti} {et~al.}(2019){Miniutti}, {Saxton}, {Giustini},
  {Alexander}, {Fender}, {Heywood}, {Monageng}, {Coriat}, {Tzioumis}, {Read},
  {Knigge}, {Gandhi}, {Pretorius}, \&
  {Ag{\'\i}s-Gonz{\'a}lez}}]{2019_Miniutti_Natur.573..381M}
{Miniutti}, G., {Saxton}, R.~D., {Giustini}, M., {et~al.} 2019, \nat, 573, 381

\bibitem[{{Morgan} {et~al.}(2012){Morgan}, {Hainline}, {Chen}, {Tewes},
  {Kochanek}, {Dai}, {Kozlowski}, {Blackburne}, {Mosquera}, {Chartas},
  {Courbin}, \& {Meylan}}]{2012_Morgan_ApJ...756...52M}
{Morgan}, C.~W., {Hainline}, L.~J., {Chen}, B., {et~al.} 2012, \apj, 756, 52

\bibitem[{{Mowlavi} {et~al.}(2023){Mowlavi}, {Holl}, {Lecoeur-Ta{\"\i}bi},
  {Barblan}, {Kochoska}, {Pr{\v{s}}a}, {Mazeh}, {Rimoldini}, {Gavras},
  {Audard}, {Jevardat de Fombelle}, {Nienartowicz}, {Garc{\'\i}a-Lario}, \&
  {Eyer}}]{2023_Mowlavi_A&A...674A..16M}
{Mowlavi}, N., {Holl}, B., {Lecoeur-Ta{\"\i}bi}, I., {et~al.} 2023, \aap, 674,
  A16

\bibitem[{{Nicholl} {et~al.}(2024){Nicholl}, {Pasham}, {Mummery}, {Guolo},
  {Gendreau}, {Dewangan}, {Ferrara}, {Remillard}, {Bonnerot}, {Chakraborty},
  {Hajela}, {Dhillon}, {Gillan}, {Greenwood}, {Huber}, {Janiuk}, {Salvesen},
  {van Velzen}, {Aamer}, {Alexander}, {Angus}, {Arzoumanian}, {Auchettl},
  {Berger}, {de Boer}, {Cendes}, {Chambers}, {Chen}, {Chornock}, {Fulton},
  {Gao}, {Gillanders}, {Gomez}, {Gompertz}, {Fabian}, {Herman}, {Ingram},
  {Kara}, {Laskar}, {Lawrence}, {Lin}, {Lowe}, {Magnier}, {Margutti}, {McGee},
  {Minguez}, {Moore}, {Nathan}, {Oates}, {Patra}, {Ramsden}, {Ravi}, {Ridley},
  {Sheng}, {Smartt}, {Smith}, {Srivastav}, {Stein}, {Stevance}, {Turner},
  {Wainscoat}, {Weston}, {Wevers}, \& {Young}}]{2024_Nicholl_Natur.634..804N}
{Nicholl}, M., {Pasham}, D.~R., {Mummery}, A., {et~al.} 2024, \nat, 634, 804

\bibitem[{{Ochsenbein} {et~al.}(2000){Ochsenbein}, {Bauer}, \&
  {Marcout}}]{2000_Ochsenbein_A&AS..143...23O}
{Ochsenbein}, F., {Bauer}, P., \& {Marcout}, J. 2000, \aaps, 143, 23

\bibitem[{{Olausen} \& {Kaspi}(2014)}]{2014_Olausen_ApJS..212....6O}
{Olausen}, S.~A. \& {Kaspi}, V.~M. 2014, \apjs, 212, 6

\bibitem[{{Oliveira} {et~al.}(2017){Oliveira}, {Rodrigues}, {Cieslinski},
  {Jablonski}, {Silva}, {Almeida}, {Rodr{\'\i}guez-Ardila}, \&
  {Palhares}}]{2017_Oliveira_AJ....153..144O}
{Oliveira}, A.~S., {Rodrigues}, C.~V., {Cieslinski}, D., {et~al.} 2017, \aj,
  153, 144

\bibitem[{{Page} \& {Shaw}(2022)}]{2022_Page_hxga.book..107P}
{Page}, K.~L. \& {Shaw}, A.~W. 2022, in Handbook of X-ray and Gamma-ray
  Astrophysics, ed. C.~{Bambi} \& A.~{Sangangelo}, 107

\bibitem[{{Page} {et~al.}(2012){Page}, {Brindle}, {Talavera}, {Still}, {Rosen},
  {Yershov}, {Ziaeepour}, {Mason}, {Cropper}, {Breeveld}, {Loiseau}, {Mignani},
  {Smith}, \& {Murdin}}]{2012_Page_MNRAS.426..903P}
{Page}, M.~J., {Brindle}, C., {Talavera}, A., {et~al.} 2012, \mnras, 426, 903

\bibitem[{{Page} {et~al.}(2023){Page}, {Brindle}, {Talavera}, {Still}, {Rosen},
  {Yershov}, {Ziaeepour}, {Mason}, {Cropper}, {Breeveld}, {Loiseau}, {Mignani},
  {Smith}, \& {Murdin}}]{2023_Page_yCat.2378....0P}
{Page}, M.~J., {Brindle}, C., {Talavera}, A., {et~al.} 2023, {VizieR Online
  Data Catalog: XMM-OM Serendipitous Source Survey Catalogue (XMM-SUSS6.0)
  (Page+, 2023)}, VizieR On-line Data Catalog: II/378. Originally published in:
  2012MNRAS.426..903P

\bibitem[{{Palmer}(2020)}]{2020_Palmer_ATel13675....1P}
{Palmer}, D.~M. 2020, The Astronomer's Telegram, 13675, 1

\bibitem[{{Pastor-Marazuela} {et~al.}(2020){Pastor-Marazuela}, {Webb},
  {Wojtowicz}, \& {van Leeuwen}}]{2020_Pastor_Marazuela_A&A...640A.124P}
{Pastor-Marazuela}, I., {Webb}, N.~A., {Wojtowicz}, D.~T., \& {van Leeuwen}, J.
  2020, \aap, 640, A124

\bibitem[{{Pye} {et~al.}(2015){Pye}, {Rosen}, {Fyfe}, \&
  {Schr{\"o}der}}]{2015_Pye_A&A...581A..28P}
{Pye}, J.~P., {Rosen}, S., {Fyfe}, D., \& {Schr{\"o}der}, A.~C. 2015, \aap,
  581, A28

\bibitem[{{Quintin} {et~al.}(2023){Quintin}, {Webb}, {Guillot}, {Miniutti},
  {Kammoun}, {Giustini}, {Arcodia}, {Soucail}, {Clerc}, {Amato}, \&
  {Markwardt}}]{2023_Quintin_A&A...675A.152Q}
{Quintin}, E., {Webb}, N.~A., {Guillot}, S., {et~al.} 2023, \aap, 675, A152

\bibitem[{{Quirola-V{\'a}squez} {et~al.}(2022){Quirola-V{\'a}squez}, {Bauer},
  {Jonker}, {Brandt}, {Yang}, {Levan}, {Xue}, {Eappachen}, {Zheng}, \&
  {Luo}}]{2022_Quirola_A&A...663A.168Q}
{Quirola-V{\'a}squez}, J., {Bauer}, F.~E., {Jonker}, P.~G., {et~al.} 2022,
  \aap, 663, A168

\bibitem[{{Rea} {et~al.}(2017){Rea}, {Coti Zelati}, {Esposito}, {D'Avanzo}, {de
  Martino}, {Israel}, {Torres}, {Campana}, {Belloni}, {Papitto}, {Masetti},
  {Carrasco}, {Possenti}, {Wieringa}, {Wilhelmi}, {Li}, {Bozzo}, {Ferrigno},
  {Linares}, {Tauris}, {Hernanz}, {Ribas}, {Monelli}, {Borghese}, {Baglio}, \&
  {Casares}}]{2017_Rea_MNRAS.471.2902R}
{Rea}, N., {Coti Zelati}, F., {Esposito}, P., {et~al.} 2017, \mnras, 471, 2902

\bibitem[{{Read} {et~al.}(2011){Read}, {Saxton}, \&
  {Esquej}}]{2011_Read_ATel.3811....1R}
{Read}, A.~M., {Saxton}, R.~D., \& {Esquej}, P. 2011, The Astronomer's
  Telegram, 3811, 1

\bibitem[{{Rees}(1988)}]{1988_Rees_Natur.333..523R}
{Rees}, M.~J. 1988, \nat, 333, 523

\bibitem[{{Ridnaia} {et~al.}(2021){Ridnaia}, {Svinkin}, {Frederiks}, {Bykov},
  {Popov}, {Aptekar}, {Golenetskii}, {Lysenko}, {Tsvetkova}, {Ulanov}, \&
  {Cline}}]{2021_Ridnaia_NatAs...5..372R}
{Ridnaia}, A., {Svinkin}, D., {Frederiks}, D., {et~al.} 2021, Nature Astronomy,
  5, 372

\bibitem[{{Romano} {et~al.}(2023){Romano}, {Evans}, {Bozzo}, {Mangano},
  {Vercellone}, {Guidorzi}, {Ducci}, {Kennea}, {Barthelmy}, {Palmer}, {Krimm},
  \& {Cenko}}]{2023_Romano_A&A...670A.127R}
{Romano}, P., {Evans}, P.~A., {Bozzo}, E., {et~al.} 2023, \aap, 670, A127

\bibitem[{{Rosen} {et~al.}(2016){Rosen}, {Webb}, {Watson}, {Ballet}, {Barret},
  {Braito}, {Carrera}, {Ceballos}, {Coriat}, {Della Ceca}, {Denkinson},
  {Esquej}, {Farrell}, {Freyberg}, {Gris{\'e}}, {Guillout}, {Heil},
  {Koliopanos}, {Law-Green}, {Lamer}, {Lin}, {Martino}, {Michel}, {Motch},
  {Nebot Gomez-Moran}, {Page}, {Page}, {Page}, {Pakull}, {Pye}, {Read},
  {Rodriguez}, {Sakano}, {Saxton}, {Schwope}, {Scott}, {Sturm}, {Traulsen},
  {Yershov}, \& {Zolotukhin}}]{2016_Rosen_A&A...590A...1R}
{Rosen}, S.~R., {Webb}, N.~A., {Watson}, M.~G., {et~al.} 2016, \aap, 590, A1

\bibitem[{{Ruiz} {et~al.}(2024){Ruiz}, {Georgakakis}, {Georgantopoulos},
  {Akylas}, {Pierre}, \& {Starck}}]{2024_Ruiz_MNRAS.527.3674R}
{Ruiz}, A., {Georgakakis}, A., {Georgantopoulos}, I., {et~al.} 2024, \mnras,
  527, 3674

\bibitem[{{Ruiz} {et~al.}(2022){Ruiz}, {Georgakakis}, {Gerakakis}, {Saxton},
  {Kretschmar}, {Akylas}, \& {Georgantopoulos}}]{2022_Ruiz_MNRAS.511.4265R}
{Ruiz}, A., {Georgakakis}, A., {Gerakakis}, S., {et~al.} 2022, \mnras, 511,
  4265

\bibitem[{{Saxton} {et~al.}(2021){Saxton}, {Komossa}, {Auchettl}, \&
  {Jonker}}]{2021_Saxton_SSRv..217...18S}
{Saxton}, R., {Komossa}, S., {Auchettl}, K., \& {Jonker}, P.~G. 2021,
  {Correction to: X-Ray Properties of TDEs}, Space Science Reviews, Volume 217,
  Issue 1, article id.18

\bibitem[{{Schwope} {et~al.}(2024){Schwope}, {Knauff}, {Kurpas}, {Salvato},
  {Stelzer}, {St{\"u}tz}, \&
  {Tub{\'\i}n-Arenas}}]{2024_Schwope_A&A...690A.243S}
{Schwope}, A.~D., {Knauff}, K., {Kurpas}, J., {et~al.} 2024, \aap, 690, A243

\bibitem[{{Schwope} {et~al.}(2001){Schwope}, {Schwarz}, {Sirk}, \&
  {Howell}}]{2001_Schwope_A&A...375..419S}
{Schwope}, A.~D., {Schwarz}, R., {Sirk}, M., \& {Howell}, S.~B. 2001, \aap,
  375, 419

\bibitem[{{Shanks} {et~al.}(2015){Shanks}, {Metcalfe}, {Chehade}, {Findlay},
  {Irwin}, {Gonzalez-Solares}, {Lewis}, {Yoldas}, {Mann}, {Read}, {Sutorius},
  \& {Voutsinas}}]{2015_Shanks_MNRAS.451.4238S}
{Shanks}, T., {Metcalfe}, N., {Chehade}, B., {et~al.} 2015, \mnras, 451, 4238

\bibitem[{{Sidoli}(2013)}]{2013_Sidoli_arXiv1301.7574S}
{Sidoli}, L. 2013, arXiv e-prints, arXiv:1301.7574

\bibitem[{{Sidoli}(2017)}]{2017_Sidoli_mbhe.confE..52S}
{Sidoli}, L. 2017, in XII Multifrequency Behaviour of High Energy Cosmic
  Sources Workshop (MULTIF2017), 52

\bibitem[{{Skrutskie} {et~al.}(2006){Skrutskie}, {Cutri}, {Stiening},
  {Weinberg}, {Schneider}, {Carpenter}, {Beichman}, {Capps}, {Chester},
  {Elias}, {Huchra}, {Liebert}, {Lonsdale}, {Monet}, {Price}, {Seitzer},
  {Jarrett}, {Kirkpatrick}, {Gizis}, {Howard}, {Evans}, {Fowler}, {Fullmer},
  {Hurt}, {Light}, {Kopan}, {Marsh}, {McCallon}, {Tam}, {Van Dyk}, \&
  {Wheelock}}]{2006_Skrutskie_AJ....131.1163S}
{Skrutskie}, M.~F., {Cutri}, R.~M., {Stiening}, R., {et~al.} 2006, \aj, 131,
  1163

\bibitem[{{Sridhar} {et~al.}(2021){Sridhar}, {Metzger}, {Beniamini},
  {Margalit}, {Renzo}, {Sironi}, \& {Kovlakas}}]{2021_Sridhar_ApJ...917...13S}
{Sridhar}, N., {Metzger}, B.~D., {Beniamini}, P., {et~al.} 2021, \apj, 917, 13

\bibitem[{{Srivastav} {et~al.}(2025){Srivastav}, {Chen}, {Gillanders},
  {Rhodes}, {Smartt}, {Huber}, {Aryan}, {Yang}, {Beri}, {Cooper}, {Nicholl},
  {Smith}, {Stevance}, {Carotenuto}, {Chambers}, {Aamer}, {Angus}, {Fulton},
  {Moore}, {Smith}, {Young}, {de Boer}, {Gao}, {Lin}, {Lowe}, {Magnier},
  {Minguez}, {Pan}, \& {Wainscoat}}]{2025_Srivastav_ApJ...978L..21S}
{Srivastav}, S., {Chen}, T.~W., {Gillanders}, J.~H., {et~al.} 2025, \apjl, 978,
  L21

\bibitem[{{Str{\"u}der} {et~al.}(2001{\natexlab{a}}){Str{\"u}der},
  {Aschenbach}, {Br{\"a}uninger}, {Drolshagen}, {Englhauser}, {Hartmann},
  {Hartner}, {Holl}, {Kemmer}, {Meidinger}, {St{\"u}big}, \&
  {Tr{\"u}mper}}]{2001_Struder_A&A...375L...5S}
{Str{\"u}der}, L., {Aschenbach}, B., {Br{\"a}uninger}, H., {et~al.}
  2001{\natexlab{a}}, \aap, 375, L5

\bibitem[{{Str{\"u}der} {et~al.}(2001{\natexlab{b}}){Str{\"u}der}, {Briel},
  {Dennerl}, {Hartmann}, {Kendziorra}, {Meidinger}, {Pfeffermann}, {Reppin},
  {Aschenbach}, {Bornemann}, {Br{\"a}uninger}, {Burkert}, {Elender},
  {Freyberg}, {Haberl}, {Hartner}, {Heuschmann}, {Hippmann}, {Kastelic},
  {Kemmer}, {Kettenring}, {Kink}, {Krause}, {M{\"u}ller}, {Oppitz}, {Pietsch},
  {Popp}, {Predehl}, {Read}, {Stephan}, {St{\"o}tter}, {Tr{\"u}mper}, {Holl},
  {Kemmer}, {Soltau}, {St{\"o}tter}, {Weber}, {Weichert}, {von Zanthier},
  {Carathanassis}, {Lutz}, {Richter}, {Solc}, {B{\"o}ttcher}, {Kuster},
  {Staubert}, {Abbey}, {Holland}, {Turner}, {Balasini}, {Bignami}, {La
  Palombara}, {Villa}, {Buttler}, {Gianini}, {Lain{\'e}}, {Lumb}, \&
  {Dhez}}]{2001_Struder_A&A...365L..18S}
{Str{\"u}der}, L., {Briel}, U., {Dennerl}, K., {et~al.} 2001{\natexlab{b}},
  \aap, 365, L18

\bibitem[{{Suleimanov} {et~al.}(2022){Suleimanov}, {Doroshenko}, \&
  {Werner}}]{2022_Suleimanov_MNRAS.511.4937S}
{Suleimanov}, V.~F., {Doroshenko}, V., \& {Werner}, K. 2022, \mnras, 511, 4937

\bibitem[{{Sun} {et~al.}(2022){Sun}, {Liu}, {Pan}, {Liu}, {Alp}, {Hu}, {Li},
  {Zhang}, \& {Yuan}}]{2022_Sun_ApJ...927..224S}
{Sun}, H., {Liu}, H.-Y., {Pan}, H.-W., {et~al.} 2022, \apj, 927, 224

\bibitem[{{Troja} {et~al.}(2017){Troja}, {Piro}, {van Eerten}, {Wollaeger},
  {Im}, {Fox}, {Butler}, {Cenko}, {Sakamoto}, {Fryer}, {Ricci}, {Lien}, {Ryan},
  {Korobkin}, {Lee}, {Burgess}, {Lee}, {Watson}, {Choi}, {Covino}, {D'Avanzo},
  {Fontes}, {Gonz{\'a}lez}, {Khandrika}, {Kim}, {Kim}, {Lee}, {Lee}, {Kutyrev},
  {Lim}, {S{\'a}nchez-Ram{\'\i}rez}, {Veilleux}, {Wieringa}, \&
  {Yoon}}]{2017_Troja_Natur.551...71T}
{Troja}, E., {Piro}, L., {van Eerten}, H., {et~al.} 2017, \nat, 551, 71

\bibitem[{{Turner} {et~al.}(2001){Turner}, {Abbey}, {Arnaud}, {Balasini},
  {Barbera}, {Belsole}, {Bennie}, {Bernard}, {Bignami}, {Boer}, {Briel},
  {Butler}, {Cara}, {Chabaud}, {Cole}, {Collura}, {Conte}, {Cros}, {Denby},
  {Dhez}, {Di Coco}, {Dowson}, {Ferrando}, {Ghizzardi}, {Gianotti}, {Goodall},
  {Gretton}, {Griffiths}, {Hainaut}, {Hochedez}, {Holland}, {Jourdain},
  {Kendziorra}, {Lagostina}, {Laine}, {La Palombara}, {Lortholary}, {Lumb},
  {Marty}, {Molendi}, {Pigot}, {Poindron}, {Pounds}, {Reeves}, {Reppin},
  {Rothenflug}, {Salvetat}, {Sauvageot}, {Schmitt}, {Sembay}, {Short},
  {Spragg}, {Stephen}, {Str{\"u}der}, {Tiengo}, {Trifoglio}, {Tr{\"u}mper},
  {Vercellone}, {Vigroux}, {Villa}, {Ward}, {Whitehead}, \&
  {Zonca}}]{2001_Turner_A&A...365L..27T}
{Turner}, M.~J.~L., {Abbey}, A., {Arnaud}, M., {et~al.} 2001, \aap, 365, L27

\bibitem[{{Turolla} {et~al.}(2015){Turolla}, {Zane}, \&
  {Watts}}]{2015_Turolla_RPPh...78k6901T}
{Turolla}, R., {Zane}, S., \& {Watts}, A.~L. 2015, Reports on Progress in
  Physics, 78, 116901

\bibitem[{{van der Klis}(2004)}]{2004_van_der_klis_astro.ph.10551V}
{van der Klis}, M. 2004, arXiv e-prints, astro

\bibitem[{{Warner}(1995)}]{1995_Warner_cvs..book.....W}
{Warner}, B. 1995, {Cataclysmic variable stars}, Vol.~28

\bibitem[{{Watson} {et~al.}(2009){Watson}, {Schr{\"o}der}, {Fyfe}, {Page},
  {Lamer}, {Mateos}, {Pye}, {Sakano}, {Rosen}, {Ballet}, {Barcons}, {Barret},
  {Boller}, {Brunner}, {Brusa}, {Caccianiga}, {Carrera}, {Ceballos}, {Della
  Ceca}, {Denby}, {Denkinson}, {Dupuy}, {Farrell}, {Fraschetti}, {Freyberg},
  {Guillout}, {Hambaryan}, {Maccacaro}, {Mathiesen}, {McMahon}, {Michel},
  {Motch}, {Osborne}, {Page}, {Pakull}, {Pietsch}, {Saxton}, {Schwope},
  {Severgnini}, {Simpson}, {Sironi}, {Stewart}, {Stewart}, {Stobbart}, {Tedds},
  {Warwick}, {Webb}, {West}, {Worrall}, \&
  {Yuan}}]{2009_Watson_A&A...493..339W}
{Watson}, M.~G., {Schr{\"o}der}, A.~C., {Fyfe}, D., {et~al.} 2009, \aap, 493,
  339

\bibitem[{{Waxman} \& {Katz}(2017)}]{2017_Waxman_hsn..book..967W}
{Waxman}, E. \& {Katz}, B. 2017, in Handbook of Supernovae, ed. A.~W. {Alsabti}
  \& P.~{Murdin}, 967

\bibitem[{{Webb} {et~al.}(2023){Webb}, {Carrera}, {Schwope}, {Motch}, {Ballet},
  {Watson}, {Page}, {Freyberg}, {Georgantopoulos}, {Coriat}, {Barret},
  {Massida}, {Gupta}, {Tranin}, {Quintin}, {Teresa Ceballos}, {Mateos},
  {Corral}, {Dominguez}, {Stiele}, {Traulsen}, {Pires}, {Nebot}, {Michel},
  {Xavier Pineau}, {Kuuttila}, {Maggi}, {Chakroborty}, {Birchall}, {Kuin},
  {Akylas}, {Ruiz}, {Pouliasis}, \& {Georgakakis}}]{2023_Webb_AN....34420102W}
{Webb}, N.~A., {Carrera}, F.~J., {Schwope}, A., {et~al.} 2023, Astronomische
  Nachrichten, 344, e20220102

\bibitem[{{Webb} {et~al.}(2020){Webb}, {Coriat}, {Traulsen}, {Ballet}, {Motch},
  {Carrera}, {Koliopanos}, {Authier}, {de la Calle}, {Ceballos}, {Colomo},
  {Chuard}, {Freyberg}, {Garcia}, {Kolehmainen}, {Lamer}, {Lin}, {Maggi},
  {Michel}, {Page}, {Page}, {Perea-Calderon}, {Pineau}, {Rodriguez}, {Rosen},
  {Santos Lleo}, {Saxton}, {Schwope}, {Tom{\'a}s}, {Watson}, \&
  {Zakardjian}}]{2020_Webb_A&A...641A.136W}
{Webb}, N.~A., {Coriat}, M., {Traulsen}, I., {et~al.} 2020, \aap, 641, A136

\bibitem[{{Webbe} {et~al.}(2025){Webbe}, {Khan}, {Webb}, \&
  {Quintin}}]{2025_Webbe_MNRAS}
{Webbe}, R., {Khan}, N., {Webb}, N.~A., \& {Quintin}. 2025, \mnras, TBC, TBC

\bibitem[{{Webbe} \& {Young}(2023)}]{2023_Webbe_MNRAS.518.3428W}
{Webbe}, R. \& {Young}, A.~J. 2023, \mnras, 518, 3428

\bibitem[{{Wenger} {et~al.}(2000){Wenger}, {Ochsenbein}, {Egret}, {Dubois},
  {Bonnarel}, {Borde}, {Genova}, {Jasniewicz}, {Lalo{\"e}}, {Lesteven}, \&
  {Monier}}]{2000_Wenger_A&AS..143....9W}
{Wenger}, M., {Ochsenbein}, F., {Egret}, D., {et~al.} 2000, \aaps, 143, 9

\bibitem[{{Zhang} {et~al.}(2024){Zhang}, {Xiong}, {Gao}, {Yang}, {Lu}, {Na}, \&
  {Qin}}]{2024_Zhang_MNRAS.529.3699Z}
{Zhang}, X., {Xiong}, D.-r., {Gao}, Q.-g., {et~al.} 2024, \mnras, 529, 3699

\end{thebibliography}
\appendix
\section{Parameter estimation}\label{app:ParamEstimate}
Let us assume that a pixel has been flagged as variable, and we want to estimate the properties of the associated burst or eclipse. The idea is that we will compute an upper limit value $UL$, which is to be larger than the peak amplitude at a given confidence level.
Let us for instance consider the median - in this case, the peak amplitude is below $UL$ at a 50\% confidence level. Computing the corresponding $UL$ for 16\% and 84\% allows building a $1\sigma$ confidence interval for the peak. In order to compute these $UL$ values, we integrate the Poisson statistics up to $UL$, so that we recovered the intended fraction of the total integral over all possible peak amplitudes:
\begin{equation}
    \int_{0}^{UL}P(N{\rm |Peak}~s)P(s)ds = {\rm f}\times\int_{0}^{\infty} P(N{\rm|Peak}~s)P(s)ds
\end{equation}
where $f$ is the intended fraction. Assuming Poisson statistics, an expected rate of $\mu$ and a flat prior for the peak amplitudes,
\begin{equation}
        \int_{0}^{UL}P({N~|}~s,\mu)ds = {\rm f}\times\int_{0}^{\infty} P({ N~|}~s,\mu)ds
\end{equation}
\begin{equation}
        \int_{0}^{UL}e^{-(\mu+s)}\frac{(\mu+s)^{N}}{N!} ds = {\rm f}\times\int_{0}^{\infty} e^{-(\mu+s)}\frac{(\mu+s)^{N}}{N!}ds
\end{equation}
\begin{equation}
        \int_{\mu}^{UL+\mu}e^{-s'}s'^{N}ds = {\rm f}\times\int_{\mu}^{\infty} e^{-s'}s'^{N}ds
\end{equation}
\begin{equation}
        \int_{0}^{UL+\mu}e^{-s'}s'^{N}ds-\int_{0}^{\mu}e^{-s'}s'^{N}ds = {\rm f}\times\int_{\mu}^{\infty} e^{-s'}s'^{N}ds
\end{equation}
\begin{equation}
        \Gamma_{lower}(N+1,UL+\mu)-\Gamma_{lower}(N+1,\mu) = {\rm f}\times\Gamma_{upper}(N+1,\mu)
\end{equation}
\begin{equation}
        \mathcal{P}(N+1,UL+\mu)-\mathcal{P}(N+1,\mu) = {\rm f}\times\mathcal{Q}(N+1,\mu)
\end{equation}
\begin{equation}
        \mathcal{P}(N+1,UL+\mu)=\mathcal{P}(N+1,\mu)+{\rm f}\times\mathcal{Q}(N+1,\mu)
\end{equation}
\begin{equation}
\label{estimate}
        UL+\mu=\mathcal{P}^{-1}\left[N+1,\mathcal{P}(N+1,\mu)+{\rm f}\times\mathcal{Q}(N+1,\mu)\right]
\end{equation}
with $\mathcal{Q}$ and $\mathcal{P}$ the regularized upper and lower incomplete gamma function. By using Eq. \ref{estimate} and taking $f=\{0.16,0.5,0.84\}$, we can retrieve the 1$\sigma$ lower and upper limits as well as the median of the estimated flux. This only requires using the expected and observed counts, $\mu$ and $N$ respectively.

It works similarly for eclipses, the main difference being that all integrals are on a finite support (between 0 and the expected rate $\mu$):

\begin{equation}
        \int_{\mu-UL}^{\mu}e^{-s'}s'^{N}ds = {\rm f}\times\int_{0}^{\mu} e^{-s'}s'^{N}ds
\end{equation}
\begin{equation}
        \Gamma_{lower}(N+1,\mu)-\Gamma_{lower}(N+1,\mu-UL) = {\rm f}\times\Gamma_{lower}(N+1,\mu)
\end{equation}
\begin{equation}
        \mathcal{P}(N+1,\mu-UL) = \mathcal{P}(N+1,\mu)-{\rm f}\times\mathcal{P}(N+1,\mu)
\end{equation}
\begin{equation}
        \mu-UL = \mathcal{P}^{-1}\left[N+1,\mathcal{P}(N+1,\mu)-{\rm f}\times\mathcal{P}(N+1,\mu)\right]
\end{equation}
\\
\\

\section{Additional benchmarking statistics}
\begin{figure}[h]
    \includegraphics[width=0.95\columnwidth]{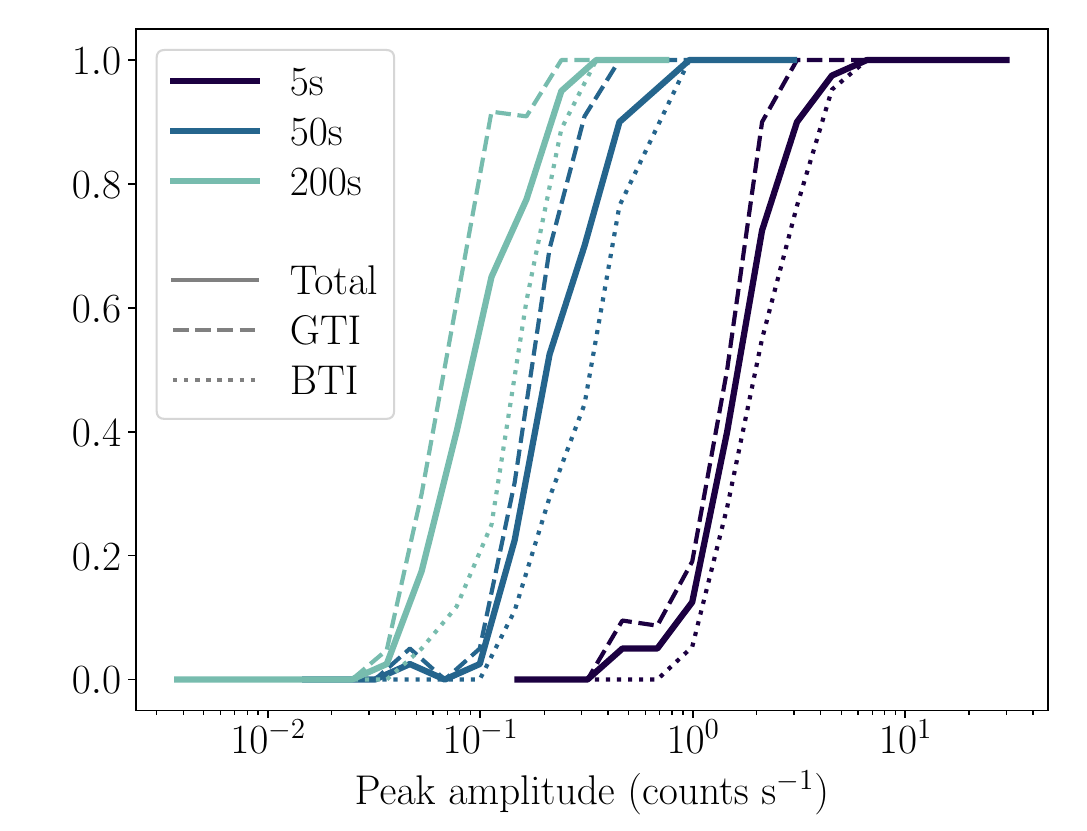}
    \caption{Detection fraction at $5\sigma$ for GTI, BTI and total frames.}
    \label{fig:benchmark_GTIvBTI}
\end{figure}

\begin{figure}[h]
    \includegraphics[width=0.95\columnwidth]{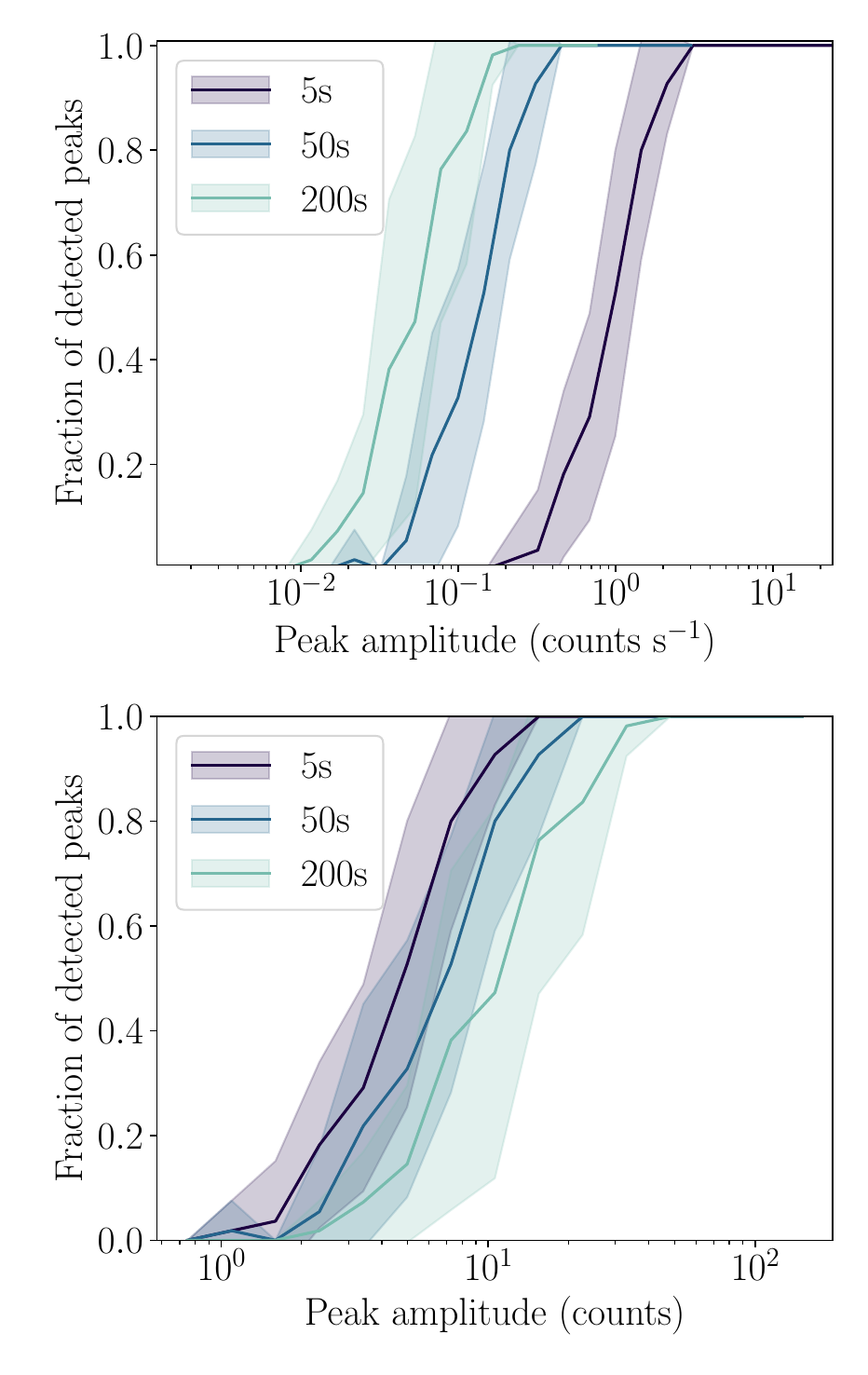}
    \caption{Detection Fraction for different time binnings evaluated at the \tSig sensitivity.}
    \label{fig:benchmark_3sig}
\end{figure}

\onecolumn
\begin{figure*}[t]
    \includegraphics[width=1.0\textwidth]{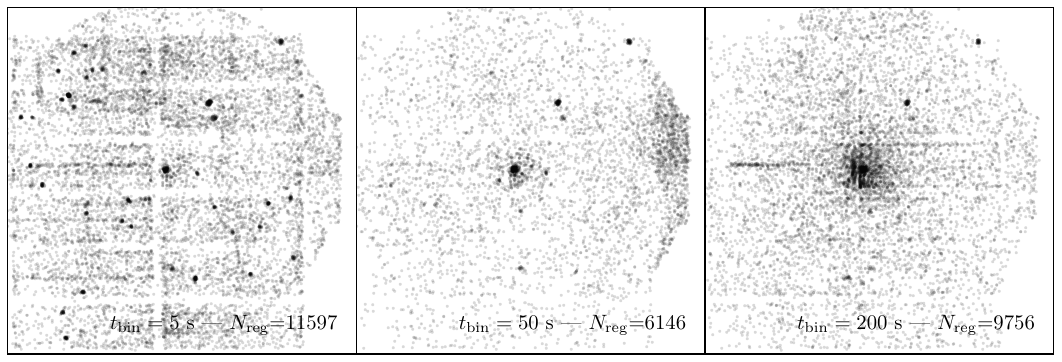}
    \caption{Spatial distribution of transient detections \tSig in the 0.2-12.0 keV Energy band in detector coordinates. The three panels correspond to the
    different time binning, 5s, 50s and 200s.
    A number of areas of over-density can be seen, detections located in these areas are flagged.}
    \label{fig:spatial_dist}
\end{figure*}

\begin{figure*}[t]
    \centering
    \includegraphics[width=1.0\textwidth, , trim=100 0 100 35, clip]{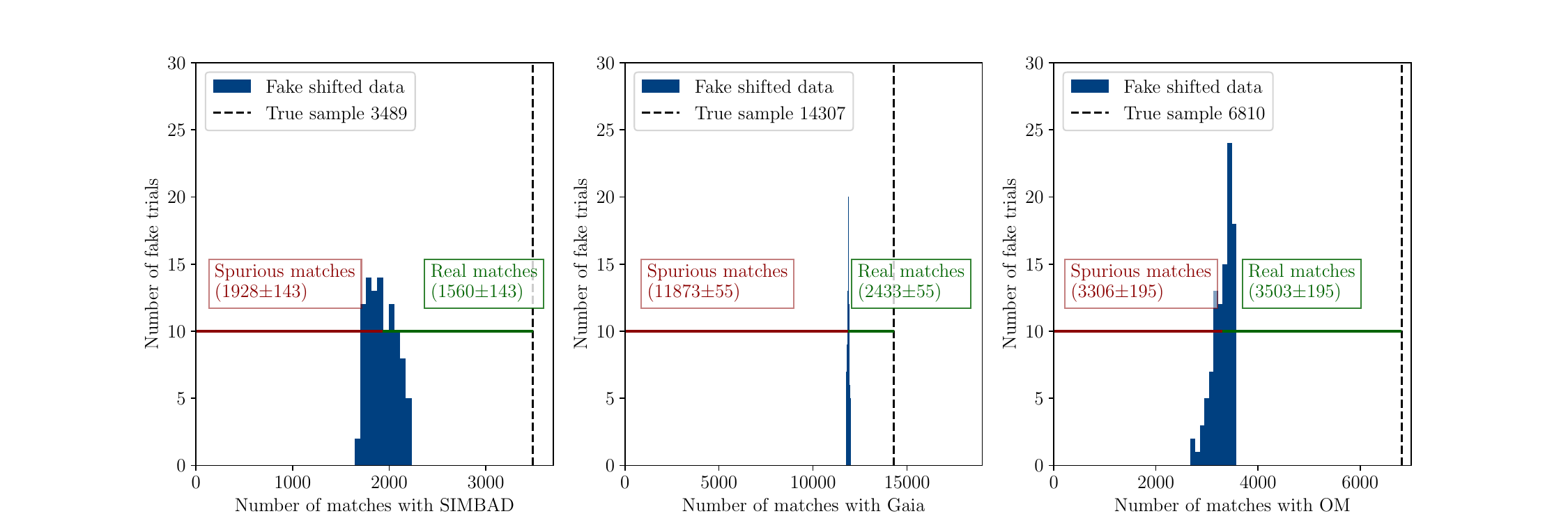}
    \caption{Estimation of the fraction of spurious associations between new EXOD sources (not in DR14) and the SIMBAD, Gaia and OM catalogues. In each panel, the blue histogram corresponds to the results of the Monte Carlo shifted cross-match method, and the black dotted line to the number of associations in the true dataset. The red and green lines show the numbers of spurious matches and real matches, respectively.}
    \label{fig:Crossmatch_Shift}
\end{figure*}

\section{Spectral fits for ``Tornado'' burst} \label{sec:tornado_burst_fits}
\begin{longtable}{@{}llccrrr@{}}
\hline
\textbf{Model}        & \textbf{Component} & \textbf{Parameter} & \textbf{Unit} & \textbf{Value} & \textbf{Error} & \textbf{$\chi^2$/DoF \& P-value} \\ 
\hline
\endhead
TBabs*Powerlaw        & TBabs         & $n_H$         & $10^{22}$    & 16.77          & $\pm 3.96$     & \multirow{1.5}{*}{50.70/36 $\sim$ 1.41} \\ 
                      & Powerlaw      & $\Gamma$      &              & 2.60           & $\pm 0.80$     & \multirow{1.5}{*}{0.0251}\\ 
                      & Powerlaw      & norm          &              & $0.155$        & $\pm 0.222$    & \\ 
\hline
TBabs*TBabs*Powerlaw  & TBabs(1)      & $n_H$         & $10^{22}$    & 1.30           & frozen         & \multirow{2}{*}{50.69/36 $\sim$ 1.41} \\ 
                      & TBabs(2)      & $n_H$         & $10^{22}$    & 15.45          & $\pm 3.95$     & \multirow{2}{*}{0.0252}\\ 
                      & Powerlaw      & $\Gamma$      &              & 2.59           & $\pm 0.80$     & \\ 
                      & Powerlaw      & norm          &              & $0.154$        & $\pm 0.220$    & \\ 
\hline
TBabs*Raymond         & TBabs         & $n_H$         & $10^{22}$    & 15.06          & $\pm 2.88$     & \multirow{2.5}{*}{51.09/36 $\sim$ 1.42} \\ 
                      & Raymond       & $kT$          & keV          & 4.09           & $\pm 2.08$     & \multirow{2.5}{*}{0.0231}\\ 
                      & Raymond       & Abundance     &              & 0.30           & frozen         & \\ 
                      & Raymond       & Redshift      &              & 0.00           & frozen         & \\ 
                      & Raymond       & norm          &              & $0.185$        & $\pm 0.112$    & \\ 
\hline
TBabs*Apec            & TBabs         & $n_H$         & $10^{22}$    & 15.28          & $\pm 2.96$     & \multirow{2.5}{*}{51.46/36 $\sim$ 1.43} \\ 
                      & Apec          & $kT$          & keV          & 3.87           & $\pm 1.90$     & \multirow{2.5}{*}{0.0213}\\ 
                      & Apec          & Abundance     &              & 0.30           & frozen         & \\ 
                      & Apec          & Redshift      &              & 0.00           & frozen         & \\ 
                      & Apec          & norm          &              & $0.193$        & $\pm 0.123$    & \\ 
\hline
\label{tab:tornado_fits}
\end{longtable}

\onecolumn
\section{Diagram of the creation of the expectation emission}
 \begin{figure*}[h]
     \centering
     \includegraphics[width=0.8\textwidth]{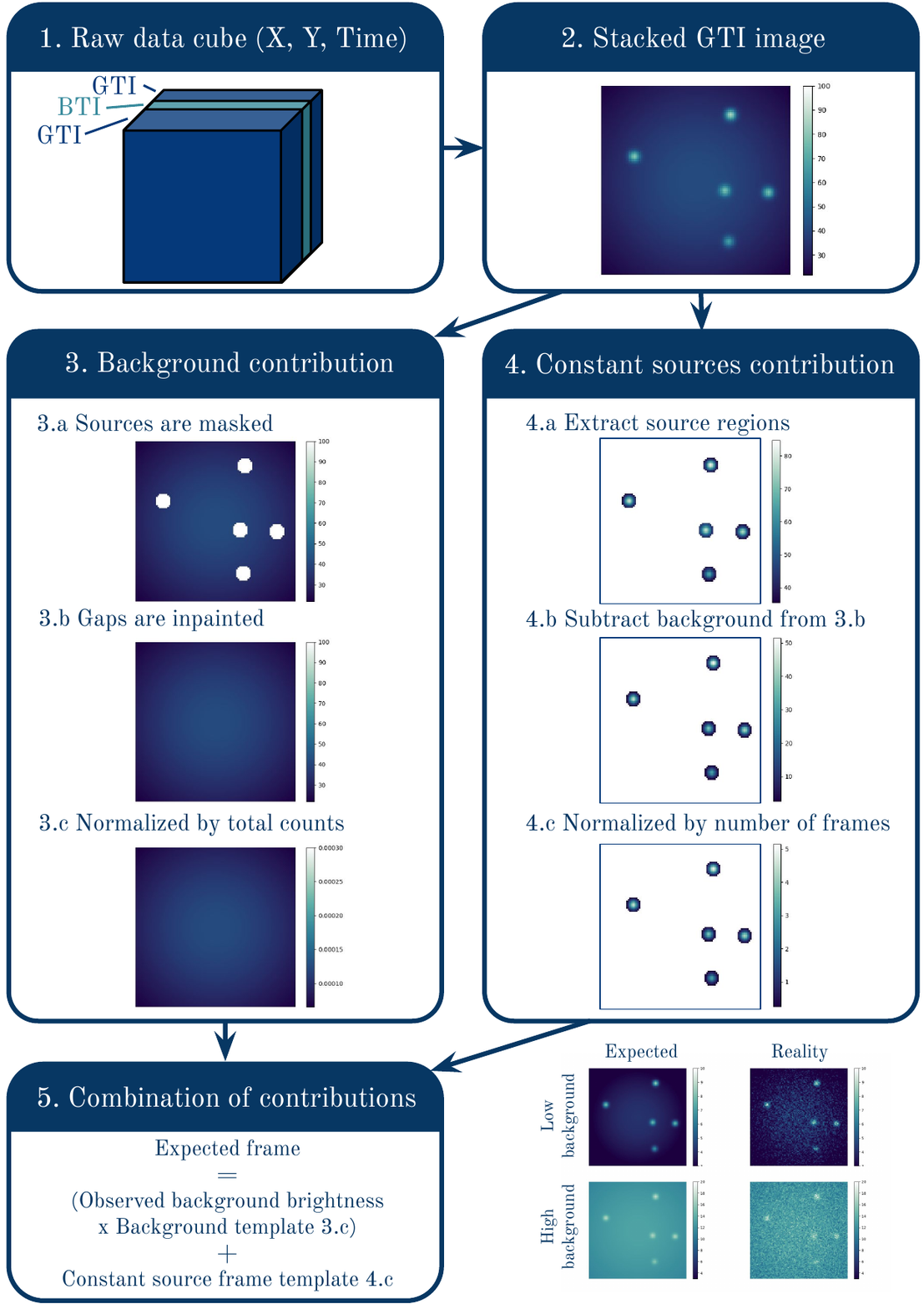}
     \caption{Diagram of the steps used to estimate the expectation
     value for each cell of the data cube, for details see sec. \ref{sec:expectation_cube_creation}}
     \label{fig:template_creation_diagram}
 \end{figure*}
\newpage
 \section{Breakdown of SIMBAD crossmatch results}
 \begin{table*}[h]
 \caption{Source classifications numbering 15 or more obtained by crossmatching
    sources in the full EXOD \tSig catalogue with SIMBAD.}
    \centering
    \resizebox{0.97\textwidth}{!}{%
    \begin{tabular}{rrrrrrr}
    \hline
    Object Type & EXOD \tSig  & DR14 (var) & DR14    & EXOD $\cap$ DR14 ($\neg$ var) & EXOD $\cap$ DR14 (var) & SIMBAD \\
          & 32,247 & 6,311      & 692,109       & 5,910                         & 2,568                  & 18,815,217 \\    
    \hline
                       No Crossmatch &  24,444 (75.8\%) &       2,992 (47.41\%) &  536,542 (77.52\%) &       3,449 (58.36\%) &          665 (25.9\%) &            0 (0.0\%) \\
                                Star &   1,235 (3.83\%) &          582 (9.22\%) &    22,383 (3.23\%) &          379 (6.41\%) &          285 (11.1\%) &  5,233,041 (27.81\%) \\
                              Galaxy &   1,210 (3.75\%) &           54 (0.86\%) &     14,552 (2.1\%) &          284 (4.81\%) &           33 (1.29\%) &  3,595,127 (19.11\%) \\
                        X-ray Source &     525 (1.63\%) &          318 (5.04\%) &    20,340 (2.94\%) &          228 (3.86\%) &           190 (7.4\%) &     160,117 (0.85\%) \\
Galaxy towards a Cluster of Galaxies &     319 (0.99\%) &           11 (0.17\%) &     2,699 (0.39\%) &            65 (1.1\%) &            1 (0.04\%) &     708,968 (3.77\%) \\
                    Eclipsing Binary &     245 (0.76\%) &           92 (1.46\%) &     6,403 (0.93\%) &            47 (0.8\%) &           48 (1.87\%) &   1,657,637 (8.81\%) \\
      Long-Period Variable Candidate &     243 (0.75\%) &           22 (0.35\%) &     3,056 (0.44\%) &           46 (0.78\%) &           12 (0.47\%) &   1,225,960 (6.52\%) \\
                    Seyfert 1 Galaxy &      225 (0.7\%) &          166 (2.63\%) &     2,117 (0.31\%) &          121 (2.05\%) &          103 (4.01\%) &      32,946 (0.18\%) \\
                Long-Period Variable &     222 (0.69\%) &           21 (0.33\%) &     2,152 (0.31\%) &           60 (1.02\%) &           16 (0.62\%) &      546,022 (2.9\%) \\
             High Proper Motion Star &     219 (0.68\%) &          204 (3.23\%) &     2,291 (0.33\%) &            59 (1.0\%) &          132 (5.14\%) &     368,565 (1.96\%) \\
      Young Stellar Object Candidate &     167 (0.52\%) &          118 (1.87\%) &     3,144 (0.45\%) &           57 (0.96\%) &           42 (1.64\%) &     184,040 (0.98\%) \\
                Young Stellar Object &      161 (0.5\%) &          245 (3.88\%) &      3,469 (0.5\%) &           41 (0.69\%) &          103 (4.01\%) &      58,532 (0.31\%) \\
                        Radio Source &     140 (0.43\%) &           14 (0.22\%) &     3,825 (0.55\%) &           51 (0.86\%) &            23 (0.9\%) &      639,367 (3.4\%) \\
                  Cataclysmic Binary &     137 (0.42\%) &          167 (2.65\%) &       373 (0.05\%) &            8 (0.14\%) &          129 (5.02\%) &       4,266 (0.02\%) \\
                       Part of Cloud &     126 (0.39\%) &           21 (0.33\%) &        92 (0.01\%) &           55 (0.93\%) &           44 (1.71\%) &       4,623 (0.02\%) \\
                    Globular Cluster &     117 (0.36\%) &           20 (0.32\%) &       599 (0.09\%) &           55 (0.93\%) &           11 (0.43\%) &      22,741 (0.12\%) \\
                              Quasar &     113 (0.35\%) &           22 (0.35\%) &    30,944 (4.47\%) &           96 (1.62\%) &            4 (0.16\%) &     804,018 (4.27\%) \\
     Red Giant Branch star Candidate &     102 (0.32\%) &            3 (0.05\%) &       464 (0.07\%) &           10 (0.17\%) &            1 (0.04\%) &     104,910 (0.56\%) \\
              High Mass X-ray Binary &      77 (0.24\%) &           145 (2.3\%) &       610 (0.09\%) &            4 (0.07\%) &           69 (2.69\%) &       1,651 (0.01\%) \\
                    Seyfert 2 Galaxy &      72 (0.22\%) &           37 (0.59\%) &     1,013 (0.15\%) &           38 (0.64\%) &           32 (1.25\%) &      25,306 (0.13\%) \\
              Sub-Millimetric Source &      71 (0.22\%) &            8 (0.13\%) &       366 (0.05\%) &           20 (0.34\%) &           10 (0.39\%) &      28,849 (0.15\%) \\
                      Orion Variable &      69 (0.21\%) &          121 (1.92\%) &       872 (0.13\%) &           10 (0.17\%) &           58 (2.26\%) &       3,098 (0.02\%) \\
                Spectroscopic Binary &      69 (0.21\%) &           58 (0.92\%) &     1,232 (0.18\%) &           16 (0.27\%) &           43 (1.67\%) &     181,697 (0.97\%) \\
                   RR Lyrae Variable &      67 (0.21\%) &             6 (0.1\%) &        709 (0.1\%) &           15 (0.25\%) &            3 (0.12\%) &     322,345 (1.71\%) \\
          Globular Cluster Candidate &       65 (0.2\%) &             6 (0.1\%) &       253 (0.04\%) &           22 (0.37\%) &            7 (0.27\%) &      17,494 (0.09\%) \\
                          HII Region &       63 (0.2\%) &           23 (0.36\%) &       405 (0.06\%) &           25 (0.42\%) &           11 (0.43\%) &      45,024 (0.24\%) \\
               Active Galaxy Nucleus &       63 (0.2\%) &           13 (0.21\%) &     7,017 (1.01\%) &           39 (0.66\%) &            3 (0.12\%) &      48,387 (0.26\%) \\
                        T Tauri Star &      61 (0.19\%) &          138 (2.19\%) &      1,352 (0.2\%) &            9 (0.15\%) &           51 (1.99\%) &       5,751 (0.03\%) \\
                    Cluster of Stars &      56 (0.17\%) &            9 (0.14\%) &       265 (0.04\%) &            12 (0.2\%) &           14 (0.55\%) &      17,336 (0.09\%) \\
                 Cluster of Galaxies &      55 (0.17\%) &            7 (0.11\%) &     1,201 (0.17\%) &           32 (0.54\%) &            5 (0.19\%) &     118,879 (0.63\%) \\
            Red Supergiant Candidate &      51 (0.16\%) &             0 (0.0\%) &       161 (0.02\%) &           19 (0.32\%) &            4 (0.16\%) &      10,128 (0.05\%) \\
                       Variable Star &      50 (0.16\%) &            19 (0.3\%) &       812 (0.12\%) &           15 (0.25\%) &           12 (0.47\%) &      98,497 (0.52\%) \\
                   Eruptive Variable &      50 (0.16\%) &           68 (1.08\%) &       367 (0.05\%) &           16 (0.27\%) &           34 (1.32\%) &       8,034 (0.04\%) \\
                     Molecular Cloud &      47 (0.15\%) &             6 (0.1\%) &       257 (0.04\%) &           15 (0.25\%) &            6 (0.23\%) &      49,961 (0.27\%) \\
 Brightest Galaxy in a Cluster (BCG) &      46 (0.14\%) &           12 (0.19\%) &     1,295 (0.19\%) &           31 (0.52\%) &           10 (0.39\%) &     165,395 (0.88\%) \\
                    Planetary Nebula &      45 (0.14\%) &            3 (0.05\%) &       203 (0.03\%) &           20 (0.34\%) &            1 (0.04\%) &      12,936 (0.07\%) \\
                     BY Dra Variable &      42 (0.13\%) &           59 (0.93\%) &     1,100 (0.16\%) &           11 (0.19\%) &           29 (1.13\%) &      85,187 (0.45\%) \\
               Low Mass X-ray Binary &      40 (0.12\%) &           51 (0.81\%) &       124 (0.02\%) &           11 (0.19\%) &           29 (1.13\%) &          657 (0.0\%) \\
                              BL Lac &      40 (0.12\%) &           16 (0.25\%) &       293 (0.04\%) &           25 (0.42\%) &           15 (0.58\%) &       5,694 (0.03\%) \\
                       Low-mass Star &      39 (0.12\%) &            38 (0.6\%) &       401 (0.06\%) &            9 (0.15\%) &           26 (1.01\%) &      80,521 (0.43\%) \\
     Active Galaxy Nucleus Candidate &      36 (0.11\%) &           11 (0.17\%) &     3,182 (0.46\%) &           24 (0.41\%) &            6 (0.23\%) &      77,240 (0.41\%) \\
                    Infra-Red Source &      36 (0.11\%) &            7 (0.11\%) &       361 (0.05\%) &            8 (0.14\%) &           16 (0.62\%) &      61,550 (0.33\%) \\
                      Classical Nova &      36 (0.11\%) &           23 (0.36\%) &       180 (0.03\%) &           19 (0.32\%) &            9 (0.35\%) &       2,041 (0.01\%) \\
          Planetary Nebula Candidate &      35 (0.11\%) &            4 (0.06\%) &       116 (0.02\%) &           15 (0.25\%) &            1 (0.04\%) &       6,512 (0.03\%) \\
               Red Giant Branch star &       32 (0.1\%) &            2 (0.03\%) &       177 (0.03\%) &             6 (0.1\%) &            4 (0.16\%) &      92,762 (0.49\%) \\
             Double or Multiple Star &      30 (0.09\%) &           23 (0.36\%) &       338 (0.05\%) &            7 (0.12\%) &           21 (0.82\%) &      65,056 (0.35\%) \\
                   Group of Galaxies &      29 (0.09\%) &            1 (0.02\%) &       287 (0.04\%) &           10 (0.17\%) &            4 (0.16\%) &      35,281 (0.19\%) \\
                  Emission-line Star &      28 (0.09\%) &           14 (0.22\%) &       637 (0.09\%) &            5 (0.08\%) &           10 (0.39\%) &      40,197 (0.21\%) \\
               White Dwarf Candidate &      26 (0.08\%) &             0 (0.0\%) &       191 (0.03\%) &            2 (0.03\%) &             0 (0.0\%) &     428,531 (2.28\%) \\
                     RS CVn Variable &      25 (0.08\%) &           35 (0.55\%) &       918 (0.13\%) &            8 (0.14\%) &           12 (0.47\%) &      83,187 (0.44\%) \\
                Emission-line galaxy &      25 (0.08\%) &            4 (0.06\%) &       960 (0.14\%) &            7 (0.12\%) &             0 (0.0\%) &      84,058 (0.45\%) \\
                        Radio Galaxy &      24 (0.07\%) &             0 (0.0\%) &       501 (0.07\%) &            12 (0.2\%) &             0 (0.0\%) &      21,335 (0.11\%) \\
                   SuperNova Remnant &      24 (0.07\%) &            8 (0.13\%) &       141 (0.02\%) &           10 (0.17\%) &            9 (0.35\%) &       1,934 (0.01\%) \\
                  Pulsating Variable &      23 (0.07\%) &             6 (0.1\%) &       184 (0.03\%) &            3 (0.05\%) &            5 (0.19\%) &      52,959 (0.28\%) \\
       Low Surface Brightness Galaxy &      23 (0.07\%) &            1 (0.02\%) &        92 (0.01\%) &            5 (0.08\%) &             0 (0.0\%) &      13,262 (0.07\%) \\
                              Pulsar &      23 (0.07\%) &           11 (0.17\%) &       169 (0.02\%) &           15 (0.25\%) &            7 (0.27\%) &       3,739 (0.02\%) \\
                        X-ray Binary &      22 (0.07\%) &            19 (0.3\%) &       178 (0.03\%) &            9 (0.15\%) &            9 (0.35\%) &       1,406 (0.01\%) \\
                       Mira Variable &      20 (0.06\%) &            5 (0.08\%) &       239 (0.03\%) &             6 (0.1\%) &            3 (0.12\%) &      86,789 (0.46\%) \\
                 Dark Cloud (nebula) &      20 (0.06\%) &            2 (0.03\%) &        64 (0.01\%) &             6 (0.1\%) &            1 (0.04\%) &      27,968 (0.15\%) \\
                          Dense Core &      18 (0.06\%) &            1 (0.02\%) &        97 (0.01\%) &            2 (0.03\%) &            5 (0.19\%) &      15,961 (0.08\%) \\
          Classical Cepheid Variable &      18 (0.06\%) &            2 (0.03\%) &       122 (0.02\%) &            4 (0.07\%) &            2 (0.08\%) &      17,552 (0.09\%) \\
                     Near-IR Source  &      18 (0.06\%) &           15 (0.24\%) &       153 (0.02\%) &            4 (0.07\%) &           10 (0.39\%) &      21,651 (0.12\%) \\
                Delta Scuti Variable &      16 (0.05\%) &            5 (0.08\%) &       109 (0.02\%) &            3 (0.05\%) &            5 (0.19\%) &      51,687 (0.27\%) \\
            Millimetric Radio Source &      16 (0.05\%) &            3 (0.05\%) &       112 (0.02\%) &            3 (0.05\%) &            2 (0.08\%) &      10,488 (0.06\%) \\
                           SuperNova &      15 (0.05\%) &            3 (0.05\%) &       125 (0.02\%) &            7 (0.12\%) &            3 (0.12\%) &       19,261 (0.1\%) \\
    \hline
    \end{tabular}%
    }
    \tablefoot{
    \textbf{EXOD}       : All Unique regions detected by EXOD. \\
    \textbf{DR14 (var)} : Variable sources in 4XMM-DR14. \\
    \textbf{DR14}       : All Sources in 4XMM-DR14. \\
    \textbf{EXOD $\cap$ DR14 ($\neg$ var)} : Sources detected in EXOD and are not variable in XMM DR14. (i.e. newly variable sources discovered with EXOD)\\
    \textbf{EXOD $\cap$ DR14 (var)} : Sources detected in EXOD and are variable in XMM DR14. \\
    \textbf{SIMBAD} : Total classifications in SIMBAD database.\\
    }
    \label{tab:simbad_cmatch}
\end{table*}

\end{document}